\documentclass[reqno,12pt]{article}
\usepackage{amsmath}
\usepackage{bm}
\usepackage{a4wide,amssymb}
\usepackage{graphicx,xcolor}
\usepackage{epstopdf}
\usepackage{bbm}

\newcommand{\qed}{ $\blacksquare$} \newcommand{\nn}{\nonumber}
\newcommand{\R}{{\mathbb R}}

\renewcommand{\a}{\alpha}

\newcommand{\g}{\gamma}\renewcommand{\d}{\delta}
           
\newcommand\om{\omega}

\newcommand{\e}{\varepsilon}

\newcommand{\pa}{\partial}

\newcommand{\z}{\zeta}
\newcommand{\la}{\lambda}

\newcommand{\Lam}{\Lambda}
\def\G{\Gamma}
\def\s{\sigma}
\def\r{{\rho}}
\def\n{\nu}
\def\p{{\pi}}
\def\f{{\varphi}}

\def\bz{{\bf z}}
\def\bx{{\bf x}}
\def\bv{{\bf v}}
\def\by{{\bf y}}

\def\bl{{\bm \lambda}}

\def\bn{{\bm \n}}
\def\bt{{\bf t}}
\def\bs{{\bm \s}}
\def\bk{{\bf k}}
\def\bze{{\bm \zeta}}
\def\bxi{{\bm \xi}}
\def\bet{{\bm \eta}}
\def\bim{\dot{\bm\imath}}
\def\T{{\mathcal T}}
\def\L{{\Lambda}}
\def\II{{\mathcal I}}

\def\RRR{{\mathbb R}}
\def\NN{{\cal N}}
\def\O{\mathit{\Omega}}

\def\MM{{\cal M}}
\def\AA{{\cal A}}
\def\LL{{\cal L}}
\def\CC{{\cal C}}
\def\SS{{\cal S}}
\def\TT{\mathtt{T}}
\def\ZZ{{\cal Z}}
\def\GG{{\cal G}}

\newtheorem{lem}{Lemma}
\newtheorem{prop}{Proposition}
\newtheorem{thm}{Theorem}
\newtheorem{defi}{Definition}

\newtheorem{hyp}{Hypothesis}

\def\be{\begin{equation}}
\def\ee{\end{equation}}
\def\bea{\begin{eqnarray}}
\def\eea{\end{eqnarray}}
\def\ni{\noindent}
\def\nn{\nonumber}

\def\ol{\overline}

\def\d{\delta}
\def\o{\omega}
\def\b{\beta}
\def\t{\tau}

\begin{document}

\begin{titlepage}

\begin{center} 
{\bf \Large{On the validity of the Boltzmann equation \\ for short range potentials}}

\vspace{1cm}
{\large M. Pulvirenti$^{\dagger 1}$, C. Saffirio$^{\dagger 2}$ and S. Simonella$^{\dagger 3}$}

\vspace{0.5cm}
{$\dagger.$\scshape {\small \ Dipartimento di Matematica, Universit\`a di Roma La Sapienza\\ 
Piazzale Aldo Moro 5, 00185 Roma -- Italy \\ \smallskip
$1.$\ International Research Center for Mathematics and Mechanics \\Êof Complex
Systems MEMOCS, Universit\`a di L'Aquila \\ Cisterna di Latina, 04012, Italy \\ \smallskip
$2.$\ Institute of Applied Mathematics, University of Bonn \\ Endenicher Allee 60, 53115 Bonn --Germany\\\smallskip
$3.$\ Zentrum Mathematik, TU M\"{u}nchen \\ Boltzmannstrasse 3, 85748 Garching -- Germany}}
\end{center}

\vspace{0.5cm}
ABSTRACT. We consider a classical system of point particles interacting by
means of a short range potential. We prove that, in the low--density
(Boltzmann--Grad) limit, the system behaves, for short times, as predicted
by the associated Boltzmann equation.
This is a revisitation and an extension of the thesis of King \cite{Ki75}
(appeared after the well known result of Lanford \cite{La75} for hard spheres) and of a
recent paper by Gallagher et al \cite {GSRT12}. Our analysis applies to any stable and smooth potential.
In the case of repulsive potentials (with no attractive parts), we estimate explicitly the rate of convergence.

\vspace{0.3cm}

\vspace{1cm}
\begin{center}
{\bf CONTENTS}
\end{center}

\vspace{4mm}

1.\ \ \ \ \ Introduction \dotfill \pageref{sec:intro}

\vspace{3mm}

2.\ \ \ \ \ The hamiltonian system \dotfill \pageref{sec:hs}

2.1\ \ \ \hspace{1mm}Statistical description \dotfill \pageref{sec:stat}

2.2\ \ \ \hspace{1mm}The two--body scattering \dotfill \pageref{sec:2bs}

\vspace{3mm}

3.\ \ \ \ \ The Grad hierarchy \dotfill \pageref{sec:Grad}

3.1\ \ \ \hspace{1mm}Series solution \dotfill \pageref{sec:series}

3.2\ \ \ \hspace{1mm}The Boltzmann hierarchy \dotfill \pageref{sec:BH}

\vspace{3mm}

4.\ \ \ \ \ Assumptions and results \dotfill \pageref{sec:results}

4.1\ \ \ \hspace{1mm}An example of initial datum \dotfill \pageref{sec:initial}

4.2\ \ \ \hspace{1mm}General strategy of the proof \dotfill \pageref{sec:strategy}

\vspace{3mm} 

5.\ \ \ \ \ Short time estimates \dotfill \pageref{sec:ste}

\vspace{3mm}

6.\ \ \ \ \ The tree expansion \dotfill \pageref{sec:tree}

6.1\ \ \ \hspace{1mm}The interacting backwards flow (IBF)\dotfill \pageref{sec:flow}

6.2\ \ \ \hspace{1mm}The Boltzmann backwards flow (BBF)\dotfill \pageref{sec:Bflow}

\vspace{3mm}

7.\ \ \ \ \ Proof of the results \dotfill \pageref{sec:proof}

7.1\ \ \ \hspace{1mm}The convergence problem: preliminary considerations \dotfill \pageref{sec:convergence}

7.2\ \ \ \hspace{1mm}Proof of Theorem \ref{thm:soft} \dotfill \pageref{sec:proofsoft}

7.3\ \ \ \hspace{1mm}Proof of Theorem \ref{thm:hard} \dotfill \pageref{sec:proofhard}

\vspace{3mm}

8.\ \ \ \ \ Stable potentials \dotfill \pageref{sec:gener}

\vspace{3mm}

9.\ \ \ \ \ Concluding remarks \dotfill \pageref{sec:conc}

\vspace{3mm}

A.\ \ \ \ Appendix (on the cross--section for the Boltzmann equation) \dotfill \pageref{sec:cs}

\vspace{3mm}

References \dotfill \pageref{sec:bib}

\thispagestyle{empty}
\end{titlepage}

%%%%%%%%%%%%%%%%%%%%%%%%%%%%%%%%%%%%%%%%%%%%%%%%%%%%%%%
%%%%%%%%%%%%%%%%%%%%%%%%%%%%%%%%%%%%%%%%%%%%%%%%%%%%%%%
%%%%%%%%%%%%%%%%%%%%%%%%%%%%%%%%%%%%%%%%%%%%%%%%%%%%%%%
%%%%%%%%%%%%%%%%%%%%%%%%%%%%%%%%%%%%%%%%%%%%%%%%%%%%%%%
%%%%%%%%%%%%%%%%%%%%%%%%%%%%%%%%%%%%%%%%%%%%%%%%%%%%%%%

\section{Introduction} \label{sec:intro}
\setcounter{equation}{0}    
\def\theequation{1.\arabic{equation}}

In a well known paper in 1975, O. Lanford presented the first mathematical proof of the validity of the Boltzmann 
equation for a system of hard spheres, for a sufficiently small time. The starting point was the series expansion describing 
the time evolution of the statistical states of a hard--sphere system. This series is the solution of a hierarchy of equations 
formally established by C. Cercignani in 1972 \cite{Ce72}, following previous ideas due to H. Grad \cite{Gr58}. 

The main idea of Lanford is to compare such a series expansion with the one arising from the solution of the 
Boltzmann equation, claiming the term by term convergence in the so called Boltzmann--Grad limit (BG limit in the sequel). 
The restriction to short times is due to the fact that the two series have been proven to converge absolutely only for a small
time interval. Actually it was remarked in \cite{Uk01} that the Lanford's approach is a Cauchy-Kowalevski kind of argument.

In \cite{La75}, although all the main ideas, as well as the strategy of the proof, were clearly discussed, the details were missing. 
The complete proof was presented later on in \cite{Ki75}, \cite{Sp84}, \cite{Uc88}, \cite{Sp91} and \cite{CIP94}. 

We mention also that the ideas of Lanford can be applied to derive the Boltzmann equation globally in time, in the special 
case of an expanding cloud of a rare gas in the vacuum \cite{IP86,IP89}.

Shortly after the appearance of the Lanford's paper, F. King in his unpublished thesis \cite{Ki75} approached the same 
validity problem for a particle system interacting by means of a positive, smooth and short range potential. In this case
the basic starting point was not the usual BBGKY hierarchy, but a variant of that due to H. Grad \cite{Gr58} (we shall 
call it the ``Grad hierarchy'' in the sequel) making the system more similar to a hard--sphere one. More precisely, 
in \cite{Gr58} only the first equation of this hierarchy was discussed, while the full hierarchy was introduced and derived 
in \cite{Ki75}. 

The Boltzmann equation considered by King was written in unusual form. Namely, calling $f=f(x,v,t)$ the distribution function,
\be
(\pa_t+v\cdot \nabla_x)f(x,v,t)=\int_{\RRR^3} dv_1 \int _{S^2_+} d\nu \ (v-v_1)\cdot\nu \Big\{f(x,v_1',t)f(x,v',t)-
f(x,v_1,t)f(x,v,t)\Big\}
\label{eq:BE}
\ee
where $S_+^2=\{\nu\in S^2 |\ (v-v_1)\cdot\nu \geq 0\},$ $S^2$ is the unit sphere in $\RRR^3$ ($d\nu$ is the 
surface measure), $(v,v_1)$ is a pair of velocities in incoming collision configuration --see also \cite{Bo64}--
and $(v',v_1')$ is the corresponding pair of outgoing velocities defined by
\be
\begin{cases}
\displaystyle v'=v-\om [\om\cdot(v-v_1)] \\
\displaystyle  v_1'=v_1+\om[\om\cdot(v-v_1)]
\end{cases}\;.\label{eq:coll}
\ee
Here $\om=\om(\nu,V)$ is the unit vector bisecting the angle between the incoming relative velocity $V=v_1-v$ and the 
outgoing relative velocity $V'=v_1'-v'$ as specified in the figure below. Note that $\nu$ is the unit vector pointing from the 
particle with velocity $v$ to the particle with velocity $v_1$, when they are about to collide (hence $\nu\cdot(v_1-v)
= \nu \cdot V \leq 0$).
\begin{figure}[htbp] 
   \centering
   \includegraphics[width=3in]{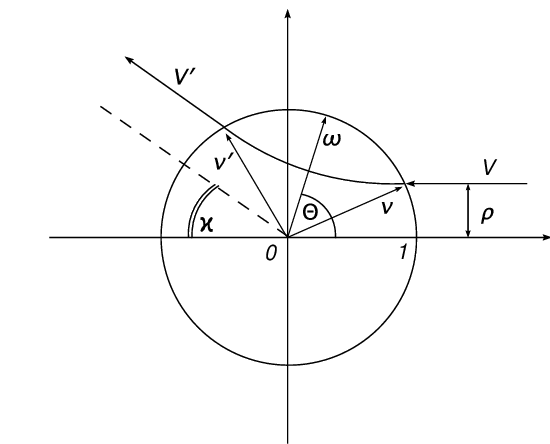} 
   \caption{The two--body scattering. We denote by $\r$ the {\em impact parameter} expressed in microscopic unities ($\r\in[-1,1]$)
   and by $\chi = \chi(\r,|V|)$ the {\em scattering angle} ($\chi\in(-\pi,\pi],$ $\chi>0$ in the figure), 
   while $\Theta$ is the angle given by the relation $\chi = \pi-2\Theta$.
   We call {\em scattering vector} the function $\o=\o(\n,V)$.}
   \label{fig:2bscatter}
\end{figure}

A more handable and usual form for the Boltzmann equation is obtained by expressing everything in terms of $\o$, namely 
\be
(\pa_t+v\cdot \nabla_x)f(x,v,t)=\int_{\RRR^3} dv_1 \int _{S^2} d\o \ B(\o,V) \Big\{f(x,v_1',t)f(x,v',t)-
f(x,v_1,t)f(x,v,t)\Big\}
\label{eq:BEs}
\ee
where $B(\o,V) / |V|$ is the {\em differential cross--section} of the potential under consideration with respect to the solid 
angle $\o$ (with $v'$ and $v_1'$ specified by \eqref{eq:coll}). In the case of hard spheres, the two formulations 
\eqref{eq:BE} and \eqref{eq:BEs} coincide since $\o=\n$.

After many years, the argument has been recently reconsidered by I. Gallagher et al in a long and self--contained paper
\cite{GSRT12} pointing out some important facts, surprisingly not discussed in the previous literature. In particular, the term
by term convergence is not innocent because $B$ is, in general, not bounded and even not defined as a single--valued function.
For instance, for smooth positive and bounded potentials (considered by King himself), $\n\rightarrow\o$ is not globally 
invertible and $B$ is unbounded. The difficulty is that one has to exclude concentration of measure on certain 
small sets in the phase space leading to an evolution
much different from the typical Boltzmann behavior. These ``bad'' events are: (i) the {\em long time} two--body {\em scattering}; 
(ii) the {\em recollisions}, i.e. the presence of a given pair of particles undergoing two or more collisions. The latter is the 
main obstacle in proving that the particle system behaves as predicted by the Boltzmann equation.

In \cite{GSRT12} the authors prove the validity of the Boltzmann equation under the hypotheses that the potential is well 
behaving in this sense, namely that the cross--section exists as a single--valued and sufficiently regular function. In the present
paper we show that, under very general assumptions on the potential, the Boltzmann equation can indeed be derived in the form 
\eqref{eq:BE}. We review the results in \cite{Ki75}, completing some parts of the proof and taking care of some inconsistencies.
Once the Boltzmann equation has been derived in the form \eqref{eq:BE}, the passage to the form \eqref{eq:BEs} is a matter 
of analysis of the two--body problem. If the cross--section is not a single--valued function, the function $B$ appearing in 
\eqref{eq:BEs} can be still expressed as a sum of the contributions arising from each monotonicity branch.

The approach discussed in \cite{GSRT12} makes explicit use of the cross--section as a tool for the control of recollisions. 
In the present paper the aim is to establish a proof that does not depend on any detail of the scattering process. In particular,
the term by term convergence (which is the most delicate point in the proof of our main results) is 
treated in a different way from the one in \cite{GSRT12} and \cite{Ki75}: see Section \ref{sec:convergence} for a
presentation of the problem, and Sections \ref{sec:proofsoft}, \ref{sec:proofhard} for a quick abstract and an explicit 
constructive proof respectively. 

In our method a very useful tool is a {\em tree expansion} describing the time evolution
of the marginals of a statistical state. This is presented in Section \ref{sec:tree}. In Section \ref{sec:hs} we introduce the mechanical
system of particles under examination and make some preliminary remark about it, while in Section \ref{sec:Grad}, along 
the lines of \cite{Ki75}, we derive the Grad hierarchy, that is the starting point of our study. In Section \ref{sec:results}
we fix the hypotheses on the initial data and state our main results. In Section \ref{sec:ste} we present the uniform short 
time estimates on the series expansion for the evolution of the marginals. The results in Sections \ref{sec:Grad} and \ref{sec:ste}
are well known by \cite{Ki75} and \cite{GSRT12}: we discuss them here briefly for the sake of completeness. Finally, in 
the Appendix we give sufficient conditions on the interaction for having a bounded or a single--valued cross--section.

One advantage of the methods developed in this paper is that they allow an explicit estimate of the error in the 
convergence to the Boltzmann equation, as soon as one has explicit estimates of the interaction time of the two--body 
process in the space of the scattering parameters. Moreover, convergence is established in a strong sense, that is 
uniformly outside a precise pathological null--measure subset of the phase space.

The analysis of sections \ref{sec:hs}--\ref{sec:proof} can be applied to any smooth and repulsive potential, enlarging the
class of interactions considered in \cite{GSRT12}. If the potential has an attractive part, there is also an
additional difficulty due to long time scattering phenomena and to the presence of trapping orbits in the two--body process. 
For the sake of clearness we treat this case separately in Section \ref{sec:gener}, where we explain how the proof 
can be adapted to extend the convergence result, assuming the stability of the interaction.

%%%%%%%%%%%%%%%%%%%%%%%%%%%%%%%%%%%%%%%%%%%%%%%%%%%%%%%
%%%%%%%%%%%%%%%%%%%%%%%%%%%%%%%%%%%%%%%%%%%%%%%%%%%%%%%
%%%%%%%%%%%%%%%%%%%%%%%%%%%%%%%%%%%%%%%%%%%%%%%%%%%%%%%
%%%%%%%%%%%%%%%%%%%%%%%%%%%%%%%%%%%%%%%%%%%%%%%%%%%%%%%
%%%%%%%%%%%%%%%%%%%%%%%%%%%%%%%%%%%%%%%%%%%%%%%%%%%%%%%

\section{The hamiltonian system} \label{sec:hs}
\setcounter{equation}{0}    
\def\theequation{2.\arabic{equation}}

We consider a system of $N$ identical classical point particles of unit mass, moving in the whole space and interacting 
by  means of a two--body, short range potential $\Phi.$ We denote by $(q_1, v_1,\cdots, q_N, v_N)$ a state of the 
system, where $q_i$ and $v_i$ indicate the position and the velocity of the particle $i,$ and $q_i(\tau)$ the position of
particle $i$ at time $\tau.$ The {\em $N-$particle Hamiltonian} is
\be
H = \frac{1}{2}\sum _{i=1}^N v_i ^2+\frac{1}{2}\sum_{\substack {i,j=1 \\ i\neq j}}^N 
\Phi(q_i -q_j)\;. \label{eq:Hamilt}
\ee

The dynamical flow is obtained by solving the {\em Newton equations}
\be
\frac{d^2q_i}{d \tau^2}(\tau)=\sum_{j\neq i} F(q_i(\tau )-q_j(\tau)) \label{eq:Newton}
\ee
where $F(q_i-q_j)=F_{i,j}=-\nabla \Phi (q_i-q_j)$ is the force due to the particle $j$, acting on the particle $i$.
We will assume $\Phi$ to be smooth enough in order to have existence and uniqueness of the solution to \eqref{eq:Newton}
for any initial datum such that $q_i\neq q_j$ (see Hypothesis 1 in Section \ref{sec:2bs}, and Section \ref{sec:gener}).

Consider now a small parameter $\e$ indicating the ratio between the macro and the micro unities. We pass to 
{\em macroscopic variables} defining
\be
x=\e q;\ \ \ t=\e \tau\;.
\ee
In these variables the equations of motion become
\be
\frac{d^2x_i}{dt^2}(t) = \frac{1}{\e}\sum_{j\neq i} F\left(\frac{x_i(t)-x_j(t)}{\e}\right)\;.
\label{eq:Newtonscaled}
\ee

\newpage
\ni {\bf The Boltzmann--Grad scaling.}
\medskip

\ni In order to have a kinetic picture, a tagged particle, say particle $1,$ must deliver a finite number of collisions in a macroscopic 
unit time. As a consequence, the density $N\e^3$ must vanish. More precisely $N$ should be $O(\e ^{-2})$. Indeed, assuming 
the characteristic interaction length of the potential $\Phi$ to be one in microscopic variables, namely $\Phi (q)=0$ if $|q|>1$, 
consider the tube spanned in the space by particle $1$ in the (macro) time $1$:
\be
\Big\{x \in \RRR^3 \ | \inf _{0\leq t \leq 1}  |x-x_1(t) | \leq \e \Big\}\;,
\ee
where $x_1(t)$ is the trajectory of particle $1$ (in macro variables).
The number of particles in the tube is the number of particles potentially interacting with particle $1$ in a macroscopic 
unit time. Hence, if $ N=O(\e ^{-2})$, such a number is expected to be finite. Therefore the scaling we will consider is
\be
N \to \infty, \qquad \e \to 0, \qquad N\e^2=\l^{-1}>0,
\label{eq:BGscaling}
\ee
for a system of $N$ particles following \eqref{eq:Newtonscaled}, where $l>0$ is proportional to the mean free path.
From now on, we shall fix $\l = 1$ for notational simplicity.

The scaling \eqref{eq:BGscaling} is usually called {\em low--density limit} and 
it is equivalent to the {\em BG limit} originally introduced for the hard--sphere system, \cite{Gr58}. In this scaling a triple collision
- namely a situation in which three or more particles are 
simultaneously interacting - will be very unlikely. Moreover a two--body collision - namely a scattering process involving 
only two particles - will take place typically on a scale of time of $O(\e)$, but since the force is $O(\e ^{-1})$ it will produce a 
finite effect. In other words, the expected dynamics is qualitatively similar to that of the hard--sphere systems.

%%%%%%%%%%%%%%%%%%%%%%%%%%%%%%%%%%%%%%%%%%%%%%%%%%%%%%%  
%%%%%%%%%%%%%%%%%%%%%%%%%%%%%%%%%%%%%%%%%%%%%%%%%%%%%%%
%%%%%%%%%%%%%%%%%%%%%%%%%%%%%%%%%%%%%%%%%%%%%%%%%%%%%%%

\subsection{Statistical description} \label{sec:stat}
\setcounter{equation}{0}    
\def\theequation{2.1.\arabic{equation}}

We want to describe here our system from a statistical viewpoint. 

\bigskip
\ni {\bf Notation.} Throughout the paper, we will use bold letters for vectors of variables, e.g.
\be
\bz_j = (z_1,\cdots,z_j),\ \ \ \ \ \bz_{j,n}=(z_{j+1}, \cdots, z_{j+n}),\ \ \ \ \ z_i = (x_i,v_i)
\ee
is the notation for the state of particles $1,\cdots,j$ and $j+1,\cdots,j+n$ respectively, having position and velocity 
$(x_i,v_i).$ 

\bigskip
\ni As usual we introduce the {\em phase space} of the $N$--particle system
\be
\MM_N = \Big\{ \bz_N\in\RRR^{6N} \ \Big|\ |x_i- x_k| > 0,\  i,k =1 \cdots N,\  k\neq i \Big\}\;. \label{eq:Npps}
\ee

Consider a probability measure with density $W^N(t)$ (with respect to the Lebesgue measure), $t \geq 0$,
evolving in time in accord to the {\bf {\em Liouville equation}}
\be
(\pa_t +{\cal L}_N) W^N=0\;,
\label{eq:Liouville}
\ee
where the {\em Liouville operator} ${\cal L}_N$ is 
\be
{\cal L}_N={\cal L}_N^0+{\cal L}_N^I
\ee
with:
\bea
&& {\cal L}_N^0=\sum_{i=1}^N v_i \cdot \nabla_{x_i}\;,\nn\\
&& {\cal L}_N^I=\frac{1}{\e} \sum_{\substack{i,j=1\\ i\neq j}}^N F_{i,j} \cdot \nabla_{v_i}
\label{eq:LN0LNI}
\eea
and $F_{i,j}=- \nabla \Phi\left(\frac{x_i - x_j}{\e}\right).$ We shall assume $W^N \in C^1(\MM_N\times\RRR^+)$,
with $v_i \cdot \nabla_{x_i}W^N$, $F_{i,j}\cdot~\nabla_{v_i}W^N \in L^1(\MM_N).$ Moreover, $W^N$ is initially 
(and hence at any positive time) symmetric in the exchange of the particles. 

\bigskip
{\em Remark.\label{rem:regul}} The smoothness and integrability properties of $W^N$ stated above will be used 
along this section and Section \ref{sec:Grad}, to write the Liouville equation in a classical sense and to perform 
partial integrations. Such assumptions will be removed in Section \ref{sec:series} (Proposition~\ref{prop:BBGKY}),
after having introduced the concept of series solution for the reduced marginals of~$W^N$.
\bigskip

\ni {\bf The BBGKY hierarchy.}
\medskip

\ni
We introduce the {\em marginals} $g^N_j(\bz_j,t)$ of the time evolved measure $W^N(\bz_N,t)$, defined by
\be
g^N_j(\bz_j,t ) =\int d\bz_{j,N-j} W^N(\bz_j,\bz_{j,N-j},t)\;, \label{eq:defmarg}
\ee
which denote the probability distributions of the first $j$ particles (or of any other fixed group of $j$ particles). 
Clearly $g^N_N=W^N$.

From \eqref{eq:Liouville} and \eqref{eq:defmarg} it follows that the family $\{ g_j^N \}_{j=1}^N$ satisfies the well known
{\em BBGKY hierarchy} (\cite{Gr58}):
\bea
&& \left(\pa_t +\sum_{i=1}^j v_i \cdot \nabla_{x_i}\right) g^N_j+\frac{1}{\e}\sum_{\substack{i,k=1\\ i\neq k}}^j 
F \left(\frac{x_i - x_k}{\e}\right)\cdot\nabla_{v_i}g^N_j \nn\\
&& = - \frac{N-j}{\e}\sum_{i=1}^j \int dx_{j+1}\int dv_{j+1} F\left(\frac{x_i - x_{j+1}}{\e}\right)\cdot\nabla_{v_i} 
g^N_{j+1}\;. \label{eq:BBGKY}
\eea

\bigskip
{\em Remark.}
Notice that, for a fixed $j$, the interaction term in the left hand side of Eq. \eqref{eq:BBGKY} is, in a sense, negligible because 
the collisions among a tagged group of particles are unlikely (the potential is indeed vanishing as soon as $\e$ is smaller than 
$|x_i-x_k|$). Moreover the integral in the right hand side is $O(\e^3)$. The right hand side, which is due to the interaction 
between the group of the first $j$ particles with the rest of the system, is $O(1)$ whenever $N=O(\e^{-2})$, which is exactly the 
reason why we perform the low--density scaling \eqref{eq:BGscaling}. 
However, instead of using the above hierarchy, not very well suited for such 
a scaling, we will introduce, in Section \ref{sec:Grad}, another set of equations.

%%%%%%%%%%%%%%%%%%%%%%%%%%%%%%%%%%%%%%%%%%%%%%%%%%%%%%%
%%%%%%%%%%%%%%%%%%%%%%%%%%%%%%%%%%%%%%%%%%%%%%%%%%%%%%%
%%%%%%%%%%%%%%%%%%%%%%%%%%%%%%%%%%%%%%%%%%%%%%%%%%%%%%%

\subsection{The two--body scattering} \label{sec:2bs}
\setcounter{equation}{0}    
\def\theequation{2.2.\arabic{equation}}

Let us discuss here the scattering process between two particles, which will play a crucial role in what follows.
We turn back to microscopic unities, where the potential is assumed to have range one.

Let $q_1, v_1, q_2, v_2$ be positions and velocities of two particles which are performing a collision. It is well 
known that this two--body problem can be reduced to a central--motion problem setting the origin in the center of mass:
\be
\frac{q_1+q_2}{2}=0,\ \ \ \ \ q=q_1-q_2\;.
\ee
Then the evolution is given by
\be
\frac{d^2q}{d \tau^2}(\tau) = 2 F\left(q(\tau)\right)\;, \label{eq:Newtoncm}
\ee
where $$F = - \nabla\Phi.$$

The above equation of motion is``almost''explicitly solvable, in the following sense. 
Fixed the relative velocity $V=v_1-v_2$ 
(hence fixed a value of the energy in the center of mass), one can restrict his attention to the control of the scattering 
function $\o = \o(\nu)$, see Eq. \eqref{eq:coll} and Fig. \ref{fig:2bscatter} in Section \nolinebreak\ref{sec:intro}.
Since the scattering takes place in a plane, this amounts to control the function $\Theta =\Theta(\rho)$ (Fig. \ref{fig:2bscatter}).
The classical integral formula expressing $\Theta$ in terms of the modulus of the incoming relative
velocity $|V|,$ the potential $\Phi$ and the impact parameter $\r$ will be written in the Appendix (see Eq. \eqref{eq:defT}).
That formula is not so easy to handle with, so it will not be employed in our work.

In what follows it will be rather crucial to have an estimate on the {\em scattering time} $$\tau_* = \mbox{measure of 
the time interval for which } |q(\tau)|<1.$$ To this purpose, we need to state our precise assumptions on the potential.
\begin{hyp} \label{hyp:pot}
The two--body potential $\Phi = \Phi(q), q\in\RRR^3,$ is radial, with support $|q|<1$ and not increasing 
in $|q|.$ We assume either $\Phi\in C^2(\RRR^3),$ or $\Phi\in C^2(\RRR^3\setminus\{0\})$ and $\Phi(q) \to +\infty$ as $q\to 0.$
\end{hyp}
The smoothness assumption is needed to ensure existence and uniqueness of the flow evolution for the system of $N$ particles,
while the monotonicity is introduced to allow a simple control on the scattering time $\tau_*$, as shown below. 
We defer the more general cases (i.e. non monotonic) to Section \ref{sec:gener}.

From now on, we will use occasionally the notational inconsistency $\Phi(r)=\Phi|_{|q|=r}.$

\bigskip
\ni {\bf A bound on the time of scattering.}
\medskip

\ni
Consider the central motion given by Eq. \eqref{eq:Newtoncm} with the initial conditions describing the two particles just 
before the interaction, namely $q(0)= \n\in S^2$, $\dot q(0) = V$ and $|V|>0,$ $V \cdot \n \leq 0.$ 
Denote
\be
L = |\n\wedge V| = |\r V| \in [0,V]
\ee
the magnitude of {\em angular momentum}, being $\r$ the {\em impact parameter} (Fig.\ref{fig:2bscatter}).
A rather general estimate on $\t_*$ is the following:
\begin{lem} \label{lem:timebound}
Under Hypothesis \ref{hyp:pot} it is
\be
\t_* \leq \frac{A}{L}  \label{eq:stbound}
\ee
for some constant $A>0$, which does not depend on $\Phi$.
\end{lem}

\bigskip
{\em Proof.} From the conservation laws one derives the well known formula expressing $\t_*$ as a function of
$V$ and $L:$
\be
\t_* = \sqrt{2} \int_{r_*}^1 dr \frac{1}{\left(\frac{V^2}{2} - \frac{L^2}{2r^2}-2\Phi(r)\right)^{1/2}}\;,
\ee
where $r_*$ is the minimum distance from the origin, $r_* = \inf_{\t\in (0,\t_*)}|q(\t)|,$ related to $V$ and $L$ by
\be
\frac{V^2}{2} = \frac{L^2}{2r_*^2}+2\Phi(r_*)\;. \label{eq:defrstar}
\ee
The effective potential, i.e. the potential of the reduced one--dimensional motion 
(which is the evolution of the radial coordinate in the system of the center of mass), is the $L-$dependent function
\be
2\Phi_{eff}(r) = \frac{L^2}{2r^2}+2\Phi(r) - \frac{L^2}{2}\;,\ \ \ \ \ \ \ \ r\in[0,1]\;.
\ee
We can write
\bea
&& \t_* = \int_{r_*}^1 dr \frac{1}{\left(\Phi_{eff}(r_*)-\Phi_{eff}(r)\right)^{1/2}}\nn\\ 
&& \ \ \ \ \leq \frac{1}{\left(\min_{[0,1]}(-\Phi_{eff}')\right)^{\frac{1}{2}}} 
\int_{r_*}^1 dr \frac{1}{\sqrt{r-r_*}}\;.
\label{eq:timerstarL}
\eea
We denoted improperly $\Phi'$ the derivative with respect to $r$ of the function $\Phi|_{|q|=r}.$ Since $\Phi' \leq 0$ and
\be
\Phi'_{eff}(r) = \Phi'(r) - \frac{L^2}{2r^3}\;, \label{eq:effpotscatt}
\ee
the result follows easily. \qed

\bigskip
The estimate in Lemma \ref{lem:timebound} tells us that $\t_*=O((\r V)^{-1}).$ This has the advantage to be general 
and sufficient to our purposes. Clearly the bound can be improved in many cases. Singularities in the scattering occur 
whenever the collision is central $(\r=0)$ and the energy corresponds exactly to a point of vanishing force 
$(V^2/4=\Phi(r), \Phi'(r)=0).$ This kind of singularities does not exist if the potential is unbounded at the origin and 
strictly repulsive: for instance for potentials diverging at the origin with a power law, formula \eqref{eq:stbound} can be 
easily replaced by $\t_*= O(V^{-1}).$ From \eqref{eq:timerstarL} it can be noticed also that the singularity for low energies 
may appear only if $\Phi$ goes to zero smoothly ($C^1$) in $r=1.$

\bigskip
\ni {\bf The scattering map.}
\medskip

\ni
We conclude by introducing a map which encodes all the properties of the two--body interaction.
The {\em scattering operator} $\II$ is defined over 
\be
\Big\{(\n,V)\in S^2\times \RRR^3\setminus\{0\} \ \Big|\  V\cdot\n \leq 0\Big\}
\ee
by:
\bea
&& \II(\n,V) = (\n',V') \nn\\
&& \begin{cases}
\displaystyle V'=V-2\om (\om\cdot V) \\
\displaystyle  \n'=-\n+2\om (\om\cdot\n)
\end{cases} \label{eq:scattop}
\eea
where $\o = \o(\n,V)$ is the scattering vector, see Fig. \ref{fig:2bscatter}.
It follows that $V\cdot\n = -V'\cdot\n'.$ In particular, $V'\cdot\n' \geq 0,$ i.e. $\II$ sends incoming into outgoing
configurations. 

\bigskip
The following property of $\II$ will be used in the validation of the Boltzmann equation.
\begin{lem} \label{lem:scattop}
$\II$ is an invertible transformation that preserves Lebesgue measure.
\end{lem}

{\em Proof.} 
Of course, the dynamics being reversible, $\o(\n',V')=\o(\n,V)$ (see Fig.\ref{fig:2bscatter}) and $\II^{-1}$ is defined 
in the same way as $\II.$ 

To see that $\II$ is measure preserving, we fix cartesian coordinates on the plane where the
scattering occurs, and call $\phi$ the angle formed by $V$ and the first axis (with $\phi$ growing when
$V$ rotates counterclockwise), $\a$ the angle formed by $V$ and $\n$ and such that $\sin\a$ is the impact 
parameter $\r$ (with the convention $\a\in[\pi/2,(3\pi)/2]$, see Fig.\ref{fig:2bscatter}).
Restricting to the plane of the scattering, we have the parametrization 
$V = (|V|\cos\phi,|V|\sin\phi),$ $\n = \a.$ In the variables $|V|,\phi,\a$ the action of $\II$ is simply described by:
\be
V' = (|V'|\cos\phi',|V'|\sin\phi')\;,\ \ \ \ \ \n'=\a'\;,\nn
\ee
where
\be
\begin{cases}
\displaystyle |V'|=|V| \\
\displaystyle \a'=\pi-\a \\
\displaystyle \phi' = \phi - \chi(\sin\a,|V|)
\end{cases}\;. \label{eq:scatterpolar}
\ee
Note that $\a'\in[-\pi/2,\pi/2].$
The first equation is conservation of energy, the second conservation of angular momentum, and the third holds
by definition of scattering angle (Fig.\ref{fig:2bscatter}). It can be shown that $\chi$ is a 
differentiable function of its arguments (see the discussion in the Appendix). 
Moreover, the determinant of the jacobian of the transformation \eqref{eq:scatterpolar} 
has modulus one, independently of the form of $\chi.$ This concludes the proof. \qed
%%%%%%%%%%%%%%%%%%%%%%%%%%%%%%%%%%%%%%%%%%%%%%%%%%%%%%%
%%%%%%%%%%%%%%%%%%%%%%%%%%%%%%%%%%%%%%%%%%%%%%%%%%%%%%%
%%%%%%%%%%%%%%%%%%%%%%%%%%%%%%%%%%%%%%%%%%%%%%%%%%%%%%%
%%%%%%%%%%%%%%%%%%%%%%%%%%%%%%%%%%%%%%%%%%%%%%%%%%%%%%%
%%%%%%%%%%%%%%%%%%%%%%%%%%%%%%%%%%%%%%%%%%%%%%%%%%%%%%%

%\newpage
\section{The Grad hierarchy} \label{sec:Grad}
\setcounter{equation}{0}    
\def\theequation{3.\arabic{equation}}

In this section we derive a hierarchy of equations for a family of quantities which are very close to the marginals 
introduced in the previous Section \ref{sec:stat}. This allows to put the dynamical problem in a form somehow 
similar to the one arising in considering hard--sphere systems, and more suitable for the study of the low--density limit. 

\begin{defi} \label{def:defrm}
Given a probability density $W^N$ on the phase space $\MM_N$, we define its
reduced marginals $f^N_j: \MM_j \to \RRR$, for $j = 1, \cdots, N$, by
\be
f^N_j(\bz_j) = \int_{S(\bx_j)^{N-j}}d\bz_{j,N-j} W^N(\bz_{j},\bz_{j,N-j})\;,\label{eq:defredmarg}
\ee
where
\be
S(\bx_j)=\Big\{z= (x,v)\in\RRR^6\ \Big|\ |x-x_k| > \e\ \text{ for all } k=1,\cdots, j\Big\}\;.
\ee
\end{defi}
Of course the functions $f_j^N$ are, for any $j$, asymptotically 
equivalent (uniformly on compact sets in $\MM_j$ in the BG limit) to the usual marginals.
The advantage of the above definition will be clear in a moment.

Consider a configuration $\bz_N=(\bz_j,\bz_{j,N-j})$ such that
\be
|x_\ell -x_k| >\e
\ee
for all $\ell =1,\cdots,j$ and $k=j+1,\cdots,N.$ Since the range of the interaction is $\e$, the interaction between the group of the
first $j$ particles and the rest of the system is vanishing. Therefore the Liouville equation \eqref{eq:Liouville}
on such configurations becomes:
\be
\pa_tW^N +{\cal L}_N^0W^N + {\cal L}_j^IW^N + {\cal L}_{j,N}^IW^N =0\;,
\label{eq:Liouvred}
\ee
where ${\cal L}_N^0$ is defined in \eqref{eq:LN0LNI} and, similarly,
\bea
&& {\cal L}_j^I=\frac{1}{\e} \sum_{\substack{i,k=1\\ i\neq k}}^j F_{i,k} \cdot \nabla_{v_i}\;,\nn\\
&& {\cal L}_{j,N}^I=\frac{1}{\e} \sum_{\substack{i,k=j+1\\ i\neq k}}^N F_{i,k} \cdot \nabla_{v_i}\;. \nn
\eea

As already said in the Remark at page \pageref{rem:regul}, we make here the regularity assumptions needed
to justify Eq. \eqref{eq:Liouvred} and all the steps in the derivation that follows. 
%The following explicit bound will be used instead throughout all the paper:

%\newpage
\bigskip
\ni {\bf Derivation of the evolution equations for the $f^N_j$.}
\bigskip

\ni Integrating Eq. \eqref{eq:Liouvred} with respect to $d\bz_{j,N-j}$ over $S(\bx_j)^{N-j}$ we obtain:
\bea
&& \left(\pa_t+{\cal L}_j^I\right)f^N_j(\bz_j,t) = -\sum_{i=j+1}^N \int_{S(\bx_j)^{N-j}} d\bz_{j,N-j} 
v_i \cdot\nabla_{x_i}W^N(\bz_N,t)\nn\\
&&\ \ \ \ \ \ \ \ \ \ \ \ \ \ \ \ \ \ \ \ \ \ \ \ \ \ \ -\sum_{i=1}^j \int_{S(\bx_j)^{N-j}} d\bz_{j,N-j} v_i \cdot\nabla_{x_i}W^N(\bz_N,t)\;,
\label{eq:Gderstart}
\eea
where we used that
\be
\int_{\RRR^{3(N-j)}} d\bv_{j,N-j} {\cal L}_{j,N}^I W^N=0\;.
\ee

The first sum is handled by the divergence theorem, yielding
\bea
&& \left(\pa_t+{\cal L}_j^I\right)f^N_j(\bz_j,t) = -\sum_{i=j+1}^N \int_{S(\bx_j)}dz_{j+1} \cdots \int_{\pa S(\bx_j)}d\sigma(x_i)dv_i \cdots \int_{S(\bx_j)} dz_{N} (v_{i}\cdot \n_{i}) W^N \nn\\
&&\ \ \ \ \ \ \ \ \ \ \ \ \ \ \ \ \ \ \ \ \ \ \ \ \ \ \ -\sum_{i=1}^j \int_{S(\bx_j)^{N-j}} d\bz_{j,N-j} v_i \cdot\nabla_{x_i}W^N(\bz_N,t)\;,
\eea
%
%
%%
%\be
%-\sum_{i=j+1}^N \int_{S(\bx_j)}dz_{j+1} \cdots \int_{\pa S(\bx_j)}d\sigma(x_i)dv_i \cdots \int_{S(\bx_j)} dz_{N} 
%(v_{i}\cdot \n_{i}) W^N\;, \label{eq:Gderfirstsum}
%\ee
%%
where $\n_i$ is the outward normal to $S(\bx_j)$ in $z_i$, and $d\sigma(x_i)dv_i$ is the surface measure. 

Using the symmetry of $W^N$, we obtain $N-j$ identical integrals in the first line of the formula. Furthermore,
the boundary $\pa S(\bx_j)$ can be naturally decomposed, as regards the $x$--dependence, in a union of $j$
pieces of spherical surfaces. Namely, setting,
for $i \in \{1,\cdots,j\}$,
\bea
\sigma_i (\bx_j) =\Big\{z= (x,v)\in\RRR^6\ \Big|\ |x-x_i| = \e \mbox{\ and\ } 
|x-x_k| > \e\ \text{ for all } k=1,\cdots, j,\ k \neq i \Big\}\;, \nn\\
\eea
and $$\n_{i,j+1}={x_i-x_{j+1}\over|x_i-x_{j+1}|}\ ,$$ we find
%Note that if $z_i \in\pa S(\bx_j)$, there exists an index $k\in\{1, \cdots, j\}$ such that $|x_k-x_i|=\e$. Moreover,
%if $\bz_j\in\MM_j,$ there is only one such an index for almost all $x_i$ with respect to the surface measure $d\sigma(x_i)$. 
%Hence $\pa S(\bx_j)$, as regards the $x$--dependence, is the disjoint union of pieces of spherical surfaces.
%We call such pieces $\s_k(\bx_j),$ that is
%%
%\be
%\pa S(\bx_j)=\bigcup_{k=1}^j \sigma_k (\bx_j)\times \R^3
%\ee
%%
%where
%%
%\be
%\sigma_k (\bx_j) \subset \{ x\ |\ |x-x_i|=\e \}\;.
%\ee
%%
%We set $\n_{k,i}={x_k-x_i\over|x_k-x_i|}.$ Using the symmetry of $W^N$ we have $N-j$ identical integrals, for which 
%\eqref{eq:Gderfirstsum} becomes:
%%
%\be
%- (N-j) \sum_{k=1}^j \int_{\sigma_k (\bx_j)}d\sigma(x_{j+1}) \int_{\RRR^3}dv_{j+1}
%(v_{j+1}\cdot\n_{k,j+1})\int _{S(\bx_j)^{N-j-1}}d\bz_{j+1,N-j-1}W^N\;. \label{eq:Gderfirstsum'}
%\ee
%%
%
\bea
&& \left(\pa_t+{\cal L}_j^I\right)f^N_j(\bz_j,t) \nn\\
&& = - (N-j) \sum_{i=1}^j \int_{\sigma_i (\bx_j)}d\sigma(x_{j+1}) \int_{\RRR^3}dv_{j+1}
(v_{j+1}\cdot\n_{i,j+1})\int _{S(\bx_j)^{N-j-1}}d\bz_{j+1,N-j-1}W^N(\bz_N,t)\nn\\
&&\ \ \ \ -\sum_{i=1}^j \int_{S(\bx_j)^{N-j}} d\bz_{j,N-j} v_i \cdot\nabla_{x_i}W^N(\bz_N,t)\;.
\label{eq:Gderfirstsum'}
\eea

The second sum above is treated in a similar way by using Definition \ref{def:defrm}, so that we get 
\bea
&& \left(\pa_t+{\cal L}_j^I\right)f^N_j(\bz_j,t) \nn\\
&& = - (N-j) \sum_{i=1}^j \int_{\sigma_i (\bx_j)}d\sigma(x_{j+1}) \int_{\RRR^3}dv_{j+1}
(v_{j+1}\cdot\n_{i,j+1})\int _{S(\bx_j)^{N-j-1}}d\bz_{j+1,N-j-1}W^N(\bz_N,t)\nn\\
&&\ \ \ \  -\sum_{i=1}^j v_i \cdot\nabla_{x_i} f^N_j(\bz_j,t)\nn\\
&&\ \ \ \ + (N-j) \sum_{i=1}^j \int_{\sigma_i (\bx_j)}d\sigma(x_{j+1}) \int_{\RRR^3}dv_{j+1}
(v_i\cdot\n_{i,j+1})\int _{S(\bx_j)^{N-j-1}}d\bz_{j+1,N-j-1}W^N(\bz_N,t)\;.\nn\\
\label{eq:Gderfirstsummy'}
\eea

The integration domain of the last integral in the second and fourth line of \eqref{eq:Gderfirstsummy'} is not $S(\bx_{j+1})^{N-j-1}$,
as it would be necessary to recover $f_{j+1}^N$ and close the equation. We could reduce this integration to
$S(\bx_{j+1})^{N-j-1},$ and this would produce a small error in the BG limit. Nevertheless, we want to establish 
an exact and closed equation, for which we need further work. 

\bigskip
\ni {\bf Grad's cluster decomposition.}
\medskip

\ni
A generic configuration in $S(\bx_j)^{N-j-1}$ differs from $S(\bx_{j+1})^{N-j-1}$ because some particle, say particle $h_1$, 
could {\em overlap} with particle $j+1,$ this meaning that $|x_{h}-x_{j+1}|\leq\e$. If this is the case, consider the 
maximal {\em cluster} of overlapping particles, with ordered indices $\bim = \{i_1,\cdots,i_m\}$, a subset of $\{j+2,\cdots,N\}$ 
with $i_1<i_2<\cdots<i_m.$ We put $\bz_{\bim} = \{ z_{i_1}, \cdots, z_{i_m}\}.$ 
The other particles are far apart the group with indices $1,\cdots, j, j+1,\bim,$ hence each 
of them is in $S(\bx_{j+1},\bx_{\bim}).$ Integrating out these particles in the last integral of the second and fourth
line of \eqref{eq:Gderfirstsummy'}, one obtains, by definition, $f^N_{j+1+m}(\bz_{j+1},\bz_{\bim})$. Note that summing over all
the possible choices of $\bim$ gives a factor $\binom{N-j-1}{m}$.

Therefore, following \cite{Ki75}, we can decompose the integration domain $S(\bx_j)^{N-j-1}$
in a union of disjoint sets to obtain
\be
\int _{S(\bx_j)^{N-j-1}}d\bz_{j+1,N-j-1}W^N(\bz_N,t) = \sum_{m=0}^{N-j-1}\binom{N-j-1}{m}
\int_{\Delta_m(\bx_{j+1})} d\bz_{j+1,m} f^N_{j+1+m}(\bz_{j+1+m})\;, \nn
\ee
where
\bea
&&\Delta_m(\bx_{j+1}):=\Big\{\bz_{j+1,m} \subset S(\bx_j)^{m} \ \Big|\ \mbox{\ for each\ }
\ell=j+2,\cdots,j+1+m,\nn\\ 
&&\ \ \ \ \ \ \ \ \ \ \ \ \ \ \ \ \ \ \ \ \mbox{there exists a choice of indices $h_1,h_2,\cdots,h_r \in \{j+2,\cdots,j+1+m\}$}\nn\\
&&\ \ \ \ \ \ \ \ \ \ \ \ \ \ \ \ \ \ \ \ \mbox{such that } |x_\ell-x_{h_1}|\leq\e\;,\ \ \ |x_{h_{k-1}}-x_{h_{k}}|\leq\e\ \ 
\mbox{for $k=2,\cdots,r$}\nn\\
&&\ \ \ \ \ \ \ \ \ \ \ \ \ \ \ \ \ \ \ \ \mbox{and }\ \ \ \min_{i\in\{\ell,h_1,\cdots,h_r\}}|x_i-x_{j+1}|\leq\e\Big\}\;.
\label{eq:Deltaix}
\eea

The result is the following hierarchy of equations, which we call {\bf {\em Grad hierarchy}}:
\be¥
\left(\pa_t+{\cal L}_j\right)f^N_j = \sum_{m=0}^{N-j-1} \AA_{j+1+m}^\e f^N_{j+1+m},\ \ \ \ \ \ \ 1\leq j\leq N\;,
\label{eq:Gradh}
\ee
where the operator $\LL_j=\LL^\e_j$ depends also on $\e$ through its interacting part \eqref{eq:LN0LNI}, and
\bea
&&\AA_{j+1+m}^\e f^N_{j+1+m}(\bz_j,t) = (N-j)(N-j-1)\cdots(N-j-m)\nn\\
&&\ \ \ \ \cdot\sum_{i=1}^j \e^2 \int_{S^2}d\n
\ \mathbbm{1}_{\mbox{$\{\min_{\ell=1,\cdots,j, \ell\neq i}|x_i+\n\e-x_\ell|>\e\}$}}(\n)
\int_{\RRR^3}dv_{j+1} (v_{j+1}-v_i)\cdot\n\nn\\
&&\ \ \ \ \cdot\int_{\Delta_m(\bx_{j+1})}\frac{d\bz_{j+1,m}}{m!} 
f^N_{j+1+m}(\bz_{j},x_i+\n\e,v_{j+1},\bz_{j+1,m},t)\;,
\label{eq:defA}
\eea
with $x_{j+1}=x_i+\n\e$ in the argument of $\Delta_m.$ We indicate with $\mathbbm{1}_{\{\cdot\}}(\cdot)$ 
the characteristic function of the set defined by the condition in the curly brackets.

In particular it is
\bea
&&\AA_{j+1}^\e f^N_{j+1}(\bz_j,t) \nn\\
&&= \e^2(N-j)\sum_{i=1}^j \int_{S^2\times\RRR^3}d\n dv_{j+1} 
\ \mathbbm{1}_{\mbox{$\{\min_{\ell=1,\cdots,j;\ell\neq i}|x_i+\n\e-x_\ell|>\e\}$}}(\n)\nn\\
&&\ \ \ \cdot(v_{j+1}-v_i)\cdot\n f^N_{j+1}(\bz_{j},x_i+\n\e,v_{j+1},t)\;,\nn\\
&& = \e^2(N-j)\CC_{j+1}^\e f^N_{j+1}(\bz_j,t)\;, \label{eq:defCop}
\eea
where \eqref{eq:defCop} defines  $\CC_{j+1}^\e,$ which is the same {\em collision operator} appearing in the hard--sphere
case, see \cite{La75}. 

\bigskip
{\em Remark.} Actually it is clear that, in the BG limit, the term $m=0$, i.e. \eqref{eq:defCop}, is the only $O(1)$ term 
in the sum in the right hand side of \eqref{eq:Gradh}. Indeed, for $m>0$ and fixed $j$, the size of $\AA_{j+1+m}^\e$ will be
\be
O(N^{m+1}\e^2\e^{3m}) = O(\e^m)\;,
\ee
the $\e^{3m}$ coming from the successive integrations in the domain $\Delta_m(\bx_{j+1}).$
This implies that we are in a situation quite similar to that of the hard--sphere system \cite{La75}, and we can 
hope to derive the Boltzmann equation in a similar manner.

%%%%%%%%%%%%%%%%%%%%%%%%%%%%%%%%%%%%%%%%%%%%%%%%%%%%%%%
%%%%%%%%%%%%%%%%%%%%%%%%%%%%%%%%%%%%%%%%%%%%%%%%%%%%%%%
%%%%%%%%%%%%%%%%%%%%%%%%%%%%%%%%%%%%%%%%%%%%%%%%%%%%%%%

\subsection{Series solution} \label{sec:series}
\setcounter{equation}{0}    
\def\theequation{3.1.\arabic{equation}}

Consider the dynamical flow obtained by solving the Newton equations \eqref{eq:Newtonscaled} for a system
of $j$ particles:
\be
\frac{d^2x_i}{dt^2}(t) = \frac{1}{\e}\sum_{k\neq i} F\left(\frac{x_i(t)-x_k(t)}{\e}\right)\;,
\label{eq:Newtonscaledj}
\ee
where $i$ and $k$ run now from $1$ to $j.$ Denote by $\TT_j^\e(t)\bz_j$ the solution of this system of equations with initial 
datum $\bz_j$. The action of this flow on the functions is given by the {\em interacting flow operator} $\SS^\e_j(t),$ defined as
\be
\SS^\e_j(t) g(\bz_j)=g(\TT^\e_j(-t)\bz_j)\;.
\ee

We may represent the solution of Eq \eqref{eq:Gradh} by means of a perturbative expansion, that is just the iteration of the 
Duhamel formula:
\bea\label{fjN}
&& f_j^N(t)= \sum_{n = 0}^{N-j}\sum_{\substack{ m_1,\cdots,m_n \geq 0 :\\  j+n+\sum_{i=1}^n m_i\leq N}}
\int_0^t dt_1 \int_0^{t_1} dt_2 \cdots \int_0^{t_{n-1}}dt_n\label{eq:fnjexp}\\
&& \ \ \cdot\SS_j^\e(t-t_1)\AA^\e _{j+1+m_1}\SS_{j+1+m_1}^\e(t_1-t_2)\cdots \AA^\e _{j+n+\sum_{i=1}^nm_i}
\SS^\e_{j+n+\sum_{i=1}^nm_i}(t_n) f^N_{j+n+\sum_{i=1}^nm_i}(0)\;,\nn
\eea
where $f^N_j(t)=f^N_j(\cdot,t),$ and $f^N_j(0)$ is the reduced marginals of the initial probability distribution.
This expansion will be our main tool.

\bigskip
\ni {\bf Rigorous validation of \eqref{eq:fnjexp}.}
\medskip

\ni We derived Eq. \eqref{eq:fnjexp} assuming sufficient smoothness of the initial distribution (see Remark on page 
\pageref{rem:regul}). However, by using a density approximation, \eqref{eq:fnjexp} can be proven to 
hold for a general class of initial measures. The argument can be found in \cite{Uc88} page 281, or \cite{Sp85} page 18
for cases of hard--sphere dynamics, and it can be applied also to general smooth potentials. 
We list the main steps in what follows. 
% A different approach based on a weak formulation may be found in \cite{GSRT12}.

Consider the collection of integration variables in the right hand side of 
\eqref{eq:fnjexp}, which we call $$\bl= (t_1,\cdots,t_n,\n_1,\cdots,\n_n,v_{j+1},\cdots,v_{j+n},\bz_{j+1,m_1},
\cdots, \bz_{j+n+\sum_{i \leq n-1} m_i, m_n}),$$
see also \eqref{eq:defA}. The reduced marginal in the integrand takes a form $$f^N_{j+n+\sum_{i=1}^nm_i}
(\by_{j+n+\sum_{i=1}^nm_i}(\bz_j,\bl),0),$$ with $\bz_j\in\MM_j$ and $\bl$ in the integration domain
(for the understanding of the detailed structure of the map,
$(\bz_j,\bl)\to\by_{j+n+\sum_{i=1}^nm_i},$ we defer to the discussion in Section \ref{sec:tree} of this paper).
Now, let us write the expansion in \eqref{eq:fnjexp} for a generic measurable probability density $W^N$ following
the Liouville equation in integral form:
\be
W^N (\bz_N,t) = \SS^\e_N(t) W^N(\bz_N,0)\;. \label{eq:liouvintfo}
\ee
To have a nice control on the integration over large velocities, we shall assume the exponential decrease
$f^N_j\leq c^j e^{-\b\sum_{i=1}^jv_i^2}$ for some $c,\b>0.$ Since  $\by$ is a Borel map 
(as follows directly from measurability of the partial mappings $(\bz_j,t)\to\TT_j^\e(t)\bz_j$), the expansion 
\eqref{eq:fnjexp} makes sense for the reduced marginals of $W^N,$ and the integrals therein are absolutely convergent.

To recover identity \eqref{eq:fnjexp}, we use that there exists a sequence of smooth densities $W^{N,\g}$ which 
evolve according to \eqref{eq:fnjexp}, satisfy the exponential bound, and approximate $W^N:$ $\underset{\g\to 0}{\lim}
W^{N,\g}=W^N$ a.e. on $\MM_N.$ Since the densities evolve according to $W^N(\bz_N,t)=
W^N(\TT^\e_N(-t)\bz_N)$ (and same equation for $W^{N,\g}$), we also have $\underset{\g\to 0}{\lim}W^{N,\g}(t)
= W^N(t)$ a.e. on $\MM_N$ and, consequently, $\underset{\g\to 0}{\lim}f^{N,\g}_j(t)=f^N_j(t)$ a.e. on $\MM_j.$
We are left with the problem of taking the limit of the right hand side of \eqref{eq:fnjexp}. Using the measure preserving 
property of the flows $\TT_j^\e(t),$ it can be shown that $\by$ is not singular, in the sense that $\by^{-1}$ maps null sets in 
$\MM_{j+n+\sum_{i=1}^nm_i}$ to null sets of values of $(\bz_j,\bl)$ in its domain. This fact, together with the gaussian 
estimate, allows to apply dominated convergence, thus concluding the proof.

\bigskip
Summarizing, we have the following result.

\begin{prop} \label{prop:BBGKY} Fix $N>0$ and consider a probability measure on the phase space $\MM_N$ with density 
$W^N$ with respect to the Lebesgue measure, evolving in accord to Equation \eqref{eq:liouvintfo}.
Let $f^N_j(t)$ be the reduced marginals of $W^N(t)$, introduced in Definition \ref{def:defrm}.
Suppose that, at time zero, $W^N$ is Borel measurable, symmetric in 
the particle labels and such that $f^N_j\leq c^j e^{-\b\sum_{i=1}^jv_i^2}$ for some constants $c,\b>0.$ 
Then, the reduced marginals $f^N_j(t)$ at time $t>0$ are given by Eq. \eqref{eq:fnjexp}, for almost all points 
in $\MM_j.$
\end{prop}

\bigskip
{\em Remark.} 
It is important to observe that the definitions of the operators $\AA_j^\e$ and $\CC_j^\e$ (respectively \eqref{eq:defA}
and \eqref{eq:defCop}) involve a trace problem, so that they are well posed if they act over functions which are at least 
continuous over a.a. points of the spheres of center $x_i$ and radius $\e$ (see definition \eqref{eq:defA}).
Nevertheless, this is not relevant to our purposes, since we will 
work only with operators of the form $\int ds\AA_j^\e\SS_j^\e(s).$ These last are indeed well defined over functions $f^N_j$
satisfying the hypotheses of Proposition \ref{prop:BBGKY}, by virtue of the nonsingularity of the map $\by.$
\medskip

\bigskip
\ni {\bf Additional notations.}
\medskip

\ni For future convenience, let us conclude this subsection by giving some more definitions. The subseries associated 
to the {\em dominant term} of \eqref{eq:fnjexp} (that with all $m_i=0$) defines a new sequence of functions which we 
call $\{\tilde f^N_j\}_{j=1}^N:$
\bea
&& \tilde f_j^N(t) = \sum_{n = 0}^{N-j}\alpha ^\e_n(j) \int_0^t dt_1 \int_0^{t_1}dt_2\cdots\int_0^{t_{n-1}}dt_n 
\label{eq:BBGKYt}\\
&&\ \ \ \ \ \ \ \ \ \ \ \ \ \ \ \ \ \ \ \ \ \ \ \ \cdot\SS_j^\e(t-t_1)\CC^\e _{j+1}\SS_{j+1}^\e(t_1-t_2)\cdots 
\CC^\e _{j+n}\SS^\e_{j+n}(t_n) f^N_{j+n}(0)\;, \nn\\
&& \alpha ^\e_n(j):=\e^{2n}(N-j)(N-j-1)\cdots(N-j-n+1)\;,\ \ \ \ \ \ \ \ \ \ \ \ \ n > 0\;,\nn\\
&& \alpha ^\e_0(j) := 1\;, \label{alpha}
\eea
where we used definition \eqref{eq:defCop}. Notice that in the BG limit $\alpha ^\e_n(j)=O(1)$.

Finally, it will be convenient to {\em decompose} the collision operator $\CC^{\e}_{j+1}$ in the following form:
\bea
&& \CC^{\e}_{j+1} =\sum_{k=1}^j \CC^{\e}_{k,j+1} \nn\\
&& \CC^\e_{k,j+1} = \CC^{\e,+}_{k,j+1} - \CC^{\e,-}_{k,j+1} \nn\\
&& \CC^{\e,+}_{k,j+1}f^N_{j+1}(\bz_j,t) = \int_{S^2_-\times\RRR^3}d\n dv_{j+1}
\ \mathbbm{1}_{ \mbox{$\{\min_{\ell=1,\cdots,j;\ell\neq k}|x_k+\n\e-x_\ell|>\e\} $}}(\n) \nn\\
&&\ \ \ \ \ \ \ \ \ \ \ \ \ \ \ \ \ \ \ \ \ \ \ \ \ \cdot|(v_k-v_{j+1})\cdot\n| f^N_{j+1}(\bz_{j},x_k+\n\e,v_{j+1},t) \nn\\
&& \CC^{\e,-}_{k,j+1}f^N_{j+1}(\bz_j,t) = \int_{S^2_+\times\RRR^3}d\n dv_{j+1}
\ \mathbbm{1}_{\mbox{$\{\min_{\ell=1,\cdots,j;\ell\neq k}|x_k+\n\e-x_\ell|>\e\}$}}(\n) \nn\\
&&\ \ \ \ \ \ \ \ \ \ \ \ \ \ \ \ \ \ \ \ \ \ \ \ \ \cdot|(v_k-v_{j+1})\cdot\n| f^N_{j+1}(\bz_{j},x_k+\n\e,v_{j+1},t) 
\label{eq:Cedec}
\eea
where
\bea
&& S^2_+ = \{\nu\ |\ (v_k - v_{j+1})\cdot\n \geq 0\}\;,\nn\\
&& S^2_- = \{\nu\ |\ (v_k- v_{j+1})\cdot\n \leq 0 \}\;. 
\eea
%

%%%%%%%%%%%%%%%%%%%%%%%%%%%%%%%%%%%%%%%%%%%%%%%%%%%%%%%
%%%%%%%%%%%%%%%%%%%%%%%%%%%%%%%%%%%%%%%%%%%%%%%%%%%%%%%
%%%%%%%%%%%%%%%%%%%%%%%%%%%%%%%%%%%%%%%%%%%%%%%%%%%%%%%

\subsection{The Boltzmann hierarchy} \label{sec:BH}
\setcounter{equation}{0}    
\def\theequation{3.2.\arabic{equation}}

In this subsection we treat formally the solution to the Boltzmann equation \eqref{eq:BE} as we did in Section \ref{sec:series}
for the interacting system of particles and compare heuristically the results.

Let $f$ be a solution to Eq. \eqref{eq:BE}. Consider the products
\be
f_j(\bz_j,t)=f(t)^{\otimes j}(\bz_j) = f(z_1,t)f(z_2,t)\cdots f(z_j,t)\;. \label{eq:fjtdef}
\ee
It is easy to show that the $f_j$ solve the hierarchy of equations
\be
\left(\pa_t+{\cal L}_j^0\right)f_j = \CC_{j+1}f_{j+1},\ \ \ \ \ \ \ 1\leq j < \infty\;, \label{eq:Bhier}
\ee
where we introduced the {\em Boltzmann collision operator}
\bea
&& \CC_{j+1} =\sum_{k=1}^j \CC_{k,j+1} \\
&& \CC_{k,j+1} = \CC^{+}_{k,j+1} - \CC^{-}_{k,j+1} \nn\\
&& \CC^{+}_{k,j+1}f_{j+1}(\bz_j,t) = \int_{S^2_+\times\RRR^3}d\n dv_{j+1}
(v_k-v_{j+1})\cdot\n f_{j+1}(z_1,\cdots,x_k,v'_k,\cdots,z_j,x_k,v'_{j+1},t) \nn\\
&& \CC^{-}_{k,j+1}f_{j+1}(\bz_j,t) = \int_{S^2_+\times\RRR^3}d\n dv_{j+1}
(v_k-v_{j+1})\cdot\n f_{j+1}(z_1,\cdots,x_k,v_k,\cdots,z_j,x_k,v_{j+1},t)\;, \nn
\eea
and
\be
\begin{cases}
\displaystyle v'_k=v_{k}-\om [\om\cdot(v_k-v_{j+1})] \\
\displaystyle  v'_{j+1}=v_{j+1}+\om[\om\cdot(v_k-v_{j+1})]
\end{cases}\;, \label{eq:scatterlaw}
\ee
$\om=\om(\nu,v_{j+1}-v_k)$ being the scattering vector (see Fig. \ref{fig:2bscatter}).

The infinite hierarchy of equations \eqref{eq:Bhier} (which does not express nothing else than the Boltzmann equation whenever the factorization
property \eqref{eq:fjtdef} holds) is called the \emph{Boltzmann hierarchy}. Proceeding as before, we may represent its solution by the perturbative 
expansion around the free flow:
\bea
&& f_j(t)= \sum_{n\geq 0}\int_0^t dt_1 \int_0^{t_1} dt_2 \cdots \int_0^{t_{n-1}}dt_n \nn\\
&& \ \ \ \ \ \ \ \ \ \ \ \ \ \ \ \ \cdot\SS_j(t-t_1)\CC _{j+1}\SS_{j+1}(t_1-t_2)\cdots \CC _{j+n}\SS_{j+n}(t_n) f_{j+n}(0)\;,
\label{eq:fjexp}
\eea
where now $\SS_j(t)$ is the {\em free flow operator}, defined as
\be
\SS_j(t)g(\bz_j) = g(x_1-v_1t,v_1,\cdots,x_j-v_jt,v_j)\;.
\ee

\bigskip{\em Remark 1.}
Note that: 

\ni - Eq. (\ref{eq:fnjexp}) is an identity which expresses $f_j^N$ (well defined by means of the $N$--particle flow) 
in terms of a finite sum of operators acting on the initial sequence $ f^N_j(0);$ 

\ni - $\tilde f^N_j,$ Eq. (\ref{eq:BBGKYt}), is just a technical definition; 

\ni - Eq. \eqref{eq:fjexp} is a series whose convergence must be proven. 

\ni As for the hard--sphere case \cite{La75}, it is possible to show that such a series is indeed convergent for a short time.
We will show it in Section \ref{sec:ste}. 
\bigskip

{\em Remark 2.} The last mentioned resul implies also local existence and uniqueness of the solution to 
the time--integrated version of the Boltzmann hierarchy 
in the class of continuous functions such that $f_j(t)\leq c^j e^{-\b\sum_{i=1}^jv_i^2}$ for some $c,\b>0$ (see e.g. \cite{CIP94}).
In particular, in the case of initial product states, factorization is propagated in time and each factor is the local solution of the time--integrated
Boltzmann equation.
\bigskip

{\em Remark 3.}
Reminding  the discussion in the remark
at the end of Section \ref{sec:Grad} and the fact that $\LL^I_j$ equals zero for $\e$ small, 
we shall guess that \eqref{eq:Bhier} is what one gets just letting $\e$ go to zero in the Grad hierarchy \eqref{eq:Gradh}. 
To do so, assume for simplicity that $f^N_{j+1}$ is continuous along trajectories of the flow $\TT^\e_{j+1}$.
Then, we may try to rewrite the action of $\CC^{\e,+}_{k,j+1}$ on $f^N_{j+1}$ in such a way that the function
is evaluated in {\em incoming} collision states. 
Call $t_*=\e\tau_*$ the time of interaction of particles $k$ and $j+1.$
Since the scattering process is, in macroscopic variables, almost instantaneous ($t_* = O(\e)$), 
we assume that the other particles do not interact in the same time interval. By the continuity of the flow it will be
\be
\TT^\e_{j+1}(-t_*)(\bz_{j},x_k+\n\e,v_{j+1}) \approx (z_1,\cdots,x_k,v'_k,\cdots,z_j,x_{k},v'_{j+1})\;,
\ee
hence 
\bea
&&\CC^{\e,+}_{k,j+1}f^N_{j+1}(\bz_j,t) \\
&&\ \ \ \ \ \approx \int_{S^2_-\times\RRR^3}d\n dv_{j+1}
|(v_k-v_{j+1})\cdot\n| f^N_{j+1}(z_1,\cdots,x_k,v'_k,\cdots,z_j,x_{k},v'_{j+1},t) \nn\\
&&\ \ \ \ \ = \int_{S^2_+\times\RRR^3}d\n dv_{j+1}(v_k-v_{j+1})\cdot\n 
f^N_{j+1}(z_1,\cdots,x_k,v'_k,\cdots,z_j,x_{k},v'_{j+1},t)\;,\nn
\eea
where in the second step we simply changed $\n\to -\n.$ 

\bigskip
We stress that the above heuristic discussion is somehow dangerous. 
In fact, the required continuity property of $f^N_{j+1},$ even when true for any fixed $N,$ is lost in the limit.
This is why we work with integral formulas instead of partial differential equations. The rigorous version
of the above (standard) argument, which will be presented in Section \ref{sec:proof}, resorts to the convergence 
of \eqref{eq:fjexp} to \eqref{eq:fnjexp}, and requires only continuity of the limiting initial data $f_j(0).$

%%%%%%%%%%%%%%%%%%%%%%%%%%%%%%%%%%%%%%%%%%%%%%%%%%%%%%%
%%%%%%%%%%%%%%%%%%%%%%%%%%%%%%%%%%%%%%%%%%%%%%%%%%%%%%%
%%%%%%%%%%%%%%%%%%%%%%%%%%%%%%%%%%%%%%%%%%%%%%%%%%%%%%%
%%%%%%%%%%%%%%%%%%%%%%%%%%%%%%%%%%%%%%%%%%%%%%%%%%%%%%%
%%%%%%%%%%%%%%%%%%%%%%%%%%%%%%%%%%%%%%%%%%%%%%%%%%%%%%%

%\newpage
\section{Assumptions and results} \label{sec:results}
\setcounter{equation}{0}    
\def\theequation{4.\arabic{equation}}

We establish here the hypotheses under which we will work. We recall that $N\in \mathbb{N}$ and $\e>0$ will be always related
through the condition
\be
N\e^2 = 1\;,
\ee
with $\e < < 1$.

Beyond Hypothesis \ref{hyp:pot} on the interaction potential stated in Section \ref{sec:2bs}, we assume
\bigskip
\begin{hyp}\label{hyp:f0bound}
Let $f _{0, j}:=f_0^{\otimes j}$ be the initial condition for the series solution to the Boltzmann equation 
\eqref{eq:fjexp}.
We assume that $f_0$ is a probability density function over $\RRR^{6},$ continuous, and satisfying the bound
\be
\sup_{(x,v)\in\RRR^{6}} e^{\frac{\b}{2} v^2} f_0 (x,v) < +\infty
\ee
for some constant $\b >0.$
\end{hyp} 
\bigskip

Furthermore, indicating by $H(\bz_j)$ the $j-$particle Hamiltonian written in macroscopic variables,
\be
H(\bz_j) = \frac{1}{2}\sum _{i=1}^j v_i ^2+\frac{1}{2}\sum_{\substack {i,k=1 \\ i\neq k}}^j 
\Phi\left(\frac{x_i -x_k}{\e}\right)\;,
\ee
we have
\bigskip
\begin{hyp}\label{hyp:f0jbound}
For any $N$, let $W^N_0$ be a probability density over the phase space $\MM_N$.
We assume that $W^N_0$ is a Borel function, symmetric in the particle labels, with reduced marginals 
$\{f^N_{0,j}\}_{j=1}^N$, given by Definition \ref{def:defrm}. Moreover, there exist two constants $\a,\b > 0$
(independent of $N$) such that 
\be
f ^N_{0, j}(\bz_j) e^{\b H(\bz_j)}\leq e^{\a j}\;, \label{eq:f0jbound}
\ee
uniformly in $N$ and $\bz_j\in\MM_j$.
\end{hyp} 
\bigskip

By Proposition \ref{prop:BBGKY}, the $f^N_{0,j}$ are good initial data for the evolutions \eqref{eq:fnjexp} and 
\eqref{eq:BBGKYt}. Note also that Hypothesis \ref{hyp:f0jbound}
implies that we are fixing correlations even at time zero. Indeed, if the interaction potential diverges at the origin,
$f ^N_{0, j}(\bz_j) \to 0$ exponentially whenever $x_k \to x_i$ for $k\neq i$. Therefore, initial product states are excluded. 
This situation is similar to that of hard--sphere 
systems, in which an overlapping of any pair of particles is not allowed. Even if the potential is bounded, but positive at
the origin (which is the case of stable interactions), product states are forbidden by Hypothesis \ref{hyp:f0jbound}. 
In fact, near the diagonal ($x_k=x_i$) the factor $e^{\b H(\bz_j)}$ can grow exponentially with $j^2.$ 

%\newpage
Our last hypothesis is
\bigskip
\begin{hyp}\label{hyp:conv}
Given $f ^N_{0, j}$ and $f_{0,j}=f_0^{\otimes j}$ as introduced in the previous hypotheses, there holds
\be
\lim_{\substack{\e\rightarrow 0\\ N\e^2=1}} f^N_{0,j} = f_{0,j}\;,
\ee
uniformly on compact sets in $\MM_j$.
\end{hyp} 
\bigskip 

We are now ready to state our first result. Let us introduce a notation for the subset of particles
that cannot collide pointwise under the free evolution:
\be
\O_j = \Big\{ \bz_j \in \MM_j \ \Big|\  \left(x_i-x_k\right)\wedge\left(v_i-v_k\right) \neq 0\Big\}\;. \label{eq:subsetsp}
\ee
\bigskip
\begin{thm} \label{thm:soft}
Assume the Hypotheses \ref{hyp:pot}--\ref{hyp:conv}. 
Let $f_j^N(t)$ be the reduced marginals at time $t>0$, evolved according to
Eq. \eqref{eq:fnjexp} and let $f_j(t)$ be defined as in \eqref{eq:fjtdef} and \eqref{eq:fjexp}.
Then, there exists $t_0>0$ such that, for any positive $t<t_0$ and $j\in \mathbb{N}$,
the series expansions \eqref{eq:fnjexp} and \eqref{eq:fjexp} are absolutely convergent (uniformly in $\e$),
and
\be
\lim_{\substack{\e\rightarrow 0\\ N\e^2=1}} f_j^N(t) = f_j(t)
\ee
uniformly on compact sets in $\O_j.$
\end{thm}
\bigskip
Theorem \ref{thm:soft} is formulated and proven in the same spirit of \cite{La75} and \cite{Ki75}. As we shall 
see in Section \ref{sec:proofsoft}, the proof, based on geometrical arguments, is abstract and does not give informations
on the rate of convergence. However, the result can be improved under quantitative assumptions on the rate of 
convergence and the continuity of the initial data, as explained in what follows.

Define
\be
\MM_j(\d) = \Big\{ \bz_j\in\RRR^{6j} \ \Big|\ |x_i- x_k| > \d,\  i,k =1 \cdots j,\  k\neq i \Big\} \label{eq:defMMjd}
\ee
for $\d>0.$ We assume
\begin{hyp} \label{hyp:convrate}
%For some $\g, C'>0$ and some $\d=\d(\e)>0$ such that $\d\underset{\e\to 0}{\longrightarrow}0, 
%\frac{\d}{\e}\underset{\e\to 0}{\longrightarrow}\infty,$
%it is
%%
%\be
%\sup_{\bz_j \in \MM_j(\d)} \left | f_{0,j}(\bz_j)-f^N _{0,j}(\bz_j) \right | \leq C' e^{\alpha j}\e^\g\;,
%\ee
%%
Given $f ^N_{0, j}$ and $f_{0,j}=f_0^{\otimes j}$ as introduced in the previous hypotheses,
for some $C'>0$,
\be
\sup_{\bz_j \in \MM_j(\e)} e^{\frac{\b}{2}\sum_{i=1}^{j}v_i^2}\left| f_{0,j}(\bz_j)-f^N _{0,j}(\bz_j) \right| \leq (C')^j \e\;.
\label{eq:hypcr1}
\ee
Moreover, for some $L>0,$
\be
e^{\frac{\b}{2}v^2}\left|f_0(x,v)-f_0(x',v)\right| \leq L \left| x-x' \right|. \label{eq:hypcr2}
\ee
\end{hyp}  
Then we have the following:
\bigskip
\begin{thm} \label{thm:hard}
Assume the Hypotheses $1-5$. Let $f^N_j(t)$ and $f_j(t)$ as in Theorem \ref{thm:soft}.
Then, for all $\bz_j \in \O_j,j\in \mathbb{N}$ and $t<t_0$, 
there exists a positive $\e_0=\e_0(\bz_j)$ and constants $C>0, \g>0$ such that, for $\e<\e_0$,
\be
|f_j^N(\bz_j,t) - f_j(\bz_j,t)| \leq C^j\e^\g\;.
\ee
\end{thm}
\bigskip

Observe that Hypotheses \ref{hyp:conv}, \ref{hyp:convrate} are a natural notion of convergence compatible with the continuity
of $f_0$ and the estimate \eqref{eq:f0jbound} (which prevents convergence on the diagonals $x_i=x_k$). To clarify this point,
we construct some explicit example in the next subsection.

%%%%%%%%%%%%%%%%%%%%%%%%%%%%%%%%%%%%%%%%%%%%%%%%%%%%%%%
%%%%%%%%%%%%%%%%%%%%%%%%%%%%%%%%%%%%%%%%%%%%%%%%%%%%%%%
%%%%%%%%%%%%%%%%%%%%%%%%%%%%%%%%%%%%%%%%%%%%%%%%%%%%%%%

\subsection{An example of initial datum} \label{sec:initial}
\setcounter{equation}{0}    
\def\theequation{4.1.\arabic{equation}}

In the following we present a sequence of probability density functions $W^N_0$ 
satisfying Hypotheses \ref{hyp:f0jbound}--\ref{hyp:convrate}. 
Set, for $N = 1,2,\cdots,$
\be
W^N_0(\bz_N)=\frac{1}{\ZZ_N}f^{\otimes N}_0(\bz_N) \prod_{1\leq i < k \leq N}
\mathbbm{1}_{\mbox{$\{|x_i-x_k|>\e\}$}}(\bz_N)\;,
\label{eq:exampleWN0}
\ee
where
\be
\ZZ_N=\int_{\RRR^{6N}} d\bz_N f^{\otimes N}_0(\bz_N)\prod_{1\leq i < k \leq N}\mathbbm{1}_{\mbox{$\{|x_i-x_k|>\e\}$}}(\bz_N)
\label{eq:pfSRGT}
\ee
is the ``partition function'' and $f_0$ is some density satisfying Hypothesis \ref{hyp:f0bound} and \eqref{eq:hypcr2}.
The corresponding reduced marginals are
\be
f_{0,j}^N(\bz_j) = \frac{F^N(\bz_j)}{\ZZ_N}f^{\otimes j}_0(\bz_j) 
\prod_{1\leq i < k \leq j}\mathbbm{1}_{\mbox{$\{|x_i-x_k|>\e\}$}}(\bz_j)
\ee
with
\bea
&& F^N(\bz_j) =\int_{\RRR^{6(N-j)}} d\bz_{j,N} 
f^{\otimes (N-j)}_0(\bz_{j,N})\left(\prod_{i=1}^j\prod_{k=j+1}^N\mathbbm{1}_{\mbox{$\{|x_i-x_k|>\e\}$}}(\bz_N)\right)\nn\\ 
&& \ \ \ \ \ \ \ \ \ \ \ \ \ \ \ \ \ \ \ \ \ \ \ \ \ \ \ \ \ \ \ \ \ \ \ \ \ \ \ \ \ \ \ \ \ \ 
\cdot\left(\prod_{j+1\leq i < k \leq N}\mathbbm{1}_{\mbox{$\{|x_i-x_k|>\e\}$}}(\bz_{j,N})\right)\;. \nn\\
\eea

\bigskip
%\newpage
\begin{prop} The sequence of functions defined by \eqref{eq:exampleWN0}--\eqref{eq:pfSRGT},
satisfies the Hypotheses \ref{hyp:f0jbound}--\ref{hyp:convrate}.
\end{prop}
%\bigskip

{\em Proof.} Let us estimate $F^N(\bz_j)\ZZ_N^{-1}$.  

First observe that, for some $C_0>0,$
\be
\ZZ_{N-j}(1-C_0N\e^3)^j \leq \ZZ_N \leq \ZZ_{N-j}\;. \label{eq:ZZNbound}
\ee
The upper bound is obvious consequence of the normalization of $f_0.$ As regards the lower bound, note that
\bea
&& \ZZ_N = \int_{\RRR^{6(N-1)}} d\bz_{N-1} f^{\otimes(N-1)}_0(\bz_{N-1})
\prod_{1\leq i < k \leq N-1}\mathbbm{1}_{\mbox{$\{|x_i-x_k|>\e\}$}}(\bz_{N-1})\nn\\
&&\ \ \ \ \ \cdot\int_{\RRR^6} dz_N f_0(z_N)\prod_{i=1}^{N-1}\mathbbm{1}_{\mbox{$\{|x_i-x_N|>\e\}$}}(z_N) \nn\\
&& \geq \int_{\RRR^{6(N-1)}} d\bz_{N-1} f^{\otimes(N-1)}_0(\bz_{N-1})
\prod_{1\leq i < k \leq N-1}\mathbbm{1}_{\mbox{$\{|x_i-x_k|>\e\}$}}(\bz_{N-1})\nn\\
&&\ \ \ \ \ \cdot\int_{\RRR^6} dz_N f_0(z_N)\left(1-\sum_{i=1}^{N-1}\mathbbm{1}_{\mbox{$\{|x_i-x_N|\leq \e\}$}}(z_N)\right)\nn\\
&& \geq \ZZ_{N-1}(1-C_0(N-1)\e^3)\;,
\eea
for instance taking $C_0=(4\pi/3)\|f_0\|_{\infty}.$ Eq. \eqref{eq:ZZNbound} follows by iteration. 

We can also show that
\be
\ZZ_{N-j}\left(1-j\frac{C_0N\e^3}{1-C_0N\e^3}\right) \leq F^N(\bz_j) \leq \ZZ_{N-j}\;. \label{eq:FNbound}
\ee
The upper bound is immediate, while the lower bound follows from
\bea
&& F^N(\bz_j) \geq \int_{\RRR^{6(N-j)}} d\bz_{j,N} 
f^{\otimes (N-j)}_0(\bz_{j,N})\left(1-\sum_{i=1}^j\sum_{k=j+1}^N
\mathbbm{1}_{\mbox{$\{|x_i-x_k| \leq \e\}$}}(\bz_N)\right)\nn\\
&& \ \ \ \ \ \ \ \ \ \ \ \ \ \ \ \ \ \ \ \ \ \ \ \ \ \ \ \ \ \ \ \ \ \ \ \ \ \ \ \ \ \ \ \ \ \ 
\cdot\left(\prod_{j+1\leq i < k \leq N}\mathbbm{1}_{\mbox{$\{|x_i-x_k| > \e\}$}}(\bz_{j,N})\right)\nn\\
&&\ \ \ \ \ \ \ \ \ \ \geq \ZZ_{N-j}-j(N-j)C_0\e^3\ZZ_{N-j-1}\;, \nn\\
&&\ \ \ \ \ \ \ \ \ \ \geq \ZZ_{N-j}\left(1-jNC_0\e^3\frac{\ZZ_{N-j-1}}{\ZZ_{N-j}}\right)\;,
\eea
noticing that \eqref{eq:ZZNbound} implies $\ZZ_{N-j-1}\ZZ_{N-j}^{-1}\leq (1-C_0N\e^3)^{-1}.$

Since $N\e^2=1$, if $N$ is sufficiently large, Equations \eqref{eq:ZZNbound} and \eqref{eq:FNbound} give in turn the bounds
\be
1-2C_0j\e \leq \frac{F^N(\bz_j)}{\ZZ_N} \leq \frac{1}{(1-C_0\e)^j} \label{eq:FNZNbound}
\ee
and, in particular,
\be
\frac{F^N(\bz_j)}{\ZZ_N} \underset{N\to\infty}{\longrightarrow}1
\ee
uniformly in $\bz_j\in\MM_j$. 

Now it is easy to check that, for $N$ sufficiently large, the Hypotheses \ref{hyp:f0jbound}--\ref{hyp:convrate} are verified.
Hypothesis \ref{hyp:f0jbound} follows from Hypothesis \ref{hyp:f0bound}. Hypothesis \ref{hyp:convrate}
(hence \ref{hyp:conv}) follows from the estimates in \eqref{eq:FNZNbound}. \qed
% choosing $\eta = \e^{\g'}$ with $2/3<\g'<1$
%and taking for instance $\d=\e^{1/2}$ (so that $\d>\eta$). 
%The rate of convergence given by the above estimates is $\e^\g$ 
%with $\g = -2+3\g'.$

\bigskip
\ni {\bf Other examples.}
\medskip

\ni In definition \eqref{eq:exampleWN0} for the initial density we could also replace the product of characteristic functions by
$e^{-\b \sum_{i<k}\Phi\left(\frac{x_i-x_k}{\e}\right)},$ see \cite{GSRT12}. This defines a sequence of states which 
are, in a sense, the maximally uncorrelated states for which the Hypotheses are satisfied.

Finally, other families of initial data exhibiting a slower rate of convergence (and implying possibly a slower convergence 
in Theorem \ref{thm:hard}) can be easily constructed, for instance 
enlarging the cut--off in \eqref{eq:exampleWN0}. If in formula \eqref{eq:exampleWN0} $\e$ is replaced by 
$\e^{\g'}$ with $\g'\in(2/3,1]$, then, proceeding as before, we obtain 
\be\label{hyp:conv'}
\sup_{\bz_j \in \MM_j(\e^{\g'})} \left | f_{0,j}(\bz_j)-f^N _{0,j}(\bz_j) \right | \leq (C')^j\e^{-2+3\g'}\;.
\ee

%%%%%%%%%%%%%%%%%%%%%%%%%%%%%%%%%%%%%%%%%%%%%%%%%%%%%%%
%%%%%%%%%%%%%%%%%%%%%%%%%%%%%%%%%%%%%%%%%%%%%%%%%%%%%%%
%%%%%%%%%%%%%%%%%%%%%%%%%%%%%%%%%%%%%%%%%%%%%%%%%%%%%%%

\subsection{General strategy of the proof} \label{sec:strategy}
\setcounter{equation}{0}    
\def\theequation{4.2.\arabic{equation}}

The proof of our results follows the main ideas of \cite{La75}, adapted to the present context.
The validity argument is based on a comparison among the series for the $N-$particle system \eqref{eq:fnjexp}, and 
the Boltzmann series \eqref{eq:fjexp}.

- First, we prove that both the expansions are absolutely convergent series, for sufficiently short times and uniformly in $\e:$
see Section \ref{sec:ste}. As a consequence of the estimates in Section \ref{sec:ste}, it follows also that 
\eqref{eq:fnjexp} and \eqref{eq:BBGKYt} are asymptotically equivalent in the BG limit.

- Then, it remains to prove the term by term convergence of \eqref{eq:BBGKYt} to \eqref{eq:fjexp}. To do this, it is 
necessary a preliminary detailed analysis of the generic term. This is presented in Sections \ref{sec:tree} and \ref{sec:flow}
for the series \eqref{eq:BBGKYt}, and in Section \ref{sec:Bflow} for the Boltzmann series \eqref{eq:fjexp}. The structure
of the generic term is described with the help of a convenient representation of formulas in terms of tree graphs. It turns
out that any given term can be expressed as an integral over a set of special backwards--in--time trajectories of clusters
of particles.

- The proof of the term by term convergence is carried out in Section \ref{sec:proof}, using in a crucial way the picture
introduced in Section \ref{sec:tree}. The issues arising from the convergence will be first discussed in Section 
\ref{sec:convergence}, while the abstract proof leading to Theorem \ref{thm:soft} and the explicit 
estimates leading to Theorem \ref{thm:hard} will be presented in Sections \ref{sec:proofsoft} and 
\ref{sec:proofhard} respectively.

%%%%%%%%%%%%%%%%%%%%%%%%%%%%%%%%%%%%%%%%%%%%%%%%%%%%%%%
%%%%%%%%%%%%%%%%%%%%%%%%%%%%%%%%%%%%%%%%%%%%%%%%%%%%%%%
%%%%%%%%%%%%%%%%%%%%%%%%%%%%%%%%%%%%%%%%%%%%%%%%%%%%%%%
%%%%%%%%%%%%%%%%%%%%%%%%%%%%%%%%%%%%%%%%%%%%%%%%%%%%%%%
%%%%%%%%%%%%%%%%%%%%%%%%%%%%%%%%%%%%%%%%%%%%%%%%%%%%%%%

\section{Short time estimates} \label{sec:ste}
\setcounter{equation}{0}    
\def\theequation{5.\arabic{equation}}

The aim of this section is to prove that, for times $t$ smaller than a certain $t_0,$ the expansion for $f^N_j(t),$ 
Eq. \eqref{eq:fnjexp}, can be bounded uniformly in $\e.$ The Boltzmann series solution Eq. \eqref{eq:fjexp} 
turns out to be an absolutely convergent series for the same values of $t.$ Moreover, the difference between
$f^N_j(t)$ and $\tilde f^N_j(t)$ (defined by \eqref{eq:BBGKYt}) is negligible in the limit.
These results are easily established by assuming the bounds on the initial data in Hypothesis \ref{hyp:f0bound} and
\ref{hyp:f0jbound}. Here we will follow \cite{Ki75} (see also \cite{GSRT12, Uc88, Sp91, CIP94}).

To begin with, we notice that our assumptions on the initial data make natural the introduction of the norms:
\bea
&& \| g_j \|_\b = \sup_{\bz_j}e^{\b H(\bz_j) }|g_j(\bz_j)|\;,\ \ \ \ \ \ \ \ \ \ g_j: \MM_j\to\RRR\;,\ \ \ \b>0\;,\nn\\
&& \| g \|_{\b, \a} = \sup_{j \geq 1} e^{-\a j}\| g_j \|_\b\;,\ \ \ \ \ \ \ \ \ \ \ \ \ \ \ \ g=\{g_j\}_{j=1}^\infty\;,\ \ \ \ \ \a>0\;.
\label{eq:defnorms}
\eea
By the energy conservation
\be
\| \SS^\e (t) g \|_{\beta, \alpha}  =\| g \|_{\beta, \alpha}\;, \label{eq:encons}
\ee
for all $\beta$ and $\alpha$ for which the right hand side makes sense.

The crucial technical estimate is the following:
\begin{lem} \label{lem:Abound}
Let $g_j^N: \MM_j\to\RRR$ be a sequence of continuous\footnote{The continuity here is required only for simplicity 
of notation, since it assures well posedness of the operator action: see the Remark after Proposition \ref{prop:BBGKY}. 
If that is not true, the lemma must be reformulated for $\int_0^t ds \frac{s^{n-1}}{(n-1)!}\AA^\e\SS^\e(s)$ 
(that is what we really need to control for the proof of Proposition \ref{prop:ste} below). This can be done in an obvious
way using Eq. (\ref{eq:encons}) and adding a factor $t^n/n!$ in the right hand side of the estimate.}
functions with $g_j^N = 0$ for $j>N$ and satisfying the estimate of Hypothesis \ref{hyp:f0jbound}. Set 
$\AA^\e g^N = \Big\{\sum_{m \geq 0}\AA_{j+1+m}^\e g_{j+1+m}^N\Big\}_{j=1}^\infty.$ Then, given $\b'<\b$ and $\a'>\a$,
there exists a pure constant $\bar C>0$ such that, for $\e$ small enough,
\be
\|\AA^\e g^N\|_{\b',\a'} \leq \bar C
\left(\frac{1}{\sqrt{(\beta-\beta')(\alpha'-\alpha)}}+\frac{1}{\alpha'-\alpha}\right)\|g^N\|_{\b,\a}\;.
\label{eq:Abound}
\ee
\end{lem}

{\em Proof.} From definition \eqref{eq:defA} we find
\bea
&& e^{\b'H(\bz_j)}|\AA^\e_{j+1+m} g_{j+1+m}^N(\bz_j)|\leq (N-j-1)\cdots (N-j-m)\sum_{i=1}^j \int_{S^2} d\n \|g^N_{j+1+m}\|_{\b}\nn\\
&& \ \ \ \ \ \ \ \ \ \ \ \ \ \ \ \ \ \ \ \ \ \ \ \ \ \ \ \ \ \ \ \ \ \ \ 
\cdot \int dv_{j+1}(|v_i|+ |v_{j+1}|) e^{-(\b-\b') H(\bz_j)}e^{-\frac{\b}{2}v_{j+1}^2}\nn\\
&& \ \ \ \ \ \ \ \ \ \ \ \ \ \ \ \ \ \ \ \ \ \ \ \ \ \ \ \ \ \ \ \ \ \ \ 
\cdot\int_{\Delta_m(\bx_{j+1})} \frac{d\bz_{j+1,m}}{m!}e^{-\frac{\b}{2}\sum_{i=j+2}^{j+1+m}v^2_i}\;. \label{eq:boundAgNj}
\eea
Here we used that $\e^2(N-j)\leq 1$ and the positivity of the interaction (Hypothesis \ref{hyp:pot}), for which
\be
H(\bz_{j+1+m}) = H(\bz_j)+H(\bz_{j,1+m})\geq H(\bz_j)+\frac{1}{2}\sum_{i=j+1}^{j+1+m} v^2_i\;. \label{eq:HgHpH}
\ee
The last integral in the right hand side is bounded by
$\left(\frac{2\pi}{\b}\right)^{\frac{3}{2}m}\left(\frac{4\pi}{3}\right)^m \e^{3m}\;,$
so that \eqref{eq:boundAgNj} implies
\bea
&& e^{\b'H(\bz_j)}|\AA^\e_{j+1+m} g_{j+1+m}^N(\bz_j)|\label{eq:boundAgNj'}\\
&& \leq
\|g^N_{j+1+m}\|_{\b}(C_\b\e)^m\sum_{i=1}^j\int dv_{j+1}(|v_i|+ |v_{j+1}|) e^{-\frac{\b-\b'}{2}\sum_{i=1}^jv^2_i }
e^{-\frac{\b}{2}v_{j+1}^2}\;, \nn
\eea
where we used again the positivity of the interaction, and $C_\b$ is a suitable constant.
A Cauchy--Schwarz inequality gives
\be
\sum_{i=1}^j (|v_i|+|v_{j+1}|) \leq \sqrt{j\sum_{i=1}^jv^2_i} + j|v_{j+1}|\;,
\ee
which inserted into \eqref{eq:boundAgNj'} leads to
\be
\|\AA^\e_{j+1+m} g_{j+1+m}^N\|_{\b'} \leq (C'_\b\e)^m\|g^N_{j+1+m}\|_{\b}
\left(\frac{\sqrt{j}}{\sqrt{\b-\b'}}+j\right)\;. \label{eq:Aboundstep}
\ee
Summing over $m$ and taking the supremum over $j$ with weight $e^{-\a' j}$ we readily get \eqref{eq:Abound},
for $\e$ smaller than a constant depending only on $\b,\a.$ \qed

\bigskip
Let us apply Lemma \ref{lem:Abound}, together with \eqref{eq:encons}, to the right hand side of \eqref{eq:fnjexp}.
We proceed by iteration. For a given $n>0,$ we partition the intervals $[\b/2,\b]$ and $[\a,2 \a]$ in $n$ intervals of 
the same length $\frac{\b}{2n}$ and $\frac{\a}{n},$ and then apply the above results $n$ times. The outcome is
\bea
&&\|f^N(t)\|_{\b/2,2\a} \leq \sum_{n\geq 0}\int_0^t dt_1 \int_0^{t_1} dt_2 \cdots \int_0^{t_{n-1}}dt_n (C_{\b,\a}n)^n
\|f^N_0\|_{\b,\a}\nn\\
&&\ \ \ \ \ \ \ \ \ \ \ \ \ \ \ \ \ \ =  \sum_{n\geq 0} \frac{t^n}{n!} (C_{\b,\a}n)^n \|f^N_0\|_{\b,\a}\nn\\
&&\ \ \ \ \ \ \ \ \ \ \ \ \ \ \ \ \ \ \leq  \|f^N_0\|_{\b,\a} \sum_{n\geq 0} (tC'_{\b,\a})^n\;, \label{eq:finalest}
\eea
for suitable constants $C_{\b,\a},C'_{\b,\a},$ having used Stirling formula in the last step. Hence we obtained a geometric
series which converges for $t$ sufficiently small (and the radius of convergence is explicitly computable in terms of the
other constants). Now the same argument can be applied in a straightforward way to the series \eqref{eq:BBGKYt} and 
\eqref{eq:fjexp}. Thus, we have proven the first statement of:
\bigskip
\begin{prop} \label{prop:ste}
In the Hypotheses \ref{hyp:f0bound} and \ref{hyp:f0jbound}, we have absolute convergence of the series \eqref{eq:fnjexp}, 
\eqref{eq:BBGKYt} (uniformly in the BG limit for $\e$ small enough) and \eqref{eq:fjexp}, for all $t<t_0 = t_0(\b,\a).$
Moreover, for some $C''>0,$ if $\e$ is small enough,
\be
\| f^N(t)-\tilde f^N(t) \|_{\b/2,2\a} \leq C'' \e\;. \label{eq:clusterest}
\ee
\end{prop}
\bigskip

{\em Proof.} 
We just need to prove Eq. \eqref{eq:clusterest}. Set $\CC^\e g^N = \Big\{\e^2(N-j)\CC_{j+1}^\e g_{j+1}^N\Big\}_{j=1}^\infty.$
With the notations of Lemma \ref{lem:Abound} and proceeding in the same way, we observe that 
\bea
&&\| (\AA^\e-\CC^\e)g^N \|_{\b',\a'}\leq \sup_{j\geq 1} e^{-\a'j}\sum_{m\geq 1}
\|\AA^\e_{j+1+m}g^N_{j+1+m}\|_{\b'}\nn\\
&&\ \ \ \ \ \ \ \ \leq \sup_{j\geq 1} e^{-\a'j}\sum_{m\geq 1}(C'_\b\e)^m\|g^N_{j+1+m}\|_{\b}
\left(\frac{\sqrt{j}}{\sqrt{\b-\b'}}+j\right) \nn\\
&&\ \ \ \ \ \ \ \ \leq C''_{\b,\a}\e
\left(\frac{1}{\sqrt{(\beta-\beta')(\alpha'-\alpha)}}+\frac{1}{\alpha'-\alpha}\right)
\|g^N\|_{\b,\a}
\eea
for suitable $C''_{\b,\a}>0,$ having used \eqref{eq:Aboundstep} in the second inequality, and $\e$ sufficiently 
small in the third. Therefore, proceeding as in \eqref{eq:finalest},
\bea
&& \| f^N(t)-\tilde f^N(t) \|_{\b/2,2\a} \leq 
\sum_{n\geq 0} \frac{t^n}{n!}\sum_{k=1}^n\binom{n}{k}\e^k (C_{\b,\a}n)^n \|f^N_0\|_{\b,\a}\nn\\
&&\ \ \ \ \ \ \ \ \ \ \ \ \ \ \ \ \ \ \ \ \ \ \ \ \ \ \ \ \ \leq \e\|f^N_0\|_{\b,\a} \sum_{n\geq 0} (t2C'_{\b,\a})^n\;,
\eea
which gives the result with $C''$ depending on $\b,\a$ and on the initial datum. \qed

%%%%%%%%%%%%%%%%%%%%%%%%%%%%%%%%%%%%%%%%%%%%%%%%%%%%%%%
%%%%%%%%%%%%%%%%%%%%%%%%%%%%%%%%%%%%%%%%%%%%%%%%%%%%%%%
%%%%%%%%%%%%%%%%%%%%%%%%%%%%%%%%%%%%%%%%%%%%%%%%%%%%%%%
%%%%%%%%%%%%%%%%%%%%%%%%%%%%%%%%%%%%%%%%%%%%%%%%%%%%%%%
%%%%%%%%%%%%%%%%%%%%%%%%%%%%%%%%%%%%%%%%%%%%%%%%%%%%%%%

\section{The tree expansion} \label{sec:tree}
\setcounter{equation}{0}    
\def\theequation{6.\arabic{equation}}

In the proof of Theorems \ref{thm:soft} and \ref{thm:hard} it is convenient to represent each term 
of the expansions \eqref{eq:BBGKYt} and \eqref{eq:fjexp} as more explicit integrals of the initial data, 
$f^N_{0,j}$ and $f_{0,j}$ respectively. As we will see in the present section, it is natural to express such terms 
by means of binary trees which help us to visualize the various contributions.

Consider first Eq. \eqref{eq:BBGKYt} which, reminding Eq. \eqref{eq:Cedec}, we rewrite as
\bea
&& \tilde f_j^N(t) = \sum_{n = 0}^{N-j}\alpha ^\e_n(j) \sum_{\bs_n} {\sum_{\bk_n}}^* \left(\prod_{i=1}^n \s_i \right)
\int_0^t dt_1 \int_0^{t_1}dt_2\cdots\int_0^{t_{n-1}}dt_n \nn\\
&&\ \ \ \ \ \ \ \ \ \ \ \ \ \ \ \ \ \cdot\SS_j^\e(t-t_1)\CC^{\e,\s_1}_{k_1,j+1}\SS_{j+1}^\e(t_1-t_2)\cdots 
\CC^{\e,\s_n} _{k_n,j+n}\SS^\e_{j+n}(t_n) f^N_{0,j+n}\;, \label{eq:tfnjexp}
\eea
where
\bea
&& \bs_n=(\s_1, \cdots, \s_n), \ \ \ \ \ \ \ \ \ \  \s_i=\pm\;,% \ \ \ \ \  |\bs_n|=\sum_{i=1}^n \d_{\s_i,-}\;,
\nn\\
&& {\sum_{\bk_n}}^* = \sum_{k_1=1}^j \sum_{k_2=1}^{j+1}\cdots \sum_{k_n=1}^{j+n-1}\;. \label{eq:specialsumk}
\eea
%
%and $\d$ is the Kronecker delta.

\begin{figure}[htbp] %  figure placement: here, top, bottom, or page
\centering
\includegraphics[width=5in]{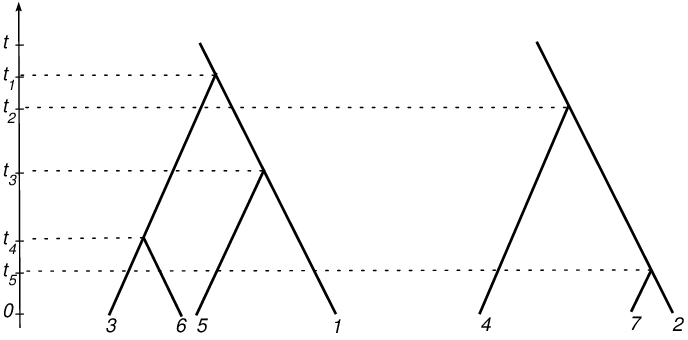} 
\caption{Tree graph $\G(2,5)=1, 2, 1, 3, 2.$ We have also drawn a time arrow in order to associate times to the nodes of the trees:
at time $t_i$ the line $j+i$ is``created''. Lines $1$ and $2$ exist for all times; they are called``root lines''.}
\label{fig:treedef}
\end{figure}

We introduce the {\bf{\em $n-$collision, $j-$particle tree graph}}, denoted $\G(j,n)$, as the collection of integers
$k_1,\cdots,k_n$ that are present in the sum \eqref{eq:specialsumk}, i.e.
\be
k_1\in I_j, k_2 \in I_{j+1}, \cdots, k_n\in I_{j+n-1}\;,\ \ \ \ \ \ \mbox{with\ \ \ \ \ \ $I_s=\{1,2,\cdots,s\},$}
\ee
so that we shall write
\be
{\sum_{\bk_n}}^* = \sum_{\G(j,n)}\;. \label{eq:defsumtrees}
\ee
Note that the number of terms in the sum is $j(j+1)\cdots(j+n-1).$
The name {\em tree graph} is justified by the fact that it has a natural graphical representation. 
This is best explained by an example: see Figure \ref{fig:treedef} which corresponds to $\G(2,5)$ given by $1, 2, 1, 3, 2.$

Given a tree graph $\G(j,n),$ and fixed a value of $\bs_n$ and of all the integration variables in the expansion 
\eqref{eq:specialsumk} (times, unit vectors, velocities), we can associate to it a special ($\e-$dependent)
trajectory of particles, which we call {\bf{\em interacting backwards flow}}              (IBF in the following), since it will be 
naturally defined by going back in time. The rules for the construction of this evolution will be explained in 
the next section. The notation for a configuration of particles in the IBF will make use of Greek alphabet i.e.
$\bze^\e(s)$,
where $s \in [0,t]$ is the time and there is {\em no} label specifying the number of particles. If
$s \in (t_{r+1},t_r)$ (with the convention $t_0=t, t_{n+1}=0$) we have $j+r$ particles:
\be
\bze^\e(s) = (\z_1^\e(s),\cdots,\z_{j+r}^\e(s)) \in \MM_{j+r}\ \ \ \ \ \mbox{for }s \in (t_{r+1},t_r)\;, \label{eq:IBFnot}
\ee
with
\be
\z_i^\e(s)=(\xi^\e_i(s),\eta^\e_i(s))\;,
\ee
the positions and velocities of all the particles being respectively
\bea
&& \bxi^\e(s) =(\xi_1^\e(s), \cdots, \xi_{j+r}^\e(s))\;, \nn\\
&& \bet^\e(s)=(\eta_1^\e(s), \dots, \eta_{j+r}^\e(s))\;. \label{eq:IBFnot'}
\eea

\bigskip
\ni {\bf Rewriting of \eqref{eq:BBGKYt} as a sum over tree graphs}
\medskip

\ni  The reason to introduce these trajectories is that we want a more explicit expression of each term of the expansion
\eqref{eq:tfnjexp}, namely our purpose is to write Eq. \eqref{eq:tfnjexp} as
\be\label{fjN-grad}
\tilde f^{N}_j (\bz_j,t)= \sum_{n=0}^{N-j}\alpha^\e_n(j)\sum _{\G (j,n)}\sum _{\bs_n}\prod_{i=1}^n \s_i 
\T_{\bs_n}^\e (\bz_j,t)\;,
\ee
where 
\be
\T_{\bs_n}^\e (\bz_j,t)= \int d \L (\bt_n , \bn_n , \bv_{j,n})\prod_{i=1}^n 
B^\e(\n_i; v_{j+i} - \eta^\e_{k_i} (t_i))f^{N}_{0,j+n}(\bze^\e(0))\;, \label{eq:Teszt}
\ee
$d\L$ is the measure on $\RRR^n\times S^{2n}\times\RRR^{3n}$ given by
\be
d\Lam ({\bf  t}_n , \bn_n ,\bv_{j,n})= \mathbbm{1}_{\{t_1>t_2 \dots >t_n\}} dt_1\dots dt_n
d\nu_1\dots d\nu_n dv_{j+1}\dots dv_{j+n}\;,
\ee
and we use the short notation
\bea
B^\e(\n_i; v_{j+i}-\eta^\e_{k_i} (t_i))=|\n_i \cdot (v_{j+i}-  \eta^\e_{k_i} (t_i))|
\mathbbm{1}_{\{\s_i\n_i \cdot (v_{j+i}-\eta^\e_{k_i} (t_i)) \geq 0\}}
\mathbbm{1}_{\{|\xi^\e_{j+i}(t_i)-\xi^\e_{k}(t_i)| > \e\  \forall k\neq k_i\}}\;.\nn\\\label{eq:defBe}
\eea
In other words, in the generic term $\T_{\bs_n}^\e (\bz_j,t),$ the initial datum $ f^{N}_{0,j+n}$ is integrated,
with the suitable weight, over all the possible time--zero states of the IBF associated to $\G(j,n), \bs_n$.

%%%%%%%%%%%%%%%%%%%%%%%%%%%%%%%%%%%%%%%%%%%%%%%%%%%%%%%
%%%%%%%%%%%%%%%%%%%%%%%%%%%%%%%%%%%%%%%%%%%%%%%%%%%%%%%
%%%%%%%%%%%%%%%%%%%%%%%%%%%%%%%%%%%%%%%%%%%%%%%%%%%%%%%

\subsection{The interacting backwards flow (IBF)} \label{sec:flow}
\setcounter{equation}{0}    
\def\theequation{6.1.\arabic{equation}}

Let us construct $\bze^\e(s)$ for a fixed collection of variables $\G(j,n), \bs_n, \bz_j,\bt_n,\bn_n,\bv_{j,n},$ with
\be
t\equiv t_0 > t_1 > t_2 > \cdots > t_n > t_{n+1}\equiv 0\;, \label{eq:ordertimes}
\ee
and $\bn_n$ satisfying a further constraint that will be specified below. The $j$ root lines of the tree graph are associated 
to the first $j$ particles, with states $\z_1^\e,\cdots,\z_j^\e.$ Each branch $j+\ell$ ($\ell=1,\cdots,n$) represents a new 
particle with the same label, and state $\z_{j+\ell}^\e.$ This new particle appears, going backwards in time, at 
time $t_\ell$ in a collision state with a previous particle (branch) $k_\ell\in\{1, \cdots, j+\ell -1\},$ with either
incoming or outgoing velocity according to $\sigma_\ell=-$ or $\sigma_\ell=+$ respectively.

More precisely, in the time interval $(t_r,t_{r-1})$ particles $1,\cdots,j+r-1$ flow according to the usual dynamics
$\TT^\e_{j+r-1}.$ This defines $\bze^\e_{j+r-1}(s)$ starting from $\bze^\e_{j+r-1}(t_{r-1}).$ At time $t_r$ the 
particle $j+r$ is``{\em created}''by particle $k_r$ in the position
\be
\xi_{j+r}^\e(t_r)= \xi_{k_r}^\e(t_r)+ \nu_r \e \label{eq:addedpos}
\ee
and with velocity $v_{j+r}$. This defines $\bze^\e(t_r) = (\z_1^\e(t_r),\cdots,\z^\e_{j+r}(t_r)).$ After that, the evolution
in $(t_{r+1},t_r)$ is contructed applying to this configuration the dynamics $\TT^\e_{j+r}$ (with negative times).
The characteristic function in the collision operator \eqref{eq:Cedec} (or the second characteristic function in \eqref{eq:defBe}),
is a constraint on $\n_r$ implying that no third particle is closer than $\e$ to the pair $k_r,j+r$ at the time $t_r.$

\begin{figure}[htbp] %  figure placement: here, top, bottom, or page
\centering
\includegraphics[width=6in]{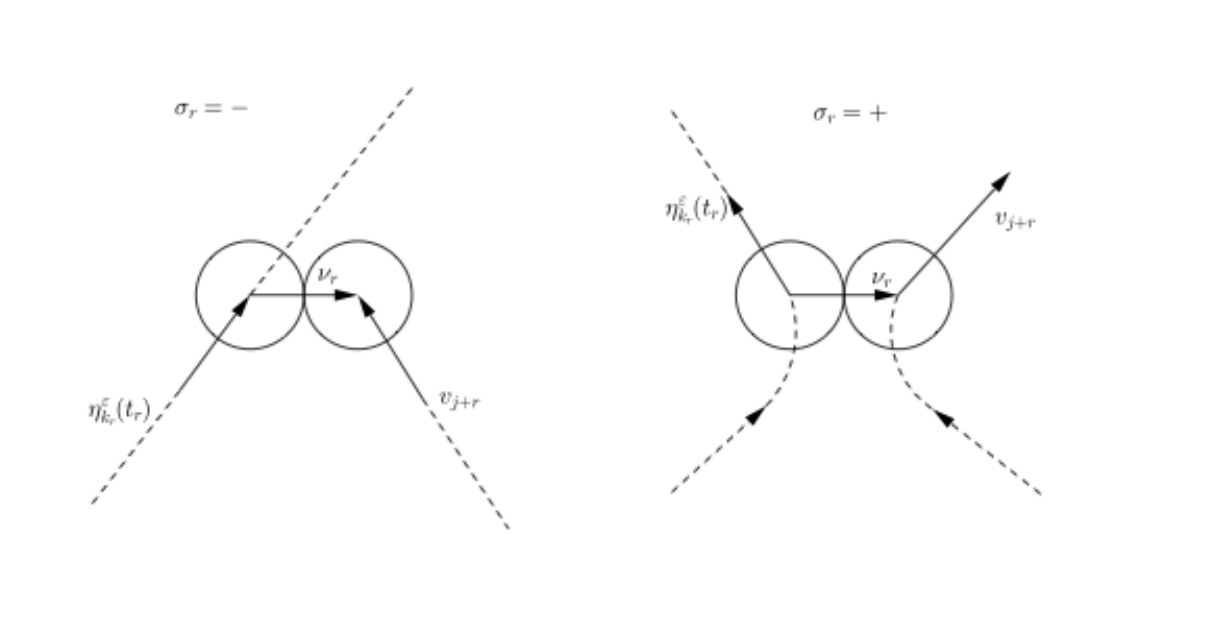} 
\caption{At time $t_r,$ particle $j+r$ is {\em created} by particle $k_r,$ either in incoming ($\s_r=-$) or in outgoing
($\s_r=+$) collision configuration. Particle $k_r$ is called the {\em progenitor} of particle $j+r.$}
\label{fig:creations}
\end{figure}

We have two cases. If $\sigma_r=-,$ then it must be $\n_r \cdot (v_{j+r}-\eta^\e_{k_r}(t_r)) \leq 0$. In this case the velocities 
are incoming and no scattering occurs, namely after $t_r$ the pair of particles moves backwards freely with velocities 
$\eta^\e_{k_r}(t_r)$ and 
$v_{j+r}$. If  $\sigma_r=+,$ we require $\n_r \cdot (v_{j+r}-\eta^\e_{k_r}(t_r)) \geq 0$ so that the pair is post--collisional. 
Then the presence of the interaction in the flow $\TT^\e_{j+r}$ forces the pair to perform a (backwards) scattering. 
The two situations are illustrated in Fig. \ref{fig:creations}.

\bigskip
{\em Remark.}\label{rem:reco}
It is very important to note that between two creation times $t_r, t_{r+1}$ any pair of particles among the $j+r,$ different
from the couple $(k_r,j+r),$ can possibly interact by reaching (or having from the beginning) a distance smaller than $\e.$ 
These interactions are called {\bf{\em recollisions}}, because they generally involve particles that have already interacted at some creation time
(in the future) with another particle of the IBF. In our language, recollisions are the ``interactions different from creations''.
Though recollisions are expected to be unlikely, we will have to analyze them with special care, since they are the main 
responsible of the different behavior of the particle dynamics from the Boltzmann evolution.

%%%%%%%%%%%%%%%%%%%%%%%%%%%%%%%%%%%%%%%%%%%%%%%%%%%%%%%
%%%%%%%%%%%%%%%%%%%%%%%%%%%%%%%%%%%%%%%%%%%%%%%%%%%%%%%
%%%%%%%%%%%%%%%%%%%%%%%%%%%%%%%%%%%%%%%%%%%%%%%%%%%%%%%

\subsection{The Boltzmann backwards flow (BBF)} \label{sec:Bflow}
\setcounter{equation}{0}    
\def\theequation{6.2.\arabic{equation}}

The discussion of the two previous sections can be repeated, with minor changes, for the case of Boltzmann series 
\eqref{eq:fjexp}. The interacting backwards flow is now substituted by the {\em Boltzmann backwards flow} (BBF)
$
\bze(s)\;,
$
for which we use the same notations of  \eqref{eq:IBFnot}--\eqref{eq:IBFnot'} with the superscript $\e$ 
omitted. The BBF is introduced exactly as the IBF, see Section \ref{sec:flow}, except for the following differences:

\ni - the interacting dynamics $\TT^\e$ is replaced by the simple free dynamics;

\ni - in the right hand side of \eqref{eq:addedpos} the second term is missing, i.e. the created particle appears at the same
position of its progenitor;

\ni - there is no constraint on $\n_r$ other than the one implied by the value of $\s_r;$

\ni - if $\s_r=+,$ to determine the state of particles in $(t_{r+1},t_r),$ {\em before} applying free evolution we have to
change velocities according to $(\eta_{k_r}(t_r^+),v_{j+r}) \to (\eta_{k_r}(t_r^-),\eta_{j+r}(t_r^-)),$ where $\to$ denotes the
scattering rule depicted in \eqref{eq:coll} and Figure \ref{fig:2bscatter}. Here $\eta_{k_r}(t_r^+)$ indicates the limit from
the future, and $\eta_{k_r}(t_r^-)$ the limit from the past.

\bigskip
{\em Remark.} \label{rem:BBF}
An important point concerning the Boltzmann backwards flow defined above is that,
in a given interval $(t_{r+1},t_r),$ the velocities $\eta_i(s)=\eta_i(t_r^-),$ besides being constant, depend only
on the velocities of particles in the future of the BBF, and on the vectors of impact $\n_1,\cdots,\n_r,$ but {\em not}
on the interaction times $t_1,\cdots,t_r.$
This simple structure of the BBF will be exploited later on (see for instance Equation \eqref{eq:softtcpr}).
\bigskip

\bigskip
\ni {\bf Rewriting of \eqref{eq:BBGKYt} as a sum over tree graphs}
\medskip

\ni Eq. \eqref{eq:BBGKYt} can be rewritten:
\be\label{fj}
f_j (\bz_j,t)= \sum_{n=0}^\infty\sum _{\G (j,n)}\sum _{\bs_n}\prod_{i=1}^n\s_i 
\T_{\bs_n}(\bz_j,t)\;,
\ee
where 
\be
\T_{\bs_n}(\bz_j,t)= \int d \L (\bt_n , \bn_n , \bv_{j,n})\prod_{i=1}^n 
B(\n_i; v_{j+i} - \eta_{k_i} (t_i^+))f_{0,j+n}(\bze(0))\;,
\ee
and
\bea
B(\n_i; v_{j+i}-\eta_{k_i} (t_i^+))=|\n_i \cdot (v_{j+i}-  \eta_{k_i} (t_i^+))|
\ \mathbbm{1}_{ \mbox{$\{\s_i\n_i \cdot (v_{j+i}-\eta_{k_i} (t_i^+)) \geq 0\} $}}\;. \label{eq:defBboltz}
\eea
%

%%%%%%%%%%%%%%%%%%%%%%%%%%%%%%%%%%%%%%%%%%%%%%%%%%%%%%%
%%%%%%%%%%%%%%%%%%%%%%%%%%%%%%%%%%%%%%%%%%%%%%%%%%%%%%%
%%%%%%%%%%%%%%%%%%%%%%%%%%%%%%%%%%%%%%%%%%%%%%%%%%%%%%%

%\subsection{A measure preserving map} \label{sec:maps}
%\setcounter{equation}{0}    
%\def\theequation{6.3.\arabic{equation}}

\bigskip
\ni {\bf A change of variables}
\bigskip

\ni In the proof of the term by term convergence it will be used a change of variables
transforming integrals over outgoing variables into integrals over incoming variables. This is simply the scattering operator
of \eqref{eq:scattop} applied to an interaction of the BBF. We introduce here such operation. 

Fix $\G(j,n),$ $1\leq r \leq n,$ $\bv_j\in\RRR^{3j}$ and define the transformation $\II^{(r)}=\II^{(r)}_{\bv_j,\G(j,n)}:$
\bea
&& \II^{(r)}: S^{2n}\times\R^{3n} \longrightarrow S^{2n}\times\R^{3n} \nn\\
&& \II^{(r)}(\bn_n,\bv_{j,n}) = (\bn_{r-1},\n'_r,\bn_{r,n-r},\bv_{j,r-1},V'_{r},\bv_{j+r,n-r})
\eea
where only the $r-$th couple $(\n_r,v_{j+r})$ is changed according to
\bea
\left\{
\begin{array}{ll}
\n'_r = -\n_r + 2 \o_r (\o_r\cdot \n_r) & \mbox{\ \ \ for $\n_r\cdot(v_{j+r}-\eta_{k_r} (t_{r}^+))>0$}\\
\n'_r = \n_r &  \mbox{\ \ \ for $\n_r\cdot(v_{j+r}-\eta_{k_r} (t_{r}^+)) \leq 0$}\\
V'_r= \eta_{j+r}(t_r^-) - \eta_{k_r}(t_r^-) \\
\end{array}
\right.\;. \label{eq:defmap2}
\eea
Here $\o_r = \o(\n_r,v_{j+r}- \eta_{k_r}(t_{r}^+)).$ 

\newpage
\begin{lem} \label{lem:map-bis}
The transformation $\II^{(r)}$ is a one--to--one, measure preserving map.
\end{lem}
\bigskip

{\em Proof.} $\II^{(r)}$ is the composition of the two transformations:
\bea
&& (\n_r,v_{j+r}) \longrightarrow (\n_r,V_r) \nn\\
&& V_r = v_{j+r} - \eta_{k_r}(t_r^+)
\eea
and
\bea
(\n_r,V_r) \longrightarrow (\n'_r,V'_r) = \II^{-1}(\n_r,V_r)\;,
\eea
where $\II^{-1}$ is the inverse scattering operator defined in Section \ref{sec:2bs}
(in the case $\n_r\cdot V_r\leq 0,$ just replace $\II^{-1}$ with the identity).
The first is a simple translation by the vector $\eta_{k_r}(t_r^+)=\eta_{k_r}(t_{r-1}^-),$ which is a function of 
$\bn_{r-1},\bv_{j,r-1}$ (see the Remark above). Therefore the result follows applying Lemma \ref{lem:scattop}. \qed

%%%%%%%%%%%%%%%%%%%%%%%%%%%%%%%%%%%%%%%%%%%%%%%%%%%%%%
%%%%%%%%%%%%%%%%%%%%%%%%%%%%%%%%%%%%%%%%%%%%%%%%%%%%%%
%%%%%%%%%%%%%%%%%%%%%%%%%%%%%%%%%%%%%%%%%%%%%%%%%%%%%%
%%%%%%%%%%%%%%%%%%%%%%%%%%%%%%%%%%%%%%%%%%%%%%%%%%%%%%
%%%%%%%%%%%%%%%%%%%%%%%%%%%%%%%%%%%%%%%%%%%%%%%%%%%%%%
\section{Proof of the results} \label{sec:proof}
\setcounter{equation}{0}    
\def\theequation{7.\arabic{equation}}

According to the strategy of Lanford, once proven the uniform convergence of the two series \eqref{eq:fnjexp} and
\eqref{eq:fjexp} for short times, we shall conclude the validity results, namely the convergence of $f_j^N(t)$
to $f_j(t)$, just proving the term by term convergence. Actually, by virtue of Proposition \ref{prop:ste} in Section \ref{sec:ste}, 
it is enough to prove the term by term convergence of the series \eqref{eq:BBGKYt} to \eqref{eq:fjexp}. 

In Section \ref{sec:tree} we have rephrased such expansions respectively in \eqref{fjN-grad} and \eqref{fj}, i.e. sums 
over binary tree graphs of integrals over the (interacting or Boltzmann) backwards flows associated to the graph.
Hence we must show convergence of the generic integral of this kind, $\T_{\bs_n}^\e(\bz_j,t),$ to its analogue in
the Boltzmann series, $\T_{\bs_n}(\bz_j,t).$ The present section is devoted to this problem.

We stress once again the importance of the formulation of Grad (introduced in Section \nolinebreak\ref{sec:Grad}) which has been
our starting point. In the language of Section \ref{sec:tree} we could say that the terms in \eqref{eq:fnjexp} that are 
absent in \eqref{eq:BBGKYt} collect all the interacting backwards flows in which two or more particles are created at some
time $t_i$ (graphically, three or more lines would emerge from a node of the tree). The use of reduced marginals 
(Definition \ref{def:defrm}) and the cluster decomposition introduced in Section \ref{sec:Grad}, allowed
to identify all these negligible terms and to isolate them from the contributions of order one, namely 
$\a_n^\e(j)\T_{\bs_n}^\e(\bz_j,t).$ Now looking at \eqref{eq:Teszt}, we see that this last object resembles very 
much the generic term in the series solution of the BBGKY hierarchy for hard spheres. Nevertheless, as we will explain 
in the next subsection, in the case of smooth interactions one has to be more careful in studying the behavior of 
$\T_{\bs_n}^\e(\bz_j,t)$ for $\e$ small.

%%%%%%%%%%%%%%%%%%%%%%%%%%%%%%%%%%%%%%%%%%%%%%%%%%%%%%
%%%%%%%%%%%%%%%%%%%%%%%%%%%%%%%%%%%%%%%%%%%%%%%%%%%%%%
%%%%%%%%%%%%%%%%%%%%%%%%%%%%%%%%%%%%%%%%%%%%%%%%%%%%%%

\subsection{The convergence problem: preliminary considerations} \label{sec:convergence}
\setcounter{equation}{0}    
\def\theequation{7.1.\arabic{equation}}

We discuss here, in an informal way, the delicate issues connected to the term by term convergence,
in order to motivate the techniques introduced in the following sections.

\bigskip
Let us focus on $\T_{\bs_n}^\e(\bz_j,t)$ and $\T_{\bs_n}(\bz_j,t).$  The integrand functions depend on the variables
$\bt_n , \bn_n , \bv_{j,n}$ completely through the trajectories of the IBF and the BBF respectively. In particular, the 
initial data $f^N_{0,j+n}$ and $f_{0,j+n}$ are integrated over the time--zero configurations of the flows. Since $f^N_{0,j+n}$
converges to $f_{0,j+n}$ by hypothesis, we must focus on the trajectories and prove that the IBF converges to the BBF for all 
values of $\bt_n , \bn_n , \bv_{j,n}$ outside a set giving a negligible contribution to the integrals.

Looking carefully at the definition of $\bze^\e(s)$ and $\bze(s)$ (see Sections \ref{sec:flow} and \ref{sec:Bflow}), we realize that a great
difference between them is generally caused by one of the following events:
\begin{enumerate}
\item a particle (say $j+i$) created in the IBF interacts for a very long time (i.e. larger than $O(\e)$) with its progenitor;
\item a couple of particles $(i,h)$ of the IBF undergoes a recollision, i.e. an interaction different from a creation 
(see the remark on page \pageref{rem:reco});
\item a particle has a very large velocity, so that small differences between the two flows 
(generated during interactions of the IBF) become large in a time of order $1.$
\end{enumerate}

Item 1, which is obviously absent in the case of hard spheres, is controlled by cutting off the variables $(\n_i, v_{j+i})$ that lead
to the singular scattering, and showing that they give a small contribution to the integrals. 
Here the main technical issue is an estimate of the time of
interaction, such as that of Lemma \ref{lem:timebound} (or its generalizations in the cases of potentials with an attractive part discussed 
in Section \ref{sec:gener}). Similarly, item 3 is controlled by cutting off the energy of the system, i.e. the large values of $|\bv_{j,n}|.$ 
Item 2 is the most delicate. It requires to demonstrate that the contribution 
of recolliding trajectories is negligible in the limit $\e \to 0$.

\begin{figure}[htbp] 
 \centering
  \includegraphics[width=6.7in]{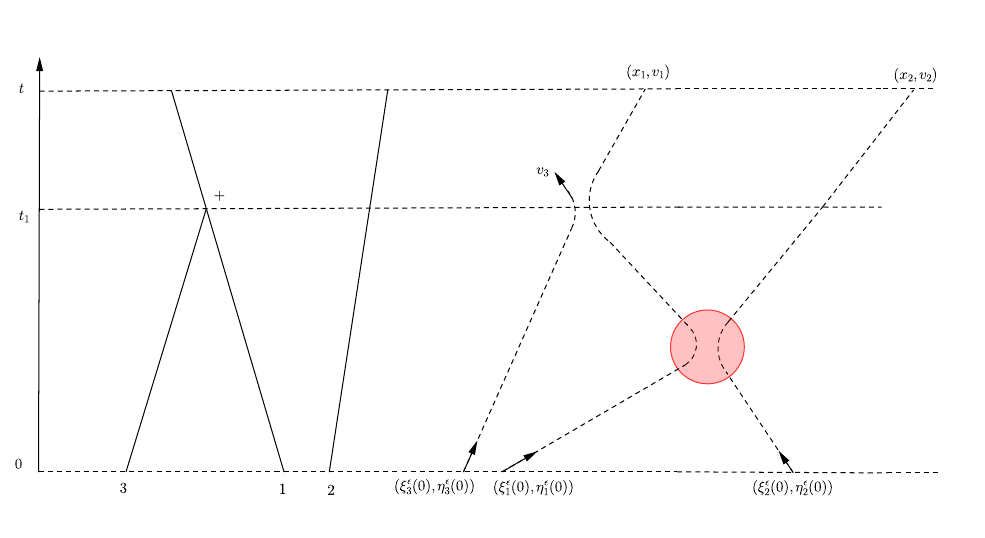} 
   \caption{On the left, a simple case: $\G(j,n)=\G(2,1)=1.$ The plus sign on the node recalls that $\s_1=+.$ We want to estimate 
the contributions to the corresponding formula $\T_+^\e(\bz_2,t),$ coming from the recolliding trajectories of the IBF. An example
of such a trajectory is symbolically represented in the figure on the right.}
   \label{fig:tree1-2}
\end{figure}

\newpage
To motivate our strategy in controlling  the recollisions, we start by the heuristic analysis of one of the simplest non--trivial cases, 
namely that in Figure \ref{fig:tree1-2}.

At time $t$ particles $1$ and $2$  are in the final configuration $\bz_2=(x_1,v_1,x_2,v_2 ) \in \mathcal{M}_2$. We assume that 
the IBF is free  up to time $t_1,$ when particle $3$ appears with velocity $v_3$ at distance $\e\n_1$ from particle 
$1,$ in outgoing ($\s_1=+$) collision configuration. After the scattering between the couple $(1,3),$ particle $1$ collides with 
particle $2.$ This is a collision which is not a creation, i.e. what we called a recollision. We shall imagine that 
$Y=\xi^\e_2(t_1)-\xi^\e_1(t_1)$ is order $1$ while $\e$ is very small. We neglect the time of scattering between the pair 
$(1,3)$ and approximate by $Y$ the relative distance between particles $1$ and $2$ just before the scattering between 
$1$ and $3.$ 

\begin{figure}[htbp] 
 \centering
  \includegraphics[width=3in]{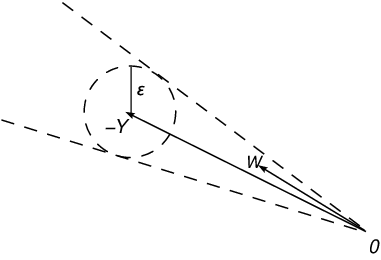} 
   \caption{The recolllision--cone $C(Y).$}
   \label{fig:sphere}
\end{figure}

Denote by $\eta_1^-$ the velocity of particle $1$ between time $t_1$ and the time of the recollision. 
Then, the recollision implies a geometrical relation between $W=v_2 - \eta_1^-$ and $Y.$ They must be chosen in such a 
way that there exists $s\in (0,t_1)$ for which $|\xi^\e_2(s)-\xi^\e_1(s)|= \e.$ This is implied by the fact that 
$W$ lies in the cone $C(Y)$ with vertex $0$, axis the direction of $Y$ and tangent to the ball of center $-Y$ and radius $\e,$ 
see Figure \ref{fig:sphere}.
Moreover, by the laws of scattering \eqref{eq:coll} and \eqref{eq:scattop}, it is easy to see that $\eta_1^-$ belongs to the 
spherical surface of the ball centered in $\frac {v_1 +v_3}2$ of diameter $\frac{|V|}2,$ where $V=v_3-v_1.$ In fact, fixed
$V$ and $v_1$ (hence at fixed total momentum), $\eta_1^-$ moves over that sphere essentially as the scattering 
vector $\o$ (see Figure \ref{fig:2bscatter}). In conclusion, $\eta_1^-$ must belong to the intersection $A$ of the cone 
$v_2-C(Y)$ and the spherical surface described above. Clearly, at a given $|V|,$ the surface measure of $A$ is $O(\e^2)$ 
once assumed $Y=O(1)$ and $\e <<1$.

Now we want to estimate
\be
\int_{|v_3|\leq R} dv_3 \int _{\nu_1\cdot (v_3 -v_1)\geq 0} d\nu_1  \nu_1\cdot (v_3 -v_1) 
\ \mathbbm{1}_{ \mbox{$\{\eta_1^- \in A\}$}}\;,
\ee
where the cutoff on $v_3$ has been added here to obtain an integral over a compact set.
By the above discussion, it follows that a rather natural way to proceed is to express the integral in terms of an integration 
with respect to $V$ and $\o,$ so that we get
\be
\int d\hat V\int_0^{2R} d|V| |V|^2  \int d\om B(\o,V)\mathbbm{1}_{ \mbox{$ \{\eta_1^- \in A\} $}}
\label{eq:exampleGSRT}
\ee
where $\hat V$ is the versor of $V,$ we assumed also $|v_1|\leq R,$ and $B$ is the function resulting from the change of variables
$\n_1\to\o(\n_1,V)$ (see Eq. \eqref{eq:BEs}). If $B$ were bounded (as in the case of hard spheres)
we would easily conclude that such a contribution is $O(\e^2).$ Unfortunately, this is not true in many physically interesting cases, 
since $B$ could not exist as a single--valued function and, even in each monotonicity branch of the scattering map $\rho \to
\Theta (\rho),$  it could diverge when the map becomes flat (see the Appendix).  Thus to control the integral \eqref {eq:exampleGSRT},
we need to know properties of the scattering map (presence and strength of singularities), depending on the details of the potential.
In this way it seems difficult to establish a unified analysis of a large class of interactions (see the discussion in the Appendix).

\bigskip
\ni {\bf A strategy for the control of recollisions}
\bigskip

\ni We propose a different method avoiding the use of the scattering cross--section (i.e. of the function $B$). This is based on two main
ideas:

- We work as much as possible on the Boltzmann flow, rather than on the interacting flow. Of course the BBF is much simpler since
the interactions ($=$ creations) are instantaneous. 
Moreover, by virtue of the property described by  the Remark on page \pageref{rem:BBF}, various parametrizations of the BBF,
different from the usual in terms of $\bt_n,\bn_n,\bv_{j,n}$,  can be conveniently used. 
In particular, the trajectories of the BBF can be 
parametrized by incoming collision variables.  For these reasons,
we find convenient to estimate the events in which some couple of particles of the BBF get closer than a certain distance (say on a scale
slightly larger than $\e$). Indeed in the complement of this set the Boltzmann trajectories are close to the particle trajectories, 
as soon as the scattering time is small and the energy is not too large (which will be assured by an additional cutoff).

- To estimate the above set of events, we use as much as possible the integration over time variables. From Figure \ref{fig:tree1-2}
one can guess that, in general, only a small ($O(\e)$) interval of values of $t_1$ will be compatible with the recollision condition.

However time integrations may produce singularities  for special configurations of 
relative velocities (see the Remark on page \pageref{rem:pr1}). 
Exploiting the global structure of the BBF, we will prove that such configurations are either excluded
by the condition on the  ``initial datum''  $\bz_j\in\O_j,$ or they correspond to small set of values of relative velocities of incoming
collisions, which will be estimated using the map $\II^{(r)}$ of Section \ref{sec:Bflow}.

%%%%%%%%%%%%%%%%%%%%%%%%%%%%%%%%%%%%%%%%%%%%%%%%%%%%%%%
%%%%%%%%%%%%%%%%%%%%%%%%%%%%%%%%%%%%%%%%%%%%%%%%%%%%%%%
%%%%%%%%%%%%%%%%%%%%%%%%%%%%%%%%%%%%%%%%%%%%%%%%%%%%%%%
%%%%%%%%%%%%%%%%%%%%%%%%%%%%%%%%%%%%%%%%%%%%%%%%%%%%%%%
%%%%%%%%%%%%%%%%%%%%%%%%%%%%%%%%%%%%%%%%%%%%%%%%%%%%%%%

\subsection{Proof of Theorem \ref{thm:soft}} \label{sec:proofsoft}
\setcounter{equation}{0}    
\def\theequation{7.2.\arabic{equation}}

By the result in Proposition \ref{prop:ste} of Section \ref{sec:ste} and the reformulations of Section \ref{sec:tree}, the proof of Theorem
\ref{thm:soft} reduces to the proof of convergence of the generic term of the expansion, i.e.

\begin{prop}\label{prop:TBT}
Under  Hypotheses \ref{hyp:pot}--\ref{hyp:conv}, for all $\G(j,n), \bs_n$ and  $(\bz_j,t)\in\O_j\times\RRR^+,$ 
\be
\lim_{\e \to 0 } \T_{\bs_n}^\e (\bz_j,t)=\T_{\bs_n}  (\bz_j,t)\;.
\ee
\end{prop}
\bigskip

The aim is to apply the dominated convergence theorem to show that the trajectories of the IBF converge
almost everywhere to those of the BBF. As already mentioned, we first 
need to``cut off away''pieces of phase space which correspond to trajectories of the IBF exhibiting recollisions,
large scattering times, or high energies, and prove that they give a negligible contribution in the limit. 
Outside this properly defined set of  ``bad events'', we will be able to estimate explicitly the distance between
the interacting  and the Boltzmann trajectories.

In all this section and in the following we will keep fixed $\bz_j\in\O_j$ and $t>0.$
Moreover the times $\bt_n$ will be always supposed to be ordered (see \eqref{eq:ordertimes}), and the $\bn_n$
to satisfy the constraint implied by $\bs_n$ (Eq. \eqref{eq:defBboltz}). In the present section we also fix $\G(j,n)$ and $\bs_n.$

\begin{figure}[htbp] 
   \centering
   \includegraphics[width=5in]{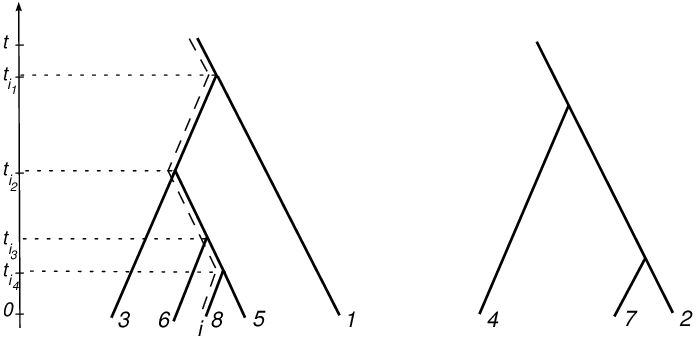} 
   \caption{The line closest to the dashed line is the path ${\cal P}_i$ in the tree $\G(2,6),$ with $i=8.$ The states of the particle associated to it 
via the BBF form the``virtual trajectory''.}
   \label{fig:ivirtual}
\end{figure}

\bigskip
We start by focusing on the BBF $\bze(s)$ and giving a new definition. Consider particle $i$ and look at the graph of $\G(j,n).$  
A polygonal path ${\cal P}_i$ is uniquely defined if we walk on the tree by going forward in time, starting from the time--zero endpoint 
of line $i$ and going up to the root--point at time $t$. See for instance Figure \ref{fig:ivirtual}. 
To ${\cal P}_i$ we may naturally associate a
one--particle piecewise--free trajectory, built up with pieces of trajectories of (different) particles of the BBF. More precisely,
fixed a BBF with parameters $(\bt_n,\bn_n,\bv_{j,n}),$ denote $t_{i_1},\cdots,t_{i_{n_i}}$ the (decreasing) subsequence of 
$t_1,\cdots,t_n$ of the times corresponding to the nodes met by following the path ${\cal P}_i$ ($n_i$ being the number of 
such nodes, with the convention $i_0=0, t_{i_0}=t$): see the figure. 
We call {\bf{\em virtual trajectory associated to particle $i$ in the BBF}}, and indicate it by 
$\z^i(s)=(\xi^i(s),\eta^i(s))\in\RRR^{6}$ with $s\in[0,t],$ the one--particle trajectory given by:
\be
\z^i(s) =
\begin{cases}
\displaystyle \z_i(s)\mbox{\ \ \ \ \ \ for\ }s\in[0,t_{i_{n_i}}) \\
\displaystyle \z_{k_{i_{r}}}(s)\mbox{\ \ \ \ for\ }s\in[t_{i_{r}},t_{i_{r-1}}),\ \ \ 0<r \leq n_i
\end{cases}\;.\label{eq:ivirtualdef}
\ee
Observe that, during the time of existence of particle $i$ in the BBF, $\z^i(s)=\z_i(s).$

Now consider a couple of particles $(i,h)$ and compare their virtual trajectories. 
Calling``root''of ${\cal P}_i$ the root line of the tree to which ${\cal P}_i$ belongs,
we have two possibilities: either the roots of ${\cal P}_i$ and ${\cal P}_h$ coincide (i.e. $i$ and $h$ belong to the same
single tree), or not. In the first case, there exists (uniquely) a node of the tree where ${\cal P}_i$ and ${\cal P}_h$ merge.
For any given couple $(i,h)$ we introduce the subsequence of $t_1,\cdots,t_n:$
\be
t \geq t^0 > t^1 > t^2 > \cdots >t^{n_{ih}}>t^{n_{ih}+1}\equiv 0\;,
\ee
defined as follows. Time $t^0$ is equal to $t$ if ${\cal P}_i$ and ${\cal P}_h$ have different roots; otherwise, it is equal
to the time (strictly smaller than $t$) of the node where ${\cal P}_i$ and ${\cal P}_h$ merge. 
The sequence $t^1,\cdots,t^{n_{ih}}$ is given by the
ordered union of the times $t_{i_1},\cdots,t_{i_{n_i}}$ and $t_{h_1},\cdots,t_{h_{n_h}}$ that belong to the interval $(0,t^0).$
Here $n_{ih}$ is the number of such times, and $t^{n_{ih}+1}$ has been put equal to zero by convention.
See also Figure \ref{fig:treeclimbing2} below.

\bigskip
We are ready to define a part of the``bad set''to be cutoffed.
\bigskip
\begin{defi} \label{def:defNNd}
Let be $\d>0.$ The
set of $\d-$overlaps, $\NN(\d) \subset \RRR^n\times S^{2n}\times\RRR^{3n},$ is
\be \label{eq:defNNd}
\NN(\d)= \Big\{\bt_n, \bn_n, \bv_{j,n}\ \Big|\ \min_{i<h}\ \min_{s\in [0,t^1]} |\xi^i(s)-\xi^h(s)| \leq \delta\Big\}\;.
\ee
\end{defi}
\bigskip
The time $t^1$ depends on the couple $(i,h)$ under consideration.
Notice that the set $\NN(\d)$ is completely defined via the BBF. Clearly, it depends also on $\bz_j,t.$

Note also that the set $\NN(\d)$ detects the $\d$--overlaps (namely when $|\xi^i(s)-\xi^h(s)| \leq \delta $) of the virtual paths
 ${\cal P}_i$ and ${\cal P}_h$ {\em excluding} the time interval $(t^1,t^0]$. 

\bigskip
In the following, $\d>\e $ will be taken as a function of $\e$ going to zero as $\e\to 0$.
Then, the first step in the proof is to show that the restriction of the integrals contained in $\T^\e_{\bs_n}(\bz_j,t)$
to the set $\NN(\d)$ is arbitrarily small with $\e.$ To do so, consider the following definition.
\bigskip
\begin{defi} The set of point--overlaps is $\NN\equiv\NN(0),$ i.e.
\be
\NN=  \Big\{\bt_n, \bn_n, \bv_{j,n}\ \Big|\ \min_{i<h}\ \min_{s\in [0,t^1]} |\xi^i(s)-\xi^h(s)|=0\Big\}\;. 
\label{eq:defNN}
\ee
\end{defi}
\bigskip
Obviously it is
\be
\lim_{\d\rightarrow 0}\mathbbm{1}_{\NN(\d)}= \mathbbm{1}_{\NN}\;. \label{eq:convchfu}
\ee
\begin{lem} \label{lem:NNdL0}
The set $\NN$ has $d\L-$measure zero. 
\end{lem}

\emph{Proof of Lemma \ref{lem:NNdL0}.}
We will show that the condition in \eqref{eq:defNN} 
implies a certain number of relations between the integration variables that can be satisfied at most for a $d\L-$null set of values.

If we are in $\NN,$ then for some couple $(i,h)$ there exists 
\be
t^* = \max\{s\in[0,t^1]\ \Big|\ |\xi^i(s)-\xi^h(s)|=0\}\;. \label{eq:deftstar}
\ee
It will be $t^*\in[t^{l+1},t^l)$ for some $l\in\{0,\cdots,n_{ih}\}.$ $l$ is the total number of interactions in the virtual trajectories
of $i$ and $h$ between the overlapping time and $t^0.$ For $q=0,\cdots,l,$ we define:
\bea
&& Y^q = \xi^h(t^q)-\xi^i(t^q)\;,\nn\\
&& \eta^i_q \equiv \eta^i(s)\;,\ \eta^h_q \equiv \eta^h(s)\ \ \ \ \ \mbox{for }\ s\in(t^{q+1},t^q)\;,\nn\\
&& W^q = \eta^h_q - \eta^i_q\;.\label{eq:defYW}
\eea
We indicate by $f\in\{0,1,\cdots,n\}$ the index such that 
\be
t^0=t_f\;. \label{eq:t0tf}
\ee
Notice that either $t^0=t$ ($f=0$) and $Y^0\neq 0$ (because $\bz_j\in\O_j$) or $t^0<t$ ($f>0$)
and $Y^0=0$ (because $t^0$ is the time of the node where ${\cal P}_i$ and ${\cal P}_h$ merge). 
A possible event in $\NN(\d)$ is pictured in Figure \ref{fig:treeclimbing2}.
\begin{figure}[htbp] 
   \centering
   \includegraphics[width=7in]{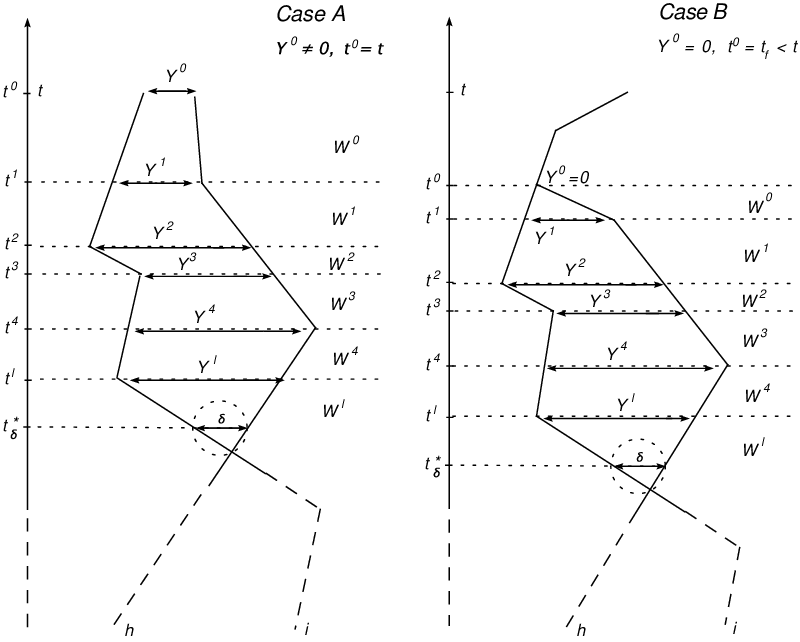} 
   \caption{
A symbolical drawing of virtual trajectories of two particles $i,h$ in the BBF showing a $\d-$overlap after $l=5$ interactions.
Relative distances and velocities are indicated as defined in \eqref{eq:defYW}.
In case A the two particles belong to different trees, while in case B they belong to the same tree.
$t^*_\d$ is the first (backwards) time when the particles reach the distance $\d.$}
\label{fig:treeclimbing2}
\end{figure}

Given a point in $\NN$, we observe preliminarily that we may assume:

\noindent (i) if $Y^0=0,$ then $W^0 \neq 0;$

\noindent (ii) $l \geq 1;$

\noindent (iii) $W^l \neq 0.$

\noindent Assumption (i) corresponds to exclude subsets of the integration domain of codimension $3.$ In fact, by the laws of the two--body
scattering, if $Y^0=0$ then $W^0 \neq 0$ except for a single value of the velocity of the particle created at time $t^0=t_f,$
namely $v_{j+f}$ must be equal to $\eta_{k_f}(t_f^+).$ Note that $l=0$ and $Y^0 \neq 0$ is impossible because $\bz_j\in\O_j.$ On the other hand
if $l=0$ and $Y^0=0,$ then necessarily $W^0=0,$ which we excluded by (i). Finally if $W^l=0,$ then the overlap takes place in the interval
$[t^l,t^{l-1})$ and this contradicts the definition of $l.$

The point--recollision condition is verified only if $\min_s|Y^l-W^ls|=0,$ which in turn implies
$Y^l \wedge \hat W^l = 0,$ where $\hat W^l = \frac{W^l}{|W^l|}.$ Since 
\be
Y^{l}=Y^{0}-\sum_{q=0}^{l-1}W^{q}(t^{q}-t^{q+1})\;,\label{eq:polygvt}
\ee
we have
\bea
&& 0=Y^{0}\wedge \hat W^{l}-\sum_{q=0}^{l-1}(W^{q}\wedge \hat W^l)(t^q-t^{q+1})\nn\\
&& \ \ =(Y^{0}-W^{0}t^{0})\wedge \hat W^{l}-\sum_{q=1}^{l}[(W^{q}-W^{q-1})\wedge \hat W^l)]t^{q}\;.
\label{eq:softtcpr}
\eea
But, by the Remark on page \pageref{rem:BBF}, all the vectors involved in this relation do not depend on time. 
Hence as soon as $[(W^{q}-W^{q-1})\wedge \hat W^l)] \neq 0$ for some $q,$ there exists at most one value of the 
time $t^{q}$ fulfilling condition \eqref{eq:softtcpr}.

Otherwise, it will be
\be
\hat W^{l}\wedge W^{q}=0 \ \ \ \ \ \ \ \ \ \ \mbox{for all\ }q=0,1,\cdots,l\;, \label{eq:WlwW0}
\ee
i.e. all (not vanishing) relative velocities are collinear. In particular, Eq. \eqref{eq:softtcpr} implies 
\be
Y^{0}\wedge\hat W^{l} = 0\;. \label{eq:Y0whWl}
\ee
As said above we have two cases, which we treat separately.
\begin{itemize}
\item {\em Case $ Y^{0}\neq 0, t^{0}=t.$} 

Both $Y^{0}$ and $W^{0}$ are collinear with $\hat W^{l}.$ Therefore 
\be
Y^{0}\wedge W^{0} = 0\;,
\ee
which is excluded since $\bz_j \in\O_j.$
\item
{\em Case $ Y^{0} = 0, t^{0}<t.$} 

If all relative velocities $W^{q}$ are coincident, then no point--recollision is possible
since $W^{0}\neq 0.$ Finally assume that $U :=W^{\ol q}-W^{\ol q-1}\neq 0$ for some $\ol q\in\{1,\cdots,l\}$
and put $\hat U=\frac{U}{|U|}.$ We have 
$\hat U \wedge \hat W^{l} = 0,$ which, together with $W^{0}\wedge \hat W^{l} = 0,$ implies
\be
\hat U \wedge W^{0} = 0\;. \label{eq:delprosoft}
\ee
Let $t^{\ol q}=t_{f'},$ where $f'>f$ (recall \eqref{eq:t0tf}). We change variables
\be
(\n_f,v_{j+f},\n_{f'},v_{j+f'}) \to (\n'_{f},V'_f,\n_{f'},V_{f'}) \label{eq:newvarrest}
\ee
according to
\be
\begin{cases}
\displaystyle (\bn_{f-1},\n'_f,\bn_{f,n-f},\bv_{j,f-1},V'_{f},\bv_{j+f,n-f})=\II^{(f)}(\bn_n,\bv_{j,n}) \\
\displaystyle V_{f'} = v_{j+f'}-\eta_{k_{f'}}(t_{f'}^+)
\end{cases}\;.
\label{eq:newvarIr}
\ee
In the first equation we used the map introduced on page \pageref{eq:defmap2}, while the second is a simple translation.
We have $W_0 = V'_{f}$ and $U= \pm[\eta_{k_{f'}}(t_{f'}^+)-\eta_{k_{f'}}(t_{f'}^-)]$ or
$U=\pm[\eta_{k_{f'}}(t_{f'}^+)-\eta_{j+f'}(t_{f'}^-)];$ see Figure \ref{fig:treeclimb}. \label{remU}From the rules of scattering it follows
that the vector $U$ depends only on $(\n_{f'},V_{f'})$\footnote{It can be $U=\pm \o (\o\cdot V_{f'})$ or 
$U=\pm[V_{f'}-\o (\o\cdot V_{f'})]$ where $\o=\o(\n_{f'},V_{f'})$ is the scattering vector.}.
Then Eq. \eqref{eq:delprosoft} defines a subset of codimension two in the space of variables $(V'_f,\n_{f'},V_{f'}).$
By Lemma \ref{lem:map-bis} of Section \ref{sec:Bflow}, the change of variables \eqref{eq:newvarrest}--\eqref{eq:newvarIr} is a 
one--to--one measure preserving map. Therefore the subset defined by \eqref{eq:delprosoft} has $d\L-$measure zero.
\begin{figure}[htbp] 
   \centering
   \includegraphics[width=3.5in]{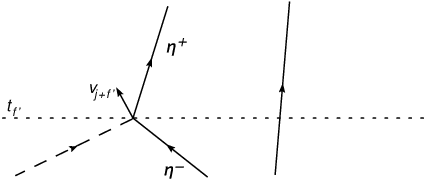} 
   \caption{Detail of the virtual trajectory of Figure \ref{fig:treeclimbing2}--Case B, for times close to $t^{\ol q}
= t_{f'}.$  In the example $\s_{f'}=+.$ The difference of relative velocities is $U = W^{\ol q}-W^{\ol q-1} = \eta^+-\eta^-$ 
(here $f'$ belongs to ${\cal P}_i;$ if $f'$ belongs to ${\cal P}_h$ then $U = \eta^--\eta ^+$), where 
$\eta^+=\eta_{k_{f'}}(t_{f'}^+)$ and $\eta^-$ can be equal to $\eta_{j+f'}(t_{f'}^-)$ or $\eta_{k_{f'}}(t_{f'}^-),$ depending 
on the structure of ${\cal P}_i.$ The variables describing the scattering are $(\n_{f'},v_{j+f'}),$ or alternatively $(\n_{f'},V_{f'}).$ 
}
   \label{fig:treeclimb}
\end{figure}
\end{itemize} \qed

It follows what we asserted before Eq. \eqref{eq:defNN}, that is
\bigskip
\begin{lem} \label{lem:restNde}
Let $\d=\d(\e)>\e $ be a function of $\e$ going to zero as $\e\to 0$. Then
\be
\lim_{\e\to 0}\int d \L (\bt_n , \bn_n , \bv_{j,n})\prod_{i=1}^n 
B^\e(\n_i; v_{j+i} - \eta^\e_{k_i}(t_i))\mathbbm{1}_{\NN(\d)}f^{N}_{0,j+n}(\bze^\e(0))=0
\label{eq:absarg}
\ee
for all $\bz_j\in \O_j$ and all $t>0.$
\end{lem}
\bigskip

{\em Proof of Lemma \ref{lem:restNde}.}
From Lemma \ref{lem:NNdL0}, it follows that the integral of any finite measure restricted to 
$\NN(\d)$ goes to zero with $\d.$ But, in our Hypothesis \ref{hyp:f0jbound}, 
$d\Lam \left(\prod B^\e\right) f^N_{0,j+n}$ is uniformly bounded by a 
finite measure for $\e$ small. Indeed, using conservation of energy, we can estimate it by
\be
d\Lam \left(\prod B^\e\right) f^N_{0,j+n} \leq 
d\Lam \left(2\sqrt{\sum_{i=1}^{j+n}v_i^2}\right)^n e^{-\frac{\b}{2}\sum_{i=1}^{j+n}v_i^2}\|f^N_{0,j+n}\|_\b\;,
\label{eq:roughboundterm}
\ee
where we used that the energy of the IBF at time $t$ is purely kinetic if $\e$ is small enough 
(having fixed $\bx_j$ outside the diagonals).\qed

\bigskip 
{\em Remark.} \label{rem:pr1} In the proof of Lemma \ref{lem:NNdL0}, we use time variables,
when possible, to show that the overlaps are rare.
Otherwise, we analyze geometrical conditions involving relative velocities of the virtual trajectories, namely Equation 
\eqref{eq:WlwW0}, which is proven to be a vanishing measure condition either by integrating in the incoming relative velocities
of a node of the tree (case $Y^0=0$), or by showing that the condition is impossible by definition of $\O_j$ (case $Y^0 \neq 0$).
In the following section, when we will deal with quantitative estimates of the recollision (or overlap) events, we will 
follow the same strategy, that is we will integrate in time unless we face the condition``$|\hat W^{l}\wedge W^{q}|$ small 
for all $q$''. Note that, in particular, this is the case of a sequence of central and grazing collisions in the virtual trajectories, for which 
the relative velocities may remain unchanged (remind that the scattering cross--section may possibly have concentrations on such 
collisions: see the Appendix). On the other hand, in this case the virtual trajectories are analogous to a free flow, for which 
the recollisions can be easily controlled. 
\bigskip

Besides $\NN(\d),$ we still have to take care of some additional subsets of the integration region. Let $\mu\in(0,1).$ Putting
\bea
&& \mathbbm{1}^\e_1 = \mathbbm{1}_{\mbox{ $ \{\frac{\b}{2}\sum_{i=1}^{j+n} v_i^2 < |\log \e | \} $}}\;,\nn\\
&& \mathbbm{1}^\e_2 = \prod_ {r=1}^n  \ \mathbbm{1}_{ \mbox{ $ \{|(v_{j+r}-\eta_{k_r}^\e(t_r))\wedge\n_r| > \e^{\mu} \}$}}\;,
\label{eq:chfue12}
\eea
a simple estimate as the one in \eqref{eq:roughboundterm} is sufficient to show that
\bea
&& \lim_{\e\to 0}\int d \L (\bt_n , \bn_n , \bv_{j,n})\prod_{i=1}^n 
B^\e(\n_i; v_{j+i} - \eta^\e_{k_i}(t_i))(1-\mathbbm{1}_{\NN(\d)})(1-\mathbbm{1}^\e_1\mathbbm{1}^\e_2)
f^{N}_{0,j+n}(\bze^\e(0))=0\;.\label{eq:remcoEL}\nn\\
\eea

Thus, to obtain the final result we are left with the proof of
\be
\lim_{\e\to 0}\int d \L (\bt_n , \bn_n , \bv_{j,n})\prod_{i=1}^n 
B^\e(\n_i; v_{j+i} - \eta^\e_{k_i}(t_i))(1-\mathbbm{1}_{\NN(\d)})\mathbbm{1}^\e_1\mathbbm{1}^\e_2
f^{N}_{0,j+n}(\bze^\e(0))= \T_{\bs_n}(\bz_j,t)\;. \label{eq:Ttildee}
\ee
Notice that up to now we did not use any property of the interacting flow but the conservation of energy.
Now we have to examine in more detail the structure of the IBF and to compare it with the Boltzmann flow.
Since we work in the complement of $\NN(\d)$ and in the sets in \eqref{eq:chfue12}, we are actually
in a favorable situation in proving that the distance between the two flows is small. Indeed, as we will show in the following
lemma, the IBF has no recollisions and its differences with the BBF are only due to the scattering time, which is absent in 
the Boltzmann flow, and to the $\e-$delocalization of the created particles (also absent in the BBF).

Choose
\be
\d = \e^{1-\mu}(\log\e)^2 \;.\label{eq:choosedelta}
\ee
Then we have:
\bigskip
\begin{lem} \label{lem:trajIBF}
If $(\bt_n,\bn_n,\bv_{j,n})$ is outside $\NN(\d)$ and inside the sets \eqref{eq:chfue12}, then for $\e$ sufficiently small
and all $i=1,\cdots,j+n$
\be
\max_{s\in[0,t]}|\xi_i(s)-\xi_i^\e (s)| \leq D\e^{1-\mu}|\log \e|^{\frac{3}{2}}
\label{eq:convxim}
\ee
for some $D>0.$ In particular, the IBF does not admit recollisions. 
Moreover, $\eta^\e_{k_r}(t_r)=\eta_{k_r}(t_r^+)$ for all $r=1,\cdots,n$, and $\eta_i^\e(0)=\eta_i(0).$
\end{lem}
\bigskip

\emph{Proof of Lemma \ref{lem:trajIBF}.} The proof is based on a simple continuity argument. 
We proceed by induction on $r$ proving that, for some $D'>0,$
\be
|\xi_i(s)-\xi_i^\e (s)| \leq D' r\e^{1-\mu}|\log \e|^{\frac{1}{2}}\;,\ \ \ \ \ s\in(t_{r+1},t_r)
\label{eq:trajproofint}
\ee
for $i=1,\cdots,j+r,$ from which \eqref{eq:convxim} follows since $n$ is smaller then a constant times $|\log\e|$ for $\e$ small. 
A byproduct will be that if particle 
$i\leq j+r-1,$ at time $t_r,$ has already completed the (possible) scattering with its progenitor (or its last son) in the IBF, 
then necessarily $\eta_i^\e(t_r)=\eta_i(t_r^+)$ (from which the last assertion of the lemma follows, because the parameters
are taken outside $\NN(\d)$).

For $r=0$ the statement is trivial.
Indeed, since we are outside $\NN(\d)$ with $\d>\e$ and the states at time $t$ of the BBF and the IBF are the same
then, for $s\in(t_1,t),$ $\z_i(s)=\z_i^\e(s).$ 

Let now $s\in(t_{r+1},t_r)$ for $r>0$ and $i=1,\cdots,j+r.$ We consider the case (the other cases being easier) in which
particle $i$ in the IBF is interacting at time $t_r$ with another particle $h.$ For $\e$ small and by inductive hypothesis, this can
be true only if the two particles coincide with the couple $(k_r,j+r),$ or if they are a couple progenitor--son, $(k_{r'},j+r')$ with $r'<r,$ 
that has not finished yet the two--body scattering. Also, note that no other particle can be at distance $\leq\d/2$ from $i$ and $h$ at time
$t_r$ (for $\e$ small and $n=O(|\log\e|)$). 

Denote by $s^*\in (t_{r+1},t_{r})$ the last (first backwards) recollision time of particle $i$ or $h$ in the IBF, that is $|\xi^\e_{h'}(s)
- \xi_i^\e(s)|>\e$ for all $h'\neq h,i$ and all $s\in(s^*,t_r),$ and same equation with $i$ replaced by $h.$
Then, for the same values of $s,$ particles $i$ and $h$ behave as they were isolated and we have that $|\xi^\e_i(s)-\xi_i(s)|$
is equal to the same quantity evaluated in $\max(s,t_r-\ol t),$ where $\ol t$ is the time of the two--body scattering (in fact,
once the backward scattering finishes, it must be $\eta^\e_i(s)=\eta_i(s)$). Hence, taking into account Lemma \ref{lem:timebound}
and \eqref{eq:chfue12}, we have
\bea
&& |\xi^\e_i(s)-\xi_i(s)| \leq |\xi^\e_i(t_r)-\xi_i(t_r)| + 2\sqrt{2/\b|\log\e|}\ol t \nn\\
&&\ \ \ \ \ \ \ \ \ \ \ \ \ \ \ \ \ \ 
\leq D'(r-1)\e^{1-\mu}|\log \e|^{\frac{1}{2}} + 2\sqrt{2/\b|\log\e|}A\e^{1-\mu}\;,
\eea
which implies \eqref{eq:trajproofint} with $D' = 2\sqrt{2/\b}A,$ for $s\in(s^*,t_r).$ The same holds of course for particle $h.$

In particular (taking $\e$ small so that $n$ is $O(|\log\e|))$ we have that $|\xi^\e_i(s)-\xi_i(s)| \leq \frac{\d}{4}$
up to the first recollision time. Since we are outside $\NN(\d),$ for all $h'\neq h,i$
\bea
&& |\xi^\e_{h'}(s) - \xi_i^\e(s)|\geq|\xi_{h'}(s)- \xi_i(s)|-|\xi^\e_{h'}(s)- \xi_{h'}(s)|-|\xi^\e_{i}(s)- \xi_i(s)|\nn\\
&& \ \ \ \ \ \ \ \ \ \ \ \ \ \ \ \ \ \ \ \ > \d - \frac{\d}{4} - \frac{\d}{4} = \frac{\d}{2}
\eea
up to the first recollision time of $i,h,h',$ and the same equation holds for particle $h.$ But $\d/2 > \e.$ Therefore, by
continuity of the flow, $|\xi^\e_{h'}(s) - \xi_i^\e(s)| > \frac{\d}{2}$ and $|\xi^\e_{h'}(s) - \xi_h^\e(s)| > \frac{\d}{2}$
for all times $s\in(t_{r+1},t_r)$ and Eq. \eqref{eq:trajproofint} holds in the full interval. \qed

\bigskip
{\em Conclusion of the proof of Proposition \ref{prop:TBT}.}
Note that Lemma \ref{lem:trajIBF}, together with Hypothesis 4,  can be also used  to replace the initial datum 
$f^N_{0,j+n}$ by $f_{0,j+n}$ in $\T_{\bs_n}^\e (\bz_j,t)$,  that is to show that
\bea
\lim_{\e\to 0}\int d \L (\bt_n , \bn_n , \bv_{j,n})\prod_{i=1}^n 
B^\e(\n_i; v_{j+i} - \eta^\e_{k_i} (t_i))\left|f^{N}_{0,j+n}(\bze^\e(0))-f_{0,j+n}(\bze^\e(0))\right|=0\;.\nn\\ \label{eq:softinitdata}
\eea
Indeed we can restrict the above integral for values of $\bze^\e(0)$ in a compact subset of $\MM_{j+n},$ producing an arbitrarily small error.

Using this last result,  estimate \eqref{eq:roughboundterm} (with $f^N_{0,j+n}$ replaced by $f_{0,j+n}$),
dominated convergence theorem and Lemma \ref{lem:trajIBF}, Eq. \eqref{eq:Ttildee} follows. \qed

%%%%%%%%%%%%%%%%%%%%%%%%%%%%%%%%%%%%%%%%%%%%%%%%%%%%%%%
%%%%%%%%%%%%%%%%%%%%%%%%%%%%%%%%%%%%%%%%%%%%%%%%%%%%%%%
%%%%%%%%%%%%%%%%%%%%%%%%%%%%%%%%%%%%%%%%%%%%%%%%%%%%%%%
%%%%%%%%%%%%%%%%%%%%%%%%%%%%%%%%%%%%%%%%%%%%%%%%%%%%%%%
%%%%%%%%%%%%%%%%%%%%%%%%%%%%%%%%%%%%%%%%%%%%%%%%%%%%%%%

\subsection{Proof of Theorem \ref{thm:hard}} \label{sec:proofhard}
\setcounter{equation}{0}    
\def\theequation{7.3.\arabic{equation}}

In this section we go through the steps in the proof of Theorem 1, estimating explicitly all the error terms
arising in the limiting procedure, under the additional Hypothesis \ref{hyp:convrate}.
We also need to take care of the dependence on $n$ to guarantee the summability. For instance the bound 
\eqref{eq:roughboundterm}  is useless in the present context because it grows as $n^{n/2}$.

%A further care is needed in these estimates in order to obtain an error that is summable in $n.$
%In fact, the simple bound \eqref{eq:roughboundterm} is prohibited in this context, since it would give an estimate of the
%$n-$th term proportional to $n!.$

In this section, for notational simplicity, we will denote by  $C$ all pure positive constants depending only on $\b$ and $\a.$
We also denote by  $\mathcal{E}_1, \mathcal{E}_2, \cdots$ all the  errors in our procedure, i.e.
it will be $$f^{N}_j (\bz_j,t)-f_j (\bz_j,t) = \sum_i\mathcal{E}_i.$$

\bigskip
\ni {\bf Preliminary error estimates}
\medskip

\ni We start noticing again that it is enough to control the difference 
%$\eqref{fjN-grad}$ and $\eqref{fj}$
%and estimate their difference
$|\tilde f^{N}_j (\bz_j,t)-f_j (\bz_j,t)|$,
since, by Proposition \ref{prop:ste}, the error due to this approximation is
\be
|\mathcal{E}_1|= |f_{j}^N-\tilde{f}_j^N| \leq C^j\e
\ee
for $t$ sufficiently small. 

\bigskip
Moreover, since $1- \a_n^\e(j)  \leq jC^n\e^2$ for $n\leq N-j,$ we have also
\be
|\mathcal{E}_2| = \sum_{n\geq 0}^{N-j}\left(1-\a_n^\e(j)\right) \Big|\sum_{\G(j,n)}\sum_{\bs_n}
\prod_{i=1}^n \s_i \mathcal{T}_{\bs_n}^\e\Big| \leq C^j\e^2\;.
\label{eq:error2}
\ee
%

%We shall notice that in the estimate of the difference between the interacting and the Boltzmann flow, see Lemma \ref{lem:trajIBF}, 
%we required $n=O(|\log\e|).$ This means that we will need a cutoff on the``dimension''of trees. Taking into account \eqref{eq:error2}
%we thus reduce to the estimate of
%
\bigskip
A third simple error arises by neglecting the rest of the series expansion so that we focus on
$\sum_{n=0}^{\ol n} \sum_{\G(j,n)}\sum_{\bs_n}| \mathcal{T}_{\bs_n}^\e-\mathcal{T}_{\bs_n}|\;,$ 
where
\be
\ol n = D''|\log\e|
\ee
with $\e$ small enough.
The error generated by this last cutoff is bounded by the remainder of the geometric series appearing in formula \eqref{eq:finalest}, 
therefore if we choose $D''=|\log(tC'_{\b,\a})|^{-1}$ it is
\be
|\mathcal{E}_3|\leq \sum_{n\geq\ol n}\sum_{\G(j,n)}\sum_{\bs_n}\left(\mathcal{T}_{\bs_n}^\e+\mathcal{T}_{\bs_n}\right) \leq C^j\e\;,
\ee
for $t$ sufficiently small. Hence,
\be
|f_{j}^N-{f}_j| \leq \mathcal{E}_1+\mathcal{E}_2+\mathcal{E}_3 + 
\sum_{n=0}^{\ol n} \sum_{\G(j,n)}\sum_{\bs_n}| \mathcal{T}_{\bs_n}^\e-\mathcal{T}_{\bs_n}|\;.\label{eq:sumdifftrees}
\ee

\bigskip
\ni {\bf Elimination of factors $B^\e, B$}
\medskip

\ni Focusing on the last term in \eqref {eq:sumdifftrees},
we will estimate first the error caused by the restriction of the integrals to suitable sets of integration variables, where 
the convergence of the flows (Lemma \ref{lem:trajIBF}) is guaranteed. 

However, first of all, to simplify the expression of the integrands
avoiding  estimate \eqref{eq:roughboundterm}, we get rid of $\left(\prod B^\e\right)$ performing the 
decomposition  $ \{ \left(\prod B^\e\right) >\e^{-\la}\} $  and  $\{ \left(\prod B^\e\right)\leq \e^{-\la}\} $ 
(and the same for $\left(\prod B\right)$) for a suitable $\la \in (0,1)$. 
%The task of this cutoff is to reduce the integrand to the simple initial datum $f^N_{0,j+n}$ (or $f_{0,j+n}$), which depends 
%only on the time--zero state of the IBF (or BBF), times the characteristic function 
%$\mathbbm{1}_{\{|\xi^\e_{j+i}(t_i)-\xi^\e_{k}(t_i)| > \e\  \forall k\neq k_i\}}$
%(in the case of $\mathcal{T}_{\bs_n}^\e$). In our hypotheses this object is easily controlled by the Gaussian of the energy:
%
Moreover we shall often use in the following the bounds 
\bea
&& f^N_{0,j+n}(\bze^\e(0)) \ \prod_{i=1}^n\ \mathbbm{1}_{ \mbox{$\{|\xi^\e_{j+i}(t_i)-\xi^\e_{k}(t_i)| > \e\  \forall k\neq k_i\} $}}
\leq C^{j+n}e^{-\frac{\b}{2}\sum_{i=1}^{j+n}v_i^2}\;,\nn\\
&& f_{0,j+n}(\bze(0)) \leq C^{j+n}e^{-\frac{\b}{2}\sum_{i=1}^{j+n}v_i^2}\;,
\label{eq:estfBsm}
\eea
consequences of Hypotheses \ref{hyp:f0bound} and \ref{hyp:f0jbound},
the conservation of energy and the fact that the energy at time $t$ is purely kinetic if $\bx_j$ is outside 
the diagonals and $\e$ is small enough. 

 We need to estimate
\bea
&& |\mathcal{E}_4| \leq \sum_{n=0}^{\ol n} \sum_{\G(j,n)}\sum_{\bs_n}\nn\\
&&\ \ \ \  \cdot\Big[
\int d \L (\bt_n , \bn_n , \bv_{j,n})\ \mathbbm{1}_{ \mbox{$ \{\left(\prod B^\e\right)>\e^{-\la}\} $}}
\prod_{i=1}^n B^\e(\n_i; v_{j+i} - \eta^\e_{k_i} (t_i))f^{N}_{0,j+n}(\bze^\e(0))\nn\\
&&\ \ \ \ + \int d \L (\bt_n , \bn_n , \bv_{j,n})\ \mathbbm{1}_{ \mbox{$\{\left(\prod B\right)>\e^{-\la}\} $}}
\prod_{i=1}^n B(\n_i; v_{j+i} - \eta_{k_i} (t_i^+))f_{0,j+n}(\bze(0))\Big]\nn\\
&&\ \ \ \ \leq \e^{\la} \sum_{n=0}^{\ol n} \sum_{\G(j,n)}\sum_{\bs_n}C^{j+n}
\int d\L \ e^{-\frac{\b}{2}\sum_{i=1}^{j+n}v_i^2}\Big[\left(\prod B^\e\right)^2 + \left(\prod B\right)^2\Big]\;.
\eea
Remind  now the expression of $B^\e$ and Eq. \eqref{eq:defsumtrees}. Since $\sum_{k_i=1}^{j+i-1}\left(\eta_{k_i}^\e(t_i)\right)^2$
is bounded by the total energy $\sum_{i=1}^{j+n}v_i^2$ (the potential being positive), it follows easily that
\be
\sum_{\G(j,n)}\left(\prod B^\e\right)^2 \leq 2^n \prod_{i=1}^n \left((j+n)v_{j+i}^2+\sum_{l=1}^{j+n} v_l^2\right)\;.
\ee
The same estimate holds for $\sum_{\G(j,n)}\left(\prod B\right)^2.$ Therefore
\begin{equation}
|\mathcal{E}_4| \leq \e^{\la}\sum_{n\geq0} C^{j+n}\int d\L \prod_{i=1}^n\left((j+n)v_{j+i}^2e^{-\frac{\b}{4}v_{j+i}^2}
+ \frac{4n}{e\b}e^{-\frac{\b}{4}v_{j+i}^2}\right) \label{eq:estBproof}
\end{equation}
where we used the bound
\be
\sum_{i=1}^{j+n} v_i^2  e^{-\frac {\beta }{4n}\sum_{i=1}^{j+n} v_i^2  } \leq \frac{4n}{e\b}\;.
\ee
The integral on the velocities in \eqref{eq:estBproof} factorizes so that
\be
|\mathcal{E}_4| \leq \e^{\la}\sum_{n\geq0} C^{j+n}\frac{t^n}{n!} (j+n)^n\;. \label{eq:estBproof1}
\ee
Since 
\be
\frac{(j+n)^n}{n!} \leq \frac{(j+n)^{j+n}}{(j+n)!} \leq e^{j+n}\;, \label{eq:nnnfact}
\ee
we have that \eqref{eq:estBproof1} is bounded by a geometric
series. Hence we conclude that
\be
|\mathcal{E}_4| \leq C^j\e^{\la}
\ee
for $t$ sufficiently small.

\bigskip
\ni {\bf Estimate of the set $\NN(\d)$ (set of $\d-$overlaps)}
\medskip

\ni At this point we just follow the lines of the proof of Proposition \ref{prop:TBT}. Using Definition \ref{def:defNNd}
and the notations introduced therein, we want to estimate 
\bea
&& |\mathcal{E}_5| \leq \sum_{n=0}^{\ol n} \sum_{\G(j,n)}\sum_{\bs_n} \Big[
\int d \L\ \mathbbm{1}_{ \mbox{ $\{\left(\prod B^\e\right)\leq\e^{-\la}\}
$}}\left(\prod B^\e\right)\mathbbm{1}_{\NN(\d)}f^{N}_{0,j+n}\nn\\
&&\ \ \ \ \ \ \ + \int d \L\ \mathbbm{1}_{ \mbox{$
\{\left(\prod B\right)\leq\e^{-\la}\} $}} \left(\prod B\right)\mathbbm{1}_{\NN(\d)}f_{0,j+n}\Big]\;.
\eea
By \eqref{eq:estfBsm} it is
\be
|\mathcal{E}_5| \leq \e^{-\la} \sum_{n=0}^{\ol n} \sum_{\G(j,n)}\sum_{\bs_n}C^{j+n}
\int d\L\ \mathbbm{1}_{\NN(\d)} e^{-\frac{\b}{2}\sum_{i=1}^{j+n}v_i^2}\;. \label{eq:E5fin}
\ee
We have  the following crucial result:
\bigskip
\begin{lem} \label{lem:explicitrec}
Let $\d=\e^{1-\mu} (\log \e)^2$ (see Eq. \eqref{eq:choosedelta}).
Given $\bz_j\in\O_j,$ if $\e$ is sufficiently small and $1 \leq n\leq \ol n,$
\be
\int d \L (\bt_n , \bn_n ,\bv_{j,n})\mathbbm{1}_{\NN(\d)} e^{-\frac{\b}{2}\sum_{i=1}^{j+n}v_i^2} \leq C^{j+n} \frac{t^n}{n!}
\d^{\frac{2}{5}}|\log\e|^{\frac{7}{2}}\;.
\label{eq:recollterm}
\ee
\end{lem}
Here the choice of $\d$ is not strict and is determined by Lemma \ref{lem:trajIBF}.

\bigskip
\emph{Proof of Lemma \ref{lem:explicitrec}.}
First notice that, given a point in $\NN(\d),$ there exists 
\be
t^*_{\d} = \max\{s\in[0,t^1]\ |\ |\xi^i(s)-\xi^h(s)|\leq \d\} \label{eq:deftstardelta}
\ee
for some couple of particles $(i,h).$ Substituting $t^*$ by $t^*_{\d},$ we may define $l, Y^q, W^q, f$ as after \eqref{eq:deftstar};
see also Figure \ref{fig:treeclimbing2}.

The case $l=0$ and $t^0=t, Y^0 \neq 0$ is made impossible by $\bz_j\in\O_j,$ as soon as $\d$ 
(i.e. $\e$) is smaller than a constant depending on 
$\bz_j.$ Conversely, the case $l=0$ and $t^0<t, Y^0 = 0$ occurs whenever a creation in the BBF is such that the two particles 
progenitor--son do not separate enough before their next (backwards) interaction. In formulas, $|W^0|(t^0-t^1)\leq\d.$ 
This case is controlled by introducing
\be
\mathbbm{1}^\d_0 = \prod_ {r=1}^n  \ \mathbbm{1}_{\mbox{ $\{|(v_{j+r}-\eta_{k_r}(t_r^+))|(t_r-t_{r+1}) >2t\d^{2/5}\}$}}
\label{eq:chard0}
\ee
(remember that the modulus of relative velocity is conserved at collisions). Clearly it is 
\be
\mathbbm{1}_{\NN(\d)} \leq \mathbbm{1}^\d_0\mathbbm{1}_{\NN(\d)} + \sum_{r=1}^n
\ \mathbbm{1}_{ \mbox{$\{|(v_{j+r}-\eta_{k_r}(t_r^+))|(t_r-t_{r+1}) \leq 2t\d^{2/5}\}$}}\;.
\ee
The reason for the choice of the threshold in definition \eqref{eq:chard0} will be clear soon.
For the moment note that we have
\bea
&& \sum_{r=1}^n\int d \L (\bt_n , \bn_n ,\bv_{j,n})
\mathbbm{1}_{ \mbox{$ \{|(v_{j+r}-\eta_{k_r}(t_r^+))|(t_r-t_{r+1}) \leq 2t\d^{2/5}\}$}}
e^{-\frac{\b}{2}\sum_{i=1}^{j+n}v_i^2} \nn\\
&& \leq C^n\frac{t^{n-1}}{(n-1)!}\d^{\frac{2}{5}}\;.  \label{eq:attpartest}
\eea
In fact, performing first the integration in $dt_rdv_{j+r},$ reminding that $\eta_{k_r}(t_r^+)=\eta_{k_r}(t_{r-1}^-)$
is independent on $t_r,v_{j+r},$ setting $s=t_r-t_{r+1}$ and $V = v_{j+r}-\eta_{k_r}(t_r^+),$ we find
\bea
&& \int_{t_{r+1}}^{t_{r-1}} dt_r \int dv_{j+r}\mathbbm{1}_{\mbox{ $\{|(v_{j+r}-\eta_{k_r}(t_r^+))|(t_r-t_{r+1}) \leq 2t\d^{2/5}\}$}}
e^{-\frac{\b}{2}v_{j+r}^2} \nn\\
&& \leq \int_{0}^{\d^{2/5}} ds \int dv_{j+r}e^{-\frac{\b}{2}v_{j+r}^2}
+  \int_{\d^{2/5}}^{t} ds \int_{0}^{\frac{2t\d^{2/5}}{s}} d|V| 4 \p |V|^2\nn\\
&& \leq C \d^{\frac{2}{5}}\;.
\eea
Since the integrals in the other variables give $C^{n-1}t^{n-1}/(n-1)!,$ we get \eqref{eq:attpartest}.

Suppose now that the considered point in $\NN(\d)$ is such that $ \mathbbm{1}^\d_0\mathbbm{1}_{\NN(\d)}=1.$
The $\d-$overlap condition is verified only if
\be
\min_s|Y^l-W^ls|\leq\d \label{eq:deltarecs}
\ee
for some $s \in [0,t^l),$ with $Y^l$ given by \eqref{eq:polygvt}. It must be $l\geq 1$ and $W^l\neq 0$. Moreover, the relative 
velocities $W^q$ cannot be all too close to each other, i.e. the two characteristic functions imply
\be
\sum_{q=1}^l|W^q-W^{q-1}| > \d^{\frac{2}{5}}\;. \label{eq:grcecoll}
\ee
Otherwise it would be $|W^q-W^0|\leq\d^{\frac{2}{5}}$ for all $q,$ thus using \eqref{eq:polygvt} and \eqref{eq:deltarecs} 
we would deduce that
\be
|Y^0-W^0s| \leq \d + \d^{\frac{2}{5}}t
\ee
for some $s>(t^0-t^l),$ which is forbidden, for $\d$ small, either by $\bz_j\in\O_j$ (case $Y^0\neq 0$) or 
by definition of $ \mathbbm{1}^\d_0$ (case $Y^0=0$).

Eq. \eqref{eq:deltarecs} implies $|Y^l \wedge \hat W^l| \leq \d,$ where $\hat W^l = \frac{W^l}{|W^l|}.$ 
Using again \eqref{eq:polygvt} we have
\bea
&& \d \geq \Big| Y^{0}\wedge \hat W^{l}-\sum_{q=0}^{l-1}(W^{q}\wedge \hat W^l)(t^q-t^{q+1}) \Big|\nn\\
&& \ \ =\Big|(Y^{0}-W^{0}t^{0})\wedge \hat W^{l}-\sum_{q=1}^{l}[(W^{q}-W^{q-1})\wedge \hat W^l)]t^{q}\Big|\;.
\label{eq:hardtcpr}
\eea
Since the vectors involved in this relation do not depend on the times (Remark on page \pageref{rem:BBF}), we can exploit the integration 
in the variables $t^q$ to estimate the set defined by this condition. However, a singularity will arise when the vector in the square brackets
is small for all $q.$ Let us focus first in this case, which is the most delicate. Assume that
\be
\label{<}
\sum_{q=1}^l | (W^q-W^{q-1} ) \wedge  \hat W^l | \leq  \d^{\frac{3}{5}}\;. 
\ee
Notice that if Eq. \eqref{<} is not satisfied, then $|(W^{q^*}-W^{q^*-1} ) \wedge  \hat W^l | >  \d^{\frac{3}{5}}/l $ for some $q^*.$

Again we may infer that condition \eqref{<} is not possible in the case $t^0=t, Y^0 \neq 0.$ Indeed, the above inequality trivially implies
\be
|W^0\wedge\hat W^l|\leq \d^{\frac{3}{5}}\;, \label{eq:WohWl}
\ee
therefore \eqref{eq:hardtcpr} gives
\be
|Y^0\wedge\hat W^l|\leq \d + 2t\d^{\frac{3}{5}}\;.
\ee
Putting together the two last equations we have
\be
|Y^0\wedge W^0| \leq C\left(|Y^0|+|W^0|\right)\d^{\frac{3}{5}}\;,
\ee
which is excluded, for $\d$ small, by $\bz_j\in\O_j.$

Let us study  \eqref{<} in the case $t^0<t, Y^0 = 0.$ By \eqref{eq:grcecoll}, there exists a $\ol q\in\{1,\cdots,l\}$ such that
\be
U \equiv U^{\ol q} := W^{\ol q}-W^{\ol q-1}
\ee
has modulus
\be
|U| > \frac{\d^{\frac{2}{5}}}{l}\;. 
\ee
But $|U\wedge \hat W^l|\leq \d^{\frac{3}{5}}$ by \eqref{<}. Then putting $\hat U=\frac{U}{|U|}$ we have
\be
|\hat U \wedge \hat W^l|\leq n \d^{\frac{1}{5}}\;. \label{eq:hUhWl}
\ee
On the other hand, by \eqref{eq:WohWl}, either
\be
|W^0|\leq \d^{\frac{2}{5}}\;, \label{eq:smallW0}
\ee
or $|\hat W^0\wedge\hat W^l|\leq \d^{\frac{1}{5}}$ which, together
with \eqref{eq:hUhWl} and the constraint $n\leq \ol n = O(|\log\e|),$ finally gives 
\be
|\hat W^0\wedge \hat U| \leq C\d^{\frac{1}{5}}|\log\e|\;. \label{eq:hW0whU}
\ee
We will use this formula to estimate the considered events, taking advantage from the fact that $U$ depends only on the impact vector and
relative velocity describing the interaction occurring at time $t^{\ol q}$ in the BBF (as already pointed out on page \pageref{remU} and Figure
\ref{fig:treeclimb}), and that $W^0$ is the (incoming) relative velocity of the interaction at time $t^0.$

We shall summarize the discussion above as follows. Denote $V_r$ and $V'_r$ respectively the outgoing and incoming relative velocities
of the collision at time $t_r$ in the BBF. If $t^{\ol q}=t_r,$ we use the notation $U_r=U^{\ol q}.$ This is a function of $\n_r, V_r$ only. We have
\be
\mathbbm{1}^\d_0\mathbbm{1}_{\NN(\d)} \leq \sum_{i,h}\sum_{l=1}^{n_{ih}}\sum_{q^*=1}^l \mathbbm{1}_{\NN_{ih}^{l,q^*}(\d)} 
+ \sum_{r=1}^n\mathbbm{1}_{\{|V_r|\leq \d^{2/5}\}} + 
\sum_{f=1}^n\sum_{f' = f+1}^n \mathbbm{1}_{\{|\hat V'_{f} \wedge \hat U_{f'}|\leq C\d^{1/5}|\log\e|\}}
\ee
where $1 \leq i<h \leq j+n$ and
\bea
&&\NN_{ih}^{l,q^*}(\d)= \Big\{\bt_n, \bn_n, \bv_{j,n}\ \Big|\ \mbox{the virtual trajectories of $i$ and $h$ satisfy \eqref{eq:hardtcpr},}\nn\\
&&\ \ \ \ \ \ \ \ \ \ \ \ \ \ \ \ \ \ \ \ \ \ \ \ \ \ \ \ \ \ \ \ \ \mbox{with}\ |(W^{q^*}-W^{q^*-1} ) \wedge  \hat W^l | >  \d^{\frac{3}{5}}/l \Big\}\;.
\eea

Once fixed all the variables but $t^{q^*},$ if Equation \eqref{eq:hardtcpr} is verified, then $t^{q^*}$ belongs to an interval of length smaller
than $\d |(W^{q^*}-W^{q^*-1} ) \wedge  \hat W^l |^{-1}.$ If we are in $\NN_{ih}^{l,q^*}(\d)$ this is bounded by $n\d^{\frac{2}{5}},$ so that
\be
\int d \L (\bt_n , \bn_n ,\bv_{j,n})\mathbbm{1}_{\NN_{ih}^{l,q^*}(\d)} e^{-\frac{\b}{2}\sum_{i=1}^{j+n}v_i^2} \leq C^n\frac{t^{n-1}}{(n-1)!}
\d^{\frac{2}{5}}\;.
\ee

Changing variable $v_{j+r}\to V_r$ we easily find
\be
\int d \L (\bt_n , \bn_n ,\bv_{j,n})\mathbbm{1}_{
\mbox{$\{|V_r|\leq \d^{2/5}\}$}} e^{-\frac{\b}{2}\sum_{i=1}^{j+n}v_i^2} \leq C^n\frac{t^{n}}{n!}
\d^{\frac{6}{5}}\;.
\ee

Furthermore, it is
\be
\int d \L (\bt_n , \bn_n ,\bv_{j,n}) \mathbbm{1}_{\mbox{$\{|\hat V'_{f} \wedge \hat U_{f'}|\leq C\d^{1/5}|\log\e|\}$}}
e^{-\frac{\b}{2}\sum_{i=1}^{j+n}v_i^2} \leq C^n\frac{t^{n}}{n!} \d^{\frac{2}{5}}|\log\e|^{\frac{7}{2}}\;. \label{eq:delhardpr}
\ee
To prove this last inequality, it is convenient first to introduce a further restriction to the set $\{\frac{\b}{2} |V_{f'}|^2 < 4 |\log \e | \}$
where $V_{f'}=(v_{j+f'}-\eta_{k_{f'}}(t_{f'}^+)).$ If the opposite inequality holds, then either $|v_{j+f'}|$ or $|\eta_{k_{f'}}(t_{f'}))|$ cannot be
smaller than $\sqrt{2/\b|\log\e|},$ hence the total energy must be larger than $1/\b|\log\e|.$ Therefore, using
the energy--cutoff $\mathbbm{1}^\e_1$ defined in \eqref{eq:chfue12}, the error produced is bounded by
\bea
&& \int d \L (\bt_n , \bn_n ,\bv_{j,n}) \left(1-\mathbbm{1}^\e_1\right)e^{-\frac{\b}{2}\sum_{i=1}^{j+n}v_i^2} \nn\\
&& \leq e^{-\frac{1}{2}|\log\e|} \int d \L (\bt_n , \bn_n ,\bv_{j,n}) e^{-\frac{\b}{4}\sum_{i=1}^{j+n}v_i^2} \leq C^n\frac{t^n}{n!}\e^{\frac{1}{2}}\;,
\label{eq:expesten}
\eea
which is in turn certainly smaller than $(Ct)^n/n! \d^{\frac{2}{5}},$ being $\d$ given by \eqref{eq:choosedelta}.
We are left with 
\be
\int d \L (\bt_n , \bn_n ,\bv_{j,n}) \mathbbm{1}_{\mbox{$\{|\hat V'_{f} \wedge \hat U_{f'}|\leq C\d^{1/5}|\log\e|\}$}}
\mathbbm{1}_{\mbox{$\{\frac{\b}{2} |V_{f'}|^2 < 4 |\log \e | \}$}} e^{-\frac{\b}{2}\sum_{i=1}^{j+n}v_i^2}\;.
\ee
We change the integration variables according to \eqref{eq:newvarrest}--\eqref{eq:newvarIr}. Note that
\be
e^{-\frac{\b}{2}v_{j+f}^2}e^{-\frac{\b}{2}v_{j+f'}^2} = e^{-\frac{\b}{2}\left(V'_f-2\o_f(\o_f\cdot V'_f)+\eta_{k_f}(t_f^+)\right)^2}
e^{-\frac{\b}{2}\left(V_{f'}+\eta_{k_{f'}}(t_{f'}^+)\right)^2}\;,
\ee
where $\o_f=\o(\n'_f,V'_f)$ is the scattering vector at the collision. Since $V_{f'}$ varies in a compact set,
we bound the second exponential simply by $1,$ while the first exponential is estimated by 
$e^{-\frac{\b}{2}\left(|V'_f|-|\eta_{k_f}(t_f^+)|\right)^2},$ where $\eta_{k_f}(t_f^+)$ depends only on the variables with index strictly smaller than $f.$
Performing first the integrations in $dV_{f'}dV'_f$ we find
\bea
&& \int dV_{f'}\mathbbm{1}_{\{\frac{\b}{2} |V_{f'}|^2 < 4 |\log \e | \}} \int dV'_{f}
\mathbbm{1}_{\{|\hat V'_{f} \wedge \hat U_{f'}|\leq C\d^{1/5}|\log\e|\}}e^{-\frac{\b}{2}\left(|V'_f|-|\eta_{k_f}(t_f^+)|\right)^2}\nn\\
&& \leq  \int dV_{f'}\mathbbm{1}_{\{\frac{\b}{2} |V_{f'}|^2 < 4 |\log \e | \}} \left(C\d^{\frac{2}{5}}|\log\e|^2\right)
\int d|V'_{f}| |V'_{f}|^2e^{-\frac{\b}{2}\left(|V'_f|-|\eta_{k_f}(t_f^+)|\right)^2}\nn\\
&& \leq \left(C |\log\e|^{\frac{3}{2}}\right) \left(C\d^{\frac{2}{5}}|\log\e|^2\right)\left(C(1+|\eta_{k_f}(t_f^+)|^2)\right)\nn\\
&& \leq C \left(1+\sum_{i=1}^{j+f-1}v_i^2\right)\d^{\frac{2}{5}}|\log\e|^{\frac{7}{2}}\;.
\eea
Integrating in the remaining variables we readily get Equation \eqref{eq:delhardpr}.

Collecting all the estimates,  Lemma \ref{lem:explicitrec} is proved. \qed

\bigskip
Substituting Eq. \eqref{eq:recollterm} in Eq. \eqref{eq:E5fin}, performing the sums and using \eqref{eq:nnnfact}, we conclude that
\be
|\mathcal{E}_5| \leq C^j \e^{-\la}\d^{\frac{2}{5}}|\log\e|^{\frac{7}{2}} \leq \e^{\frac{2}{5}-\frac{2}{5}\mu-\la}|\log\e|^{\frac{43}{10}}
\ee
for $t$ small enough.

\bigskip
\ni {\bf High energies, low angular momenta}
\bigskip

\ni We turn now to the estimates of the errors coming from the truncations defined in \eqref{eq:chfue12}. Proceeding as above we have
\bea
&& |\mathcal{E}_6| \leq \sum_{n=0}^{\ol n} \sum_{\G(j,n)}\sum_{\bs_n} \Big[
\int d \L\ \mathbbm{1}_{\{\left(\prod B^\e\right)\leq\e^{-\la}\}}\left(\prod B^\e\right)\left(1-\mathbbm{1}_{\NN(\d)}\right) 
\left(1-\mathbbm{1}^\e_1\mathbbm{1}^\e_2\right) f^{N}_{0,j+n}\nn\\
&&\ \ \ \ \ \ \ + \int d \L\ \mathbbm{1}_{\{\left(\prod B\right)\leq\e^{-\la}\}} \left(\prod B\right)\left(1-\mathbbm{1}_{\NN(\d)}\right)
\left(1-\mathbbm{1}^\e_1\mathbbm{1}^\e_2\right) f_{0,j+n}\Big] \nn\\
&& \leq \e^{-\la} \sum_{n=0}^{\ol n} \sum_{\G(j,n)}\sum_{\bs_n}C^{j+n} \int d\L\ \left(1-\mathbbm{1}_1^\e\mathbbm{1}_2^\e\right) 
e^{-\frac{\b}{2}\sum_{i=1}^{j+n}v_i^2}\nn\\
&& \leq C^j \e^{\frac{1}{2}-\la} + 
\e^{-\la} \sum_{n=0}^{\ol n} \sum_{\G(j,n)}\sum_{\bs_n}C^{j+n} \int d\L\ \left(1-\mathbbm{1}_2^\e\right) 
e^{-\frac{\b}{2}\sum_{i=1}^{j+n}v_i^2}\;,
\eea
where in the last step we used \eqref{eq:expesten}.  Moreover
\bea
\int d\L\left(1-\mathbbm{1}_2^\e\right) e^{-\frac{\b}{2}\sum_{i=1}^{j+n}v_i^2}
\leq\sum_{r=1}^n \int d\L\ \mathbbm{1}_{ \mbox{$ \{|(v_{j+r}-\eta_{k_r}^\e(t_r))\wedge\n_r| \leq \e^{\mu} \}
$}} e^{-\frac{\b}{2}\sum_{i=1}^{j+n}v_i^2}\;. \nn\\
\eea
We perform first the integrations in $dv_{j+r}d\n_r.$  Setting $V_r = (v_{j+r}-\eta_{k_r}^\e(t_r)),$ $\a$ the 
angle between $V_r$ and $\n_r$  and using the parametrization $\n_r \to (\r,\phi)$ where $\r = \sin\a$ and $\phi\in[0,2\p)$, we have
\bea
&& \int dv_{j+r} d\n_r \mathbbm{1}_{\mbox{$  \{|(v_{j+r}-\eta_{k_r}^\e(t_r))\wedge\n_r| \leq \e^{\mu} \}$}} 
e^{-\frac{\b}{2}v_{j+r}^2}\nn\\
&& = \int dv_{j+r} (2\p)2\int_0^1 d\r \frac{\r}{\sqrt{1-\r^2}}
\mathbbm{1}_{ \mbox{$ \{|(v_{j+r}-\eta_{k_r}^\e(t_r))| \r \leq \e^{\mu} \}$}} e^{-\frac{\b}{2}v_{j+r}^2}\nn\\
&& \leq  \int dv_{j+r} (2\p)2\int_0^{\e^{\mu}} d\r \frac{\r}{\sqrt{1-\r^2}} e^{-\frac{\b}{2}v_{j+r}^2}
+ (2\p)2\int_{\e^{\mu}}^1 d\r \frac{\r}{\sqrt{1-\r^2}}\int_0^{\frac{\e^{\mu}}{\r}} d|V_r| 4\p |V_r|^2\nn\\
&& \leq C\e^{2\mu}\;.
\eea
The integrals in the other variables give $C^{n-1} t^{n}/n!,$ therefore performing the sums as above we obtain
\be
|\mathcal{E}_6| \leq C^j\left(\e^{\frac{1}{2}-\la}+\e^{2\mu-\la}\right)\;.
\ee

\bigskip
\ni {\bf Conclusion (convergence of initial data and backwards flows)}
\bigskip

\ni Now we shall estimate what is left of the last term in \eqref{eq:sumdifftrees}. This gives two errors: one is due to the convergence
of the initial data (formula \eqref{eq:hypcr1} in Hypothesis \ref{hyp:convrate}) and the other is due to the convergence of the 
IBF to the BBF (Lemma \ref{lem:trajIBF}). The first is
\bea
&& |\mathcal{E}_7| \leq \sum_{n=0}^{\ol n} \sum_{\G(j,n)}\sum_{\bs_n}
\int d \L\ \mathbbm{1}_{\{\left(\prod B^\e\right)\leq\e^{-\la}\}}\left(\prod B^\e\right)\left(1-\mathbbm{1}_{\NN(\d)}\right) 
\mathbbm{1}^\e_1\mathbbm{1}^\e_2\nn\\
&& \ \ \ \ \ \ \ \ \ \ \ \ \ \ \ \ \ \ \ \ \ \ \ \ \ \ \ \ \ \ \ \ \ \ \ \ \ \ \ \ \ \ \ \cdot\Big| f^{N}_{0,j+n}(\bze^\e(0)) -  f_{0,j+n}(\bze^\e(0))\Big|\;.
\eea
Since we integrate outside $\NN(\d),$ the BBF satisfies $\bze(0)\in\MM_j(\d).$ But $\d = \e^{1-\mu}(\log\e)^2.$ Thus, applying Lemma
\ref{lem:trajIBF}, the IBF must satisfy $\bze^\e(0)\in\MM_j(\e)$ for $\e$ small enough. Hence Hypothesis \ref{hyp:convrate} together
with conservation of energy lead to
\bea
&& |\mathcal{E}_7| \leq \e^{1-\la} \sum_{n=0}^{\ol n} \sum_{\G(j,n)}\sum_{\bs_n} C^{j+n}
\int d \L \ e^{-\frac{\b}{2}\sum_{i=1}^{j+n}v_i^2} \nn\\
&&\ \ \ \ \leq C^j \e^{1-\la}\;,
\eea
having performed the sums in the usual way.

\bigskip
Finally, the last error is
\bea
&& |\mathcal{E}_8| \leq \sum_{n=0}^{\ol n} \sum_{\G(j,n)}\sum_{\bs_n}
\int d \L\ \mathbbm{1}_{\{\left(\prod B^\e\right)\leq\e^{-\la}\}}\left(\prod B^\e\right)\left(1-\mathbbm{1}_{\NN(\d)}\right) 
\mathbbm{1}^\e_1\mathbbm{1}^\e_2\nn\\
&& \ \ \ \ \ \ \ \ \ \ \ \ \ \ \ \ \ \ \ \ \ \ \ \ \ \ \ \ \ \ \ \ \ \ \ \ \ \ \ \ \ \ \ \cdot\Big| f_{0,j+n}(\bze^\e(0)) -  f_{0,j+n}(\bze(0))\Big|\;.
\eea
Lemma \ref{lem:trajIBF}, the regularity assumption \eqref{eq:hypcr2} in Hypothesis \ref{hyp:convrate} and conservation of energy 
at collisions imply that we can bound the modulus in the last line by
\be
C^{j+n}e^{-\frac{\b}{2}\sum_{i=1}^{j+n}v_i^2}\left(\sum_{k=1}^{j+n}\left|\xi^{\e}_k(0)-\xi_k(0)\right|^2\right)^{\frac{1}{2}} 
\leq C^{j+n}e^{-\frac{\b}{2}\sum_{i=1}^{j+n}v_i^2} (j+n)^{\frac{1}{2}}D\e^{1-\mu}|\log\e|^{\frac{3}{2}}\;, 
\ee
for $\e$ sufficiently small. Therefore, proceeding as above we have
\be
|\mathcal{E}_8| \leq C^j\e^{1-\mu-\la}|\log\e|^{\frac{3}{2}}\;.
\ee

\bigskip
Putting together all the errors $\mathcal{E}_1,\cdots,\mathcal{E}_8$ and optimizing on $\mu$ and $\la$ we conclude that
\be
|f_j^N(\bz_j,t) - f_j(\bz_j,t)| \leq C^j\e^\g \mbox{\ \ \ \ \ for any\ \ \ \ \  }\g<\frac{1}{6}\;.
\ee
\qed

%%%%%%%%%%%%%%%%%%%%%%%%%%%%%%%%%%%%%%%%%%%%%%%%%%%%%%
%%%%%%%%%%%%%%%%%%%%%%%%%%%%%%%%%%%%%%%%%%%%%%%%%%%%%%
%%%%%%%%%%%%%%%%%%%%%%%%%%%%%%%%%%%%%%%%%%%%%%%%%%%%%%
%%%%%%%%%%%%%%%%%%%%%%%%%%%%%%%%%%%%%%%%%%%%%%%%%%%%%%
%%%%%%%%%%%%%%%%%%%%%%%%%%%%%%%%%%%%%%%%%%%%%%%%%%%%%%

\section{Stable potentials} \label{sec:gener}
\setcounter{equation}{0}    
\def\theequation{8.\arabic{equation}}

In this section we show how the techniques used in proving Theorem \ref{thm:soft} can be extended to treat a fairly larger class of potentials, 
including those with an  attractive part.

The potentials $\Phi$ considered in the present section satisfy the following conditions.

\bigskip
\noindent {\bf Hypothesis $\mathbf{1'}$\ } {\em $\Phi = \Phi(q), q\in\RRR^3$ is radial, with support $|q|<1$. 
We further assume 

\noindent 1)  either $\Phi\in C^2(\RRR^3),$ or $\Phi\in C^2(\RRR^3\setminus\{0\})$ and $\Phi(q) \to +\infty$ as $q\to 0$;

\noindent 2) $\Phi$ is stable.}

\bigskip
\noindent In what follows we will use the usual notational inconsistency $\Phi(r)=\Phi|_{|q|=r}.$

We remind (see e.g. \cite{Ru69}) that an interaction is stable if it fulfills the following condition:
\begin{equation}
\label{sup}
U(q_1,\cdots, q_j)=\sum_{i<h}  \Phi (|q_i-q_h|) \geq -Bj
\end{equation}
for some constant  $B>0$ . In particular, $\Phi$ is positive (possibly diverging) at the origin.
We also remark that condition \eqref{sup} ensures the existence of the Partition Function and hence the existence of an equilibrium measure.
In our context the stability will be used to guarantee that Hypothesis \ref{hyp:f0jbound} implies the bound $f^N_{0,j}\leq c^je^{-\frac{\b}{2}\sum_iv_i^2}$
(where $c=e^{\a+\b B}$), which is crucial in our proof.

The potentials $\Phi$  we are considering  include a sort of truncated Lennard-Jones potential (see Fig. \ref{fig:cutLJ} below).
\begin{figure}[htbp] 
 \centering
  \includegraphics[width=2.5in]{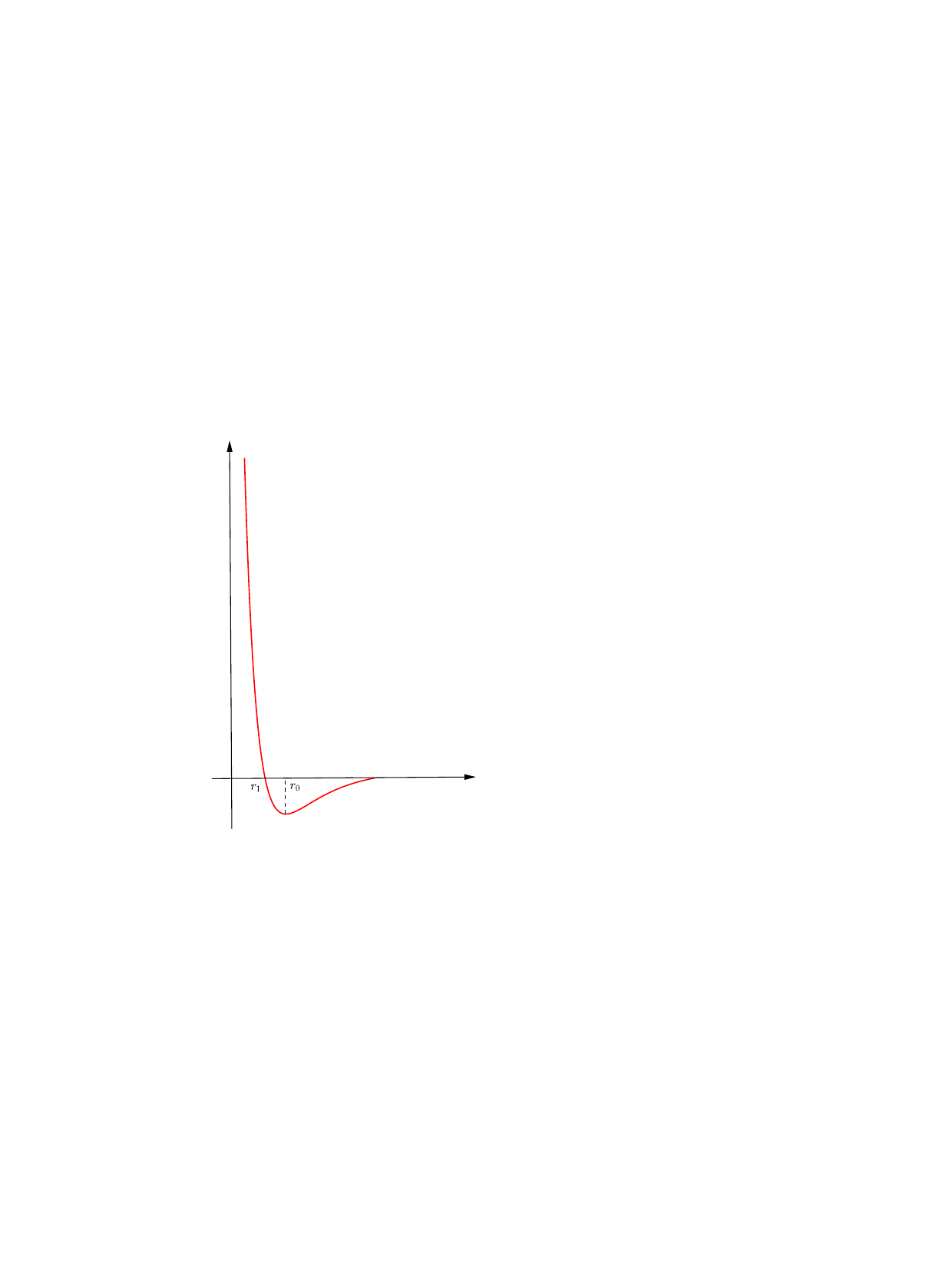} 
  \caption{A cutoffed Lennard--Jones potential.}
  \label{fig:cutLJ}
\end{figure}

We note that the proof presented in Section \ref{sec:proofsoft} depends on $\Phi$ {\em only} 
through the scattering time $\t_*$. Therefore,
in trying to extend it to the present situation, the crucial point will be the control of $\t_*,$ in absence of  Lemma \ref{lem:timebound} 
of Sec. \ref{sec:2bs} which is not valid anymore. 

According to Section \ref{sec:2bs}, we reduce the two--body particle system to the central 
motion of a single particle, with mass $\frac 12$ and velocity $V$ (the relative velocity of the two interacting particles).

We recall the  formula yielding the interaction time:
\be
\label{int}
\t_* = \sqrt{2} \int_{r_*}^1 dr \frac{1}{\left(\frac{V^2}{2} - \frac{L^2}{2r^2}-2\Phi(r)\right)^{1/2}}\;,
\ee
where $|V|>0$ is the modulus of  the relative velocity before the collision, $\rho$ is the impact parameter,  $L=|\rho V|\in[0,|V|]$ is the modulus of the 
angular momentum and $r_*$ is  the infimum of the distance from the origin during the scattering process.
$r_*$ is given by
\be
r_* = \max\Big\{ x \in [0,1) \ \Big|\  \frac{V^2}{2} = \frac{L^2}{2x^2}+2\Phi(x)\Big\}\;. \label{eq:defrsbis}
\ee

\begin{figure}[htbp] 
 \centering
  \includegraphics[width=3in]{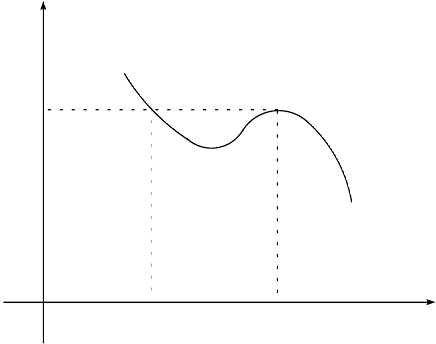} 
  \caption{When the (real) interaction has attractive parts, the effective potential (drawn in figure) can have local maxima for a given value of $L$.}
  \label{fig:effpot}
\end{figure}

Before establishing the following lemma in which we control the scattering time, we discuss the new difficulties we face in presence 
of an attractive part. 
Consider for instance the potentials described in Fig. \ref{fig:cutLJ},  with a single negative well. The effective potential
\begin{equation}
\label{epot}
\Phi(r)+\frac {L^2}{4r^2}-\frac {L^2}{4}
\end{equation}
may  have two critical points, $r_m$ and $r_M$ (minimum and maximum respectively), when $L$ is sufficiently small; see Figure \ref{fig:effpot}.
Fixing a value of $L$ for which such critical points do exist, there are values of $V$ for which
\be
\frac{V^2}{2} \approx 2\Phi (r_M)+ \frac {L^2} {2r_M^2}.
\ee
In this case the trajectory is close to an unstable periodic orbit and $\t_*$ is very  large. The two particles turn around each other many times 
and remain trapped for a long time. Clearly such situations are pathological and must be excluded  in order to have a kinetic picture. 
Actually the following lemma says that the set of such pathological events has a small measure, although we do not give explicit 
estimates.
\bigskip
\begin{lem} \label{lem:timeatt}
Given $\eta \in (0,1)$ and $K\in\RRR^+,$ there exists a set $\cal {B} (\eta)$ of pairs $(\nu, V)$ such that
\begin{equation}
\label{propo}
\int _{S^2} d \nu  \int _{\{\nu\cdot V \leq 0\} } dV  |\nu\cdot V|  \mathbbm{1}_{\cal {B} (\eta)} \to 0
\end{equation}
as $\eta \to 0$ and  such that, for $(\nu,V)\notin \cal {B} (\eta)$ and $|V| < K$,
\be
\tau_*(\n,V) <\varphi (\eta,K)\;,
\ee
where $\varphi (\eta,K) $ is a positive function which may possibly diverge as $\eta \to 0$ or $K\to+\infty.$ 
\end{lem}
\bigskip

{\em Proof.}
We can easily find the set of pairs $(L^2,V^2)$ for which $\tau_*$ diverges. Such a set is included in the set for which there exist
local maxima of the effective potential \eqref{epot}. The critical points satisfy
\begin{equation}
\label{c1}
\Phi'(y)-\frac {L^2}{2y^3}=0\;.
\end{equation}
Therefore the pairs $(L^2,V^2)$ corresponding to a divergence of the scattering time $\tau_*$ must satisfy \eqref{c1} and
\begin{equation}
\label{c2}
\frac{V^2}{2} =2\Phi(y)+\frac {L^2}{2y^2}\;.
\end{equation}
This last condition is due to Eq. \eqref{eq:defrsbis}, while \eqref{c1} ensures that the orbit reaches $r_* = y$ in an infinite time.

Consider now the curve  ${\cal C}$ in the plane, $y\in (0,1) \to (X,Y)$, whose parametric equations are
\be
\begin{cases}
\displaystyle X=2\Phi'(y)y^3 \\
\displaystyle Y=4\Phi(y) +2\Phi'(y)y 
\end{cases}\;.
\label{curve}
\ee
Then the set of singular values of $(L^2,V^2)$ lies inside the restriction $\tilde{\cal C}$ of this curve $(X,Y)$ to the``physical''subset 
\be
\Big\{(X,Y) \ \Big|\  Y>0,\  0\leq X \leq Y\Big\}\;.
\label{eq:physsubset}
\ee
Clearly when $\Phi$ is bounded the curve ${\cal C}$ is extended by continuity to $y=0$ (for which $L^2=X(0)=0, V^2 = Y(0)=4\Phi(0)$
are indeed singular points of $\t_*$). Note that, when $\Phi$ is unbounded, the parameter $y$ spanning $\tilde{\cal C}$ is bounded away from
zero, since it cannot be smaller than  $r_0 := \min \{ x \in (0,1] \ |\ \Phi'(x) \geq 0\}.$

Denoting by  $B((X,Y);\eta)$ the disk of center  $(X,Y)$ and radius $\eta$, we introduce the tube
\begin{equation}
{ \cal T} (\eta)= \bigcup_{(X,Y)\in {\cal C} }
B((X,Y);\eta)
\end{equation} 
and its restriction to the physical region
\begin{equation}
\tilde{ \cal T} (\eta)= \bigcup_{(X,Y)\in \tilde{\cal C} }
B((X,Y);\eta)\;.
\end{equation} 
Now observe that, due to the smoothness of $\Phi$,  the set $\tilde{\cal C}$ has finite length so that
\be
|\tilde{ \cal T}  (\eta)| \leq C \eta\;,
\ee
where $|A|$ denotes the Lebesgue measure of the set $A$.

Consider the set 
\be
\tilde{ \cal T} (\eta)\cup B((0,0);\eta)\cup\{|V| \geq K\}\;. \label{eq:exttube}
\ee
Its complement $\GG(\eta,K)$ in the physical region \eqref{eq:physsubset} is relatively compact. Therefore, by continuity of 
$\t_*(L^2, V^2)$ in the set \eqref{eq:physsubset} deprived of $\tilde{\cal C}$ (and hence in the closure of $\GG$), we have
\be
\tau_* <\varphi (\eta,K)
\ee
in $\GG(\eta,K),$ for a suitable positive function $\varphi,$ possibly diverging as $\eta \to 0$ or $K\to+\infty.$ 

To prove the required continuity of $\t_*,$ we observe first that $r_*=r_*(L^2,V^2)$ is continuous outside $\tilde{\cal C}$. 
Fix a point $(L_0^2,V_0^2)\notin\tilde{\cal C}$ in the set \eqref{eq:physsubset}. Then, for any $\g\in\left(0,\frac{1-r_*(L_0^2,V_0^2)}{2}\right),$ the integral
\be
\sqrt{2} \int_{r_*+\g}^1 dr \frac{1}{\left(\frac{V^2}{2} - \frac{L^2}{2r^2}-2\Phi(r)\right)^{1/2}}
\ee
is continuous in $(L_0^2,V_0^2),$ the integrand being bounded. On the other hand, 
\be
\sqrt{2} \int_{r_*}^{r_*+\g} dr \frac{1}{\left(\frac{V^2}{2} - \frac{L^2}{2r^2}-2\Phi(r)\right)^{1/2}} \longrightarrow 0
\ee
as $\g\to 0$, uniformly for $(L^2,V^2) \in B((L_0^2,V_0^2);\d),$ if $\d$ is small enough. This can be seen by an argument as the one in Lemma
\ref{lem:timebound}, namely using the estimate \eqref{eq:timerstarL} replacing the integration interval $(r_*,1)$ with $(r_*,r_*+\g)$.

To conclude the proof of the lemma, we introduce the set
\begin{equation}
{\cal B} (\eta)=\Big\{ (\nu, V)\in S^2\times\RRR^3 \ \Big|\  \left(L^2, V^2 \right)\in \tilde{ \cal T}  (\eta)\cup B((0,0);\eta)\Big\}
\end{equation}
where $L=|\n\wedge V|.$
Setting $\cos \alpha= \nu\cdot \hat V,$ $\hat V = V/|V|$ and noticing that $L=|V\sin\a|,$ the left hand side of \eqref{propo} is bounded by
\begin{eqnarray}
&& \int_{S^2}  d \nu \int_{S^2}  d\hat V \int_0^\infty d|V| |V|^3 |\cos \alpha| \mathbbm{1}_{\cal {B} (\eta)} 
= 8\pi^2 \int_0^\infty d|V| |V|^3 \int_{0}^{\pi} d\alpha \sin \alpha  |\cos \alpha|  \mathbbm{1}_{\cal {B} (\eta)}\nn \\ 
&&\ \ \ \ \ \ \ \ \ \ \ \ \leq 4\pi^2 \int_0^\infty dV^2 \int_0 ^{V^2}  dL^2  \mathbbm{1}_{\cal {B} (\eta)} \leq C\eta
\end{eqnarray}
for $\eta$ sufficiently small.
\qed

We are now in a position to establish and prove the main result of the present section.

\noindent {\bf Theorem $\mathbf{1'}$} {\em (Improved)
Under the Hypotheses $1'$ and \ref {hyp:f0bound}--\ref{hyp:conv} of Section \ref{sec:results},
there exists $t_0 >0$ such that, for any positive $t<t_0$ and $j\in \mathbb{N}$,
the series expansions \eqref{eq:fnjexp} and \eqref{eq:fjexp} are absolutely convergent (uniformly in $\e$),
and
\be
\lim_{\substack{\e\rightarrow 0\\ N\e^2=1}} f_j^N(t) = f_j(t)
\ee
uniformly on compact sets in $\O_j.$
}

\bigskip
{\em Proof.}
We just mention where  the previous proof of Theorem \ref{thm:soft} requires modifications and how to do them.

The proof consists of two parts, namely the short time estimate and the term by term convergence.

As regards the short time bound in Section \ref{sec:ste}, by virtue of the stability property, it is natural to modify the definition of the Hamiltonian 
by setting
\be
H_B ( {\bf z}_j)=H ( {\bf z}_j)+jB \geq \frac{1}{2}\sum _{i=1}^j v_i ^2 \geq 0\;. \label{eq:propHB}
\ee
Consequently we introduce the norms \eqref{eq:defnorms} replacing $H$ by $H_B$.
Next we deduce estimate \eqref {eq:Abound} by using \eqref{eq:propHB} and the fact that $H_B$ satisfies the inequality \eqref{eq:HgHpH}, i.e.
\be
H_B ( {\bf z}_{j+1+m} ) = H_B ( {\bf z}_j) +H_B ( {\bf z}_{j,1+m} ) 
\geq H_B ( {\bf z}_j) + \frac 12 \sum_{i=j+1}^{j+1+m} v_i^2\;.
\ee
We use the stability of the potential only in this part of the proof.

Now we pass  to analyze the term by term convergence. Everything is going on as in Section 7.2, with the only difference that we replace

\be
\mathbbm{1}^\e_2 = \prod_ {r=1}^n  \mathbbm{1}_{\mbox{$  \{|(v_{j+r}-\eta_{k_r}^\e(t_r))\wedge\n_r| > \e^{\mu} \}$}}
\ee
by
\be
\mathbbm{1}^\e_2 = \prod_ {r=1}^n \mathbbm{1}_{
\mbox{$\{(\n_r, v_{j+r}-\eta_{k_r}^\e(t_r)) \notin {\cal B} (\eta), |v_{j+r}-\eta_{k_r}^\e(t_r)|<K\}$}}\;.
\ee
According to the form of the function $\varphi$, we can choose $\eta=\eta(\e)$ and $K=K(\e)$ such that $\eta \to 0$ and $K\to \infty$ with 
$\e$ and the scattering time of the collisions associated to the nodes of the tree (in macroscopic variables) is bounded by 
$\e \varphi (\eta,K)\leq A\e^{1-\mu}$ whenever  $\mathbbm{1}^\e_2=1$ (by virtue of Lemma \ref{lem:timeatt}). 
Therefore Lemma \ref{lem:trajIBF} (hence Eq. \eqref{eq:Ttildee}) is still valid and the proof is completed by observing that, by Lemma \ref{lem:timeatt},
${\cal } B(\eta)$ is a set of vanishing measure (so Eq. \eqref{eq:remcoEL} holds).
\qed

%%%%%%%%%%%%%%%%%%%%%%%%%%%%%%%%%%%%%%%%%%%%%%%%%%%%%%
%%%%%%%%%%%%%%%%%%%%%%%%%%%%%%%%%%%%%%%%%%%%%%%%%%%%%%
%%%%%%%%%%%%%%%%%%%%%%%%%%%%%%%%%%%%%%%%%%%%%%%%%%%%%%
%%%%%%%%%%%%%%%%%%%%%%%%%%%%%%%%%%%%%%%%%%%%%%%%%%%%%%
%%%%%%%%%%%%%%%%%%%%%%%%%%%%%%%%%%%%%%%%%%%%%%%%%%%%%%

\section{Concluding remarks} \label{sec:conc}
\setcounter{equation}{0}    
\def\theequation{9.\arabic{equation}}

We conclude by discussing some additional remarks.

\bigskip
1.  The potentials we have considered are fairly general, but the basic hypothesis is the short-range assumption. 

From the very beginning of the Kinetic Theory, Boltzmann himself (see \cite{Bo64}), 
following Maxwell \cite{Ma67, Ma95}, considered only inverse power law potentials, besides the hard--sphere system originally
investigated in deriving his famous equation. This  is probably due to the good scaling properties of such potentials. 
Moreover the differential cross--section is well defined, even though the total cross-section is diverging because of the long range of the interaction.
On the other hand, it is not clear whether the Boltzmann equation associated to these potentials can indeed be derived under 
the low--density scaling. 

A simpler problem would be to consider a sequence of potentials  with range gently diverging with $N$. 
This problem eludes the present techniques so that  we consider it as an interesting, open problem.

For an analysis concerning the much easier problem of the validity of the linear Boltzmann equation 
for long range potentials, see reference \cite{DP99}.

\bigskip
2. In the present paper we give an explicit estimate of the error in case of a completely 
repulsive potential (Theorem 2), while, for stable potentials, we only show the convergence. It would be interesting
to develop a constructive proof of  convergence also in this last case. This would require a more precise estimate of the 
scattering time to improve Lemma \ref{lem:timeatt}.

\bigskip
3. The present validity results, as the ones in the previous literature, are formulated in a canonical context, namely, for any $\e >0,$ 
the number of particles $N$ is authomatically fixed.
An equivalent formalism is the grand--canonical one. Here the number of particles is random but the density is fixed.

More precisely consider, for a given $\e$,  the phase space of the system as
\begin{equation}
{\cal M}=\bigcup_{N\geq 0} {\cal M}_N
\end{equation}  
where ${\cal M}_N$ is the $N$--particle phase space (see \eqref{eq:Npps}).
For $z \in {\cal M}$ we define the dynamical flow by solving the Newton equations in each ${\cal M}_N$.
Similarly we define a symmetric probability measure $W^\e$ on ${\cal M}$ by means of a sequence of
symmetric probability measures $W^N$ in each ${\cal M}_N$:
\begin{equation}
W^\e|_{{\cal M}_N} =e^{-\mu_\e} \frac {\mu_\e^N}{N!} W^N.
\end{equation}

We define the sequence $\{ g_j ^\e\}_{j=1}^{\infty}$ by
\begin{equation}
  g_j ^\e =\sum_{N \geq j}e^{-\mu_\e} \frac {\mu_\e^N}{N!} g_j^N
\end{equation}
where $g^N_j$ are the marginals of $W^N$. Therefore $ g_j ^\e({\bf z}_j)$ are the probability densities of finding
the first $j$ particles in ${\bf z}_j$. Their normalizations are
\be
\sum_{N\geq j} e^{-\mu_\e} \frac {\mu_\e^N}{N!}
\ee
which is the probability of finding more than $j$ particles.
Then one defines the reduced marginals accordingly and it is easy to derive the equivalent
of the Grad hierarchy for them.

Note now that the average number of particles is $\langle N \rangle= \mu_\e$. Therefore the low--density limit will correspond to
$\mu_\e \to \infty$ and $\e^2\mu_\e \to l^{-1}>0.$ 

It is not difficult to realize that the validity result of this paper can be formulated and proven also in this context. 

\bigskip
4. We have considered the particle system in the whole space. If we want
the system to be confined in a bounded box, we have to specify the boundary conditions.
Assuming specular reflections, there are additional difficulties which we have to overcome. 
The dynamical flow is only almost everywhere defined
(see \cite {MPPP76}), but this (as for the hard--sphere systems) does not create real difficulties.
However the analysis of the recollisions requires some extra geometrical arguments.

%%%%%%%%%%%%%%%%%%%%%%%%%%%%%%%%%%%%%%%%%%%%%%%%%%%%%%
%%%%%%%%%%%%%%%%%%%%%%%%%%%%%%%%%%%%%%%%%%%%%%%%%%%%%%
%%%%%%%%%%%%%%%%%%%%%%%%%%%%%%%%%%%%%%%%%%%%%%%%%%%%%%
%%%%%%%%%%%%%%%%%%%%%%%%%%%%%%%%%%%%%%%%%%%%%%%%%%%%
%%%%%%%%%%%%%%%%%%%%%%%%%%%%%%%%%%%%%%%%%%%%%%%%%%%%%
\bigskip
{\bf Acknowledgements.} We thank K. Aoki, I. Gallagher, L. Saint--Raymond, H. Spohn, B. Texier and T. Tsuji
for stimulating discussions. 
C. Saffirio has been partially supported by ERC Grant MAQD 240518.
S. Simonella has been partially supported by PRIN 2009 ``Teorie cinetiche e applicazioni'' and by 
Indam--COFUND Marie Curie fellowship 2012, call 10.
%This research has been supported 
%by the International Research Center for Mathematics and Mechanics of Complex Systems MEMOCS, 
%Universit\`{a} di L'Aquila, Cisterna di Latina, 04012, Italy. 
\bigskip
%%%%%%%%%%%%%%%%%%%%%%%%%%%%%%%%%%%%%%%%%%%%%%%%%%%%%%%
%%%%%%%%%%%%%%%%%%%%%%%%%%%%%%%%%%%%%%%%%%%%%%%%%%%%%%%
%%%%%%%%%%%%%%%%%%%%%%%%%%%%%%%%%%%%%%%%%%%%%%%%%%%%%%%
%%%%%%%%%%%%%%%%%%%%%%%%%%%%%%%%%%%%%%%%%%%%%%%%%%%%%%%
%%%%%%%%%%%%%%%%%%%%%%%%%%%%%%%%%%%%%%%%%%%%%%%%%%%%%%%
%%%%%%%%%%%%%%%%%%%%%%%%%%%%%%%%%%%%%%%%%%%%%%%%%%%%%%%

\appendix
\section{Appendix (on the cross--section for the Boltzmann equation)} \label{sec:cs}
\setcounter{equation}{0}    
\def\theequation{A.\arabic{equation}}

In this appendix we give sufficient conditions on the interaction for having a single--valued differential cross--section (and we show
some counterexample). We also study the boundedness properties of the cross--section.
The issue is relevant both to motivate our strategy and to know whether $B$ is a well behaved single--valued function
in the usual form of the Boltzmann equation, Eq. \eqref{eq:BEs}. 

The assumptions on $\Phi$ are those established in Hypothesis \ref{hyp:pot}$'$, but possibly allowing a discontinuity of the first derivative at $|q|=1$.

Consider the planar scattering process of a particle of unit mass.
We use the notations of Section \ref{sec:2bs} and of Figure \ref{fig:2bscatter}. In particular, we denote by $\r$ the impact parameter (by symmetry
we may focus on $0 \leq  \r\leq 1$)  while the scattering angle is $\chi=\p-2\Theta$ and the energy in the laboratory $V^2/2>0.$

The differential cross--section is defined through the map $\r = \r(\Theta,|V|),$ by 
\be
\s_{\Phi} = \frac{\r}{2|\sin(2\Theta)|}\Big|\frac{d\r}{d\Theta}\Big|\;.
\ee
Therefore we need to analyze the invertibility of the map $\Theta(\r).$ 

The classical formula for $\Theta$ is
\bea
\Theta(\r) = \arcsin\r + \r\int_{r_*}^1 dr\frac{1}{r^2\sqrt{1-\frac{2\Phi(r)}{V^2}-\frac{\r^2}{r^2}}}\;, \label{eq:defT}
\eea
where $r_*$ is the minimum distance of the central motion from the origin, satisfying
\be
1-\frac{2\Phi(r_*)}{V^2}-\frac{\r^2}{r_*^2}=0\;.
\ee
For purely repulsive potentials with a singularity at the origin, the limiting values are $\Theta(0)=0$ and $\Theta(1)=\pi/2.$ In general, it is $\Theta = n\p + \Theta'$
for some $\Theta'\in(0,\p),$ where $n$ is the total number of counterclockwise turns that the trajectory makes around the origin (see Section \ref{sec:gener}).

% comment on relevance of theta for our purposes
% regularity

While the first term in the right hand side of (\ref{eq:defT}) is an increasing function of $\r,$ the second term is clearly non
monotonic (in fact it goes smoothly from $0$ to $0$ when $ \r \to 0$ or $\r \to 1$ and hence $r_* \to 1$). 
Following \cite {DP99}, we set  $y=\r/r$ and perform the change of variables
\be
\frac{2\Phi(\frac{\r}{y})}{V^2}+y^2 = \sin^2\f\;,\label{eq:deff1}
\ee
to get
\be
\Theta (\r)= \arcsin\r + \int_{\arcsin\r}^{\pi/2}d\f \frac{\sin\f}{y-\frac{\r \Phi'(\frac{\r}{y})}{V^2y^2}}\;.
\label{eq:defTf}
\ee
The advantage of this formula is that the integrand is
not singular in the integration region and we can easily compute the derivative with respect to $\r$.

%\footnote{
%A more compact expression which is valid (at least) for potentials such that $\r^4\Phi(\r)\rightarrow+\infty$ as $\r\rightarrow 0,$
%is the following:
%%
%\bea
%&& \Theta() = \int_{0}^\a d\r\frac{1}{\r^2\sqrt{\frac{2\Phi(\r)}{\EE_0}-1-\frac{^2}{\r^2}}}\;,\nn\\
%&& \frac{2\Phi(\a)}{\EE_0}-1-\frac{^2}{\a^2} = 0\;. \nn\label{eq:altT}
%\eea
%%
%This formula can be derived from (\ref{eq:defT}) by performing a contour integral in the complex plane of $\r;$ we omit details.
%In turn Eq. (\ref{eq:altT}) can be rewritten by the change of variables $y=/\r,$
%%
%\be
%\frac{2\Phi(\frac{}{y})}{\EE_0}-y^2 = \cosh^2\psi\;,\nn
%\ee
%%
%in the simpler form
%%
%\bea
%\Theta() = -\int_0^{\infty}d\psi \frac{\cosh\psi}{y+\frac{\Phi'(\frac{}{y})}{\EE_0 y^2}}\;.\nn
%\eea
%%
%}.
%None of this formulas makes clear an increasing behaviour (hence invertibility) of $\Theta().$

A straightforward calculation leads to
\bea
&& \frac{d\Theta}{d\r} = \frac{1}{\sqrt{1-\r^2}}\left(1-\frac{1}{1-\frac{\Phi'(1^-)}{V^2\r^2}}\right)\label{eq:dTdI}\\ 
&&\ \ \ \ \ \  + \int_{\arcsin{\r}}^{\p/2} d\f
\frac{\sin\f}{\left(y-\frac{\r}{V^2 y^2}\Phi'(\frac{\r}{y})\right)^3}\Big[\frac{\r}{V^2y^2}\Phi''\left(\frac{\r}{y}\right)+
\frac{2}{V^2y}\Phi'\left(\frac{\r}{y}\right)+\frac{\r}{V^4y^4}\left(\Phi'\left(\frac{\r}{y}\right)\right)^2\Big] \nn
\eea
for $0<\r<1,$ where $\Phi'(1^-)$ indicates the limit of the derivative as $|q|\rightarrow 1$ from below. 

In formula \eqref {eq:dTdI} we are also considering the case in which $\Phi$ has a discontinuity of
the first derivative in $|q|=1$ as it is the case of the inverse power law potential restricted to the unitary interval
treated in \cite{DP99}.
However, for the case of smooth potentials as the ones considered in the present paper, 
the first term in the right hand side of Eq. \eqref {eq:dTdI} is absent.

\bigskip
The following considerations can be deduced from Eq. (\ref{eq:dTdI}).

\bigskip
\noindent {\em 1)} The ratio $\r/y\rightarrow g(\f)$ as $\r\rightarrow 0,$ where $g$ is a positive function of $\f$
which form depends on $\Phi$ and $V^2.$ Then the extremal values of our derivative are:
\bea
&& \frac{d\Theta}{d\r} \underset{\r\rightarrow 0}{\longrightarrow}(1-\d_{\Phi'(1^-),0}) + \int_0^{\pi/2}d\f \frac{g(\f)\sin\f}{
\left(\frac{g(\f)}{V^2}|\Phi'(g(\f))|\right)^3}\frac{(\Phi'(g(\f)))^2}{V^4} \in (0,+\infty]\;; \nn\\
&& \frac{d\Theta}{d\r} \underset{\r\rightarrow 1}{\longrightarrow} \left\{\begin{array}{ccc}
+\infty, & & \Phi'(1^-) \neq 0 \\ 0, & & \Phi'(1)=0 \end{array}\right.\;. \label{eq:limdT}
\eea
%
%Given $\in(0,1),$ it is $ \frac{d\Theta}{d}>0$ if the square bracket in Eq. (\ref{eq:dTdI}) is positive for any $\f\in(\arcsin,\pi/2).$

\bigskip
\noindent {\em 2)} The monotonicity property $ \frac{d\Theta}{d\r}>0$ translates in a quite complicated condition on the potential $\Phi.$
A convenient sufficient condition is given by the following assertion: 
{\em In the considered class of potentials, if for all $q$ with $|q|\in(0,1)$
\be
|q|\Phi''(|q|) + 2\Phi'(|q|) \geq 0\;, \label{eq:suffcond}
\ee
then $ \frac{d\Theta}{d\r}>0$ for all $\r\in(0,1), V^2>0.$} 
This condition is derived also in \cite{AS94}.

\begin{figure}[htbp]
\centering
\includegraphics[width=0.6\textwidth]{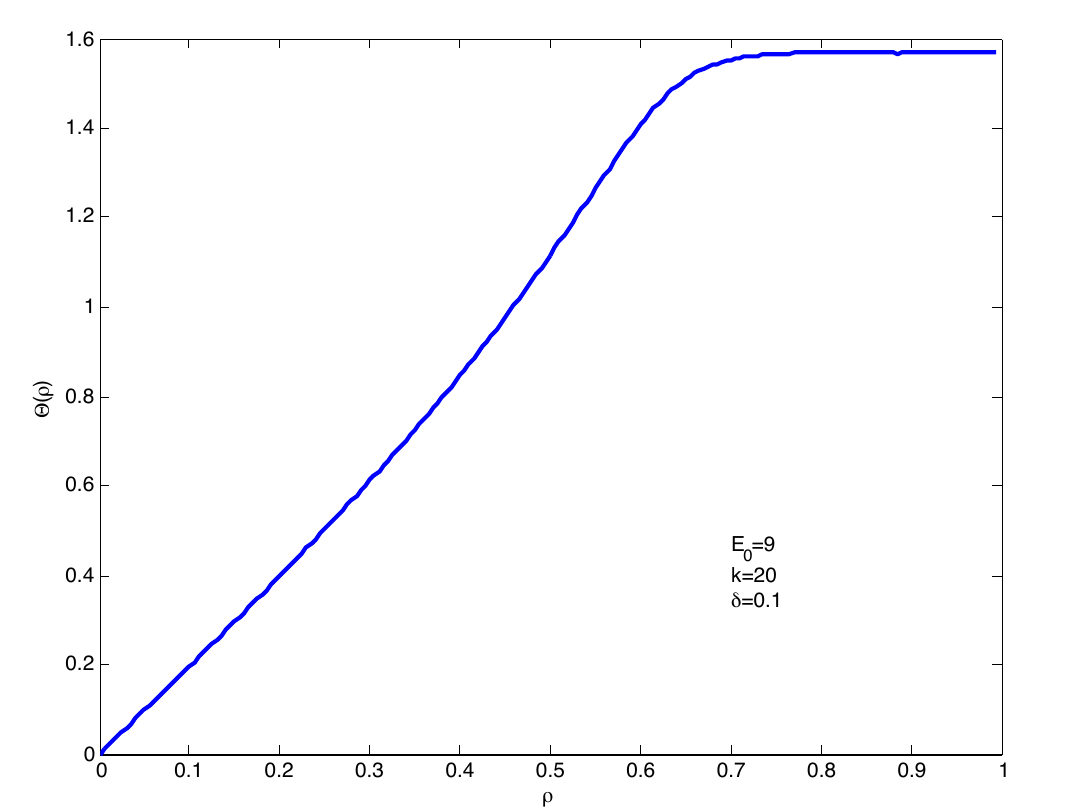}
\caption{\small{Map $\Theta(\r)$ for the potential given by Eq. (\ref{eq:smoothex}), with $\d=0.1, k=20$ and $E_0=V^2/2=9.$}}
\label{fig:smoothex}
\end{figure}

Condition (\ref{eq:suffcond}) can be easily checked for a large subset of potentials. 
For instance any potential of the form $\Phi(q) = \left(\frac{1}{|q|^k}-1\right)\d_{|q|<1}, k \geq 1,$ satisfies the 
condition, hence has strictly monotonic map. Cases which are smooth in $|q|=1$ can be constructed from the previous
by using a smooth junction. For instance\footnote{Observe that this is a function $C^1(\RRR^d)$ with a jump in the second derivative
for $|q|=1-\d.$ The parameters $k$ and $\d$ can be arranged in order to eliminate this discontinuity (e.g. $\d=1/10, k=71$). Nevertheless,
all our discussions are still valid if in the initial assumptions on the potential we require that $\Phi''$ is just piecewise continuous 
and bounded outside any ball centered in the origin.}
\be
\Phi(q) = 
\left\{
\begin{array}{cc}
e^{-\frac{1}{\d}}\left(\frac{(1-\d)^{k+1}}{\d^2 k}\frac{1}{|q|^k}+1-\frac{1-\d}{\d^2k}\right)  &  0<|q|<1-\d    \\
e^{-\frac{1}{1-|q|}}  &  1-\d \leq |q| <1 \;,    \\
0  &    |q| \geq 1  
\end{array}
\right. \label{eq:smoothex}
\ee
where $k\geq 1$ and $0<\d<1/3$ (see Fig. \ref{fig:smoothex}).

\bigskip
\noindent {\em 3)} The monotonicity property $ \frac{d\Theta}{d\r}>0$ is in general not true
when condition (\ref{eq:suffcond}) is violated. A first example is any smooth, positive, decreasing and bounded potential (for which $\Theta(0) = \Theta (1) = 0,$
that implies the existence of at least two monotonicity branches).

\begin{figure}[htbp] 
\centering
\includegraphics[width=0.55\textwidth]{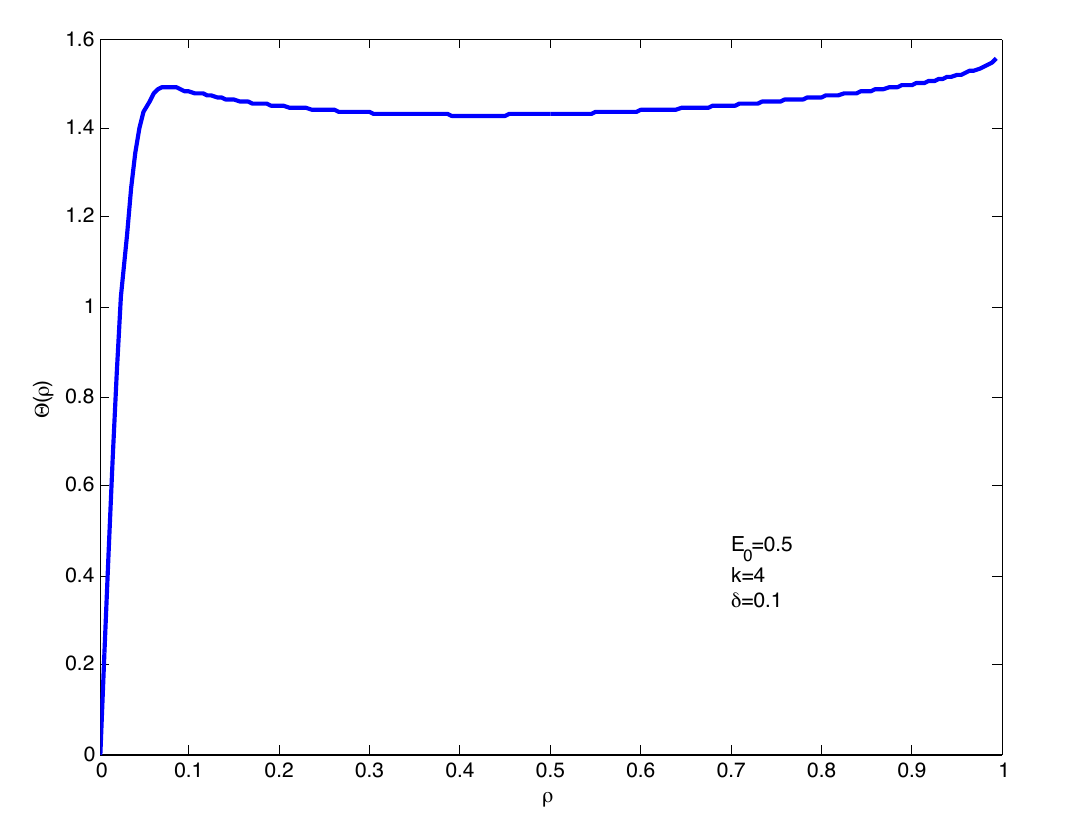}
\caption{\small{Map $\Theta(\r)$ for the potential given by Eq. (\ref{eq:countpot}), with $\d=0.1,\ k=4$ and $E_0=V^2/2=0.5.$}}
\label{fig:countpot}
\end{figure}

\bigskip
We give two different examples of potentials singular at the origin.
\begin{itemize}
\item Formula (\ref{eq:dTdI}) indicates that the sign of the second derivative of $\Phi$ is relevant when we ask about monotonicity 
of the map. In fact, examples of non monotonic maps can be constructed when $\Phi''$ is not 
always positive, for instance by taking $\Phi$ very close to the characteristic function of $|q|<1.$
If we consider the function 
\begin{equation}
\Phi(q)=-\e \tan \left(\left(\arctan \frac{1}{\e}+\frac{\pi}{2}\right)|q|-\frac{\pi}{2}\right)+1\;,\label{eq:arctg}
\end{equation}
numerical simulations show that the map $\Theta(\r)$ is non monotonic for $\e<<1$ 
as shown in Fig. \ref{fig:arctg}.
\begin{figure}[htbp]\centering
\includegraphics[width=0.55\textwidth]{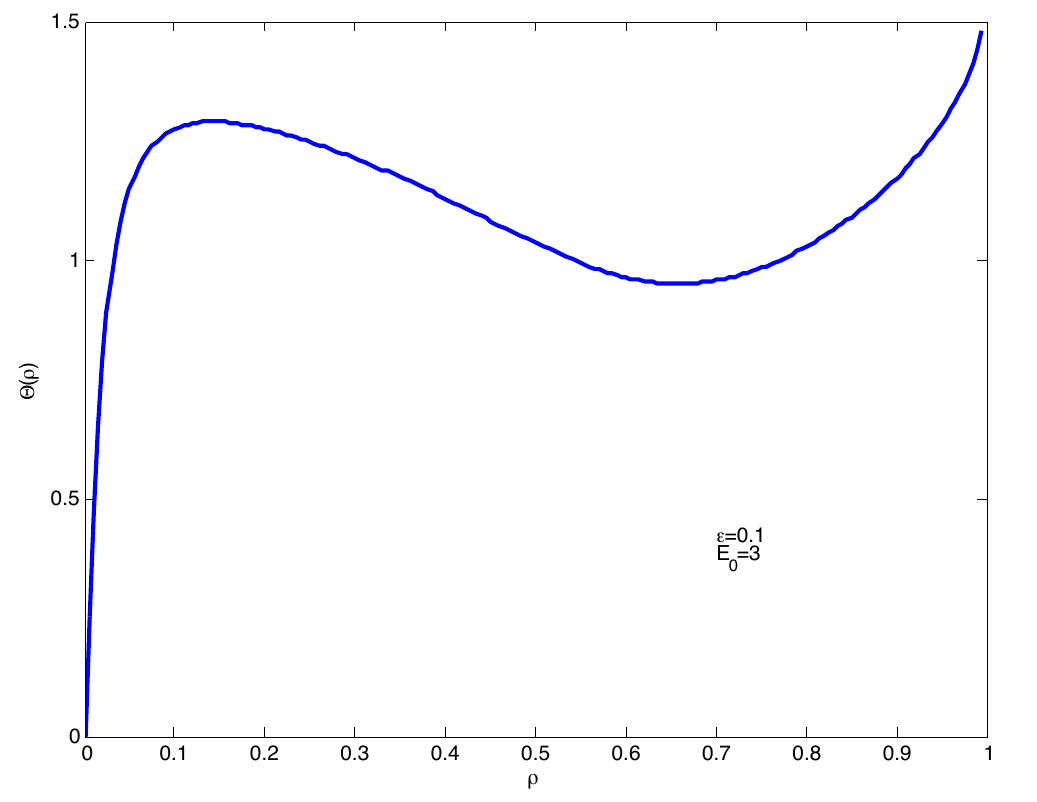} 
\caption{\small{Map $\Theta(\r)$ for the potential given by Eq. (\ref{eq:arctg}), with $\e=0.1$ and $E_0=V^2/2=3.$}}
\label{fig:arctg}
\end{figure}

\item Even when $\Phi''$ is nonnegative, the mapping can be non monotonic if Eq. (\ref{eq:suffcond}) fails, an example
being
\be
\Phi(q)=\left\{
\begin{array}{cc}
\frac{\d^{k+2}}{k |q|^k}+\d-\d^2(1+\frac{1}{k})  &  0<|q|<\d    \\
\d(1-|q|)  &  \d \leq |q| <1 \;,    \\
0  &    |q| \geq 1  
\end{array}
\right. \label{eq:countpot}
\ee
We checked numerically the non monotonicity of $\Theta(\r)$ in the case $\d = 0.1, k=4$ (see Fig. \ref{fig:countpot}).
Another example similar to the previous one but with continuous derivative in $|q|=1$ can be constructed again
by using a smooth junction.
\end{itemize}

\medskip
\noindent {\em 4)} Even when single--valued, we may have a singularity of the cross--section any time the 
condition $ \frac{d\Theta}{d\r}>0$ is violated. 
In particular, if $\Phi'(1)=0$ (i.e. the force is smooth) and $\Theta(\r)$ is strictly monotonic, we still have a divergence of $\frac{d\r}{d\Theta}$ 
for $\Theta$ near to $\pi/2$ ($\r=1$), that is
\be
\Big|\Big|\sin(2\Theta) \s_{\Phi}\Big|\Big|_{\infty}=+\infty\;.
\ee
%

%%%%%%%%%%%%%%%%%%%%%%%%%%%%%%%%%%%%%%%%%%%%%%%%%%%%%%%
%%%%%%%%%%%%%%%%%%%%%%%%%%%%%%%%%%%%%%%%%%%%%%%%%%%%%%%
%%%%%%%%%%%%%%%%%%%%%%%%%%%%%%%%%%%%%%%%%%%%%%%%%%%%%%%
%%%%%%%%%%%%%%%%%%%%%%%%%%%%%%%%%%%%%%%%%%%%%%%%%%%%%%%
%%%%%%%%%%%%%%%%%%%%%%%%%%%%%%%%%%%%%%%%%%%%%%%%%%%%%%%
%%%%%%%%%%%%%%%%%%%%%%%%%%%%%%%%%%%%%%%%%%%%%%%%%%%%%%%

\end{document}